\documentclass[preprint,12pt]{emulateapj}
\usepackage{epsfig}
\usepackage{natbib}
\usepackage{amssymb}
\usepackage{amsbsy}
\usepackage{amsmath}
\usepackage{natbib}
\usepackage[mathcal]{euscript}
\usepackage{lscape}
\bibpunct{(}{)}{,}{a}{}{,}

\begin{document}
 
\def\simlt{\vcenter{\hbox{$<$}\offinterlineskip\hbox{$\sim$}}}
\def\simgt{\vcenter{\hbox{$>$}\offinterlineskip\hbox{$\sim$}}}
\def\etal{et al.\ }

\title{CSI 2264: Simultaneous optical and infrared light curves of
young disk-bearing stars in NGC~2264 with {\em CoRoT} and {\em Spitzer}-- evidence for multiple origins of
variability\footnotemark[*]} \footnotetext[*]{Based on data
from the {\em Spitzer} and {\em CoRoT} missions. The {\em CoRoT} space mission was
developed and is operated by the French space agency CNES, with particpiation of ESA's RSSD and Science Programmes, Austria,
Belgium, Brazil, Germany, and Spain.}

\author{Ann Marie Cody\altaffilmark{1}, John Stauffer\altaffilmark{1},
 Annie Baglin\altaffilmark{2}, Giuseppina Micela\altaffilmark{3}, Luisa M. Rebull\altaffilmark{1}, Ettore
 Flaccomio\altaffilmark{3}, Mar\'{i}a Morales-Calder\'{o}n\altaffilmark{4}, Suzanne Aigrain\altaffilmark{5}, J\`{e}r\^{o}me
 Bouvier\altaffilmark{6}, Lynne A. Hillenbrand\altaffilmark{7}, Robert Gutermuth\altaffilmark{8}, Inseok Song\altaffilmark{9}, Neal Turner\altaffilmark{10},
 Silvia H.\ P.\ Alencar\altaffilmark{11}, Konstanze Zwintz\altaffilmark{12}, Peter Plavchan\altaffilmark{13}, John Carpenter\altaffilmark{7},
 Krzysztof Findeisen\altaffilmark{7}, Sean Carey\altaffilmark{1}, Susan Terebey\altaffilmark{14}, Lee Hartmann\altaffilmark{15}, Nuria Calvet\altaffilmark{15},
 Paula Teixeira\altaffilmark{16}, Frederick J. Vrba\altaffilmark{17}, Scott Wolk\altaffilmark{18}, Kevin Covey\altaffilmark{19}, Katja Poppenhaeger\altaffilmark{18}, 
 Hans Moritz G\"unther\altaffilmark{18}, Jan Forbrich\altaffilmark{16,18},
 Barbara Whitney\altaffilmark{20}, Laura Affer\altaffilmark{3}, William Herbst\altaffilmark{21}, 
 Joseph Hora\altaffilmark{18}, David Barrado\altaffilmark{4}, Jon Holtzman\altaffilmark{22}, Franck Marchis\altaffilmark{23},  
 Kenneth Wood\altaffilmark{24}, Marcelo Medeiros Guimar\~aes\altaffilmark{25}, Jorge Lillo Box\altaffilmark{4}, Ed Gillen\altaffilmark{5},
 Amy McQuillan\altaffilmark{26}, Catherine Espaillat\altaffilmark{27}, Lori Allen\altaffilmark{28},
 Paola D'Alessio\altaffilmark{29}, Fabio Favata\altaffilmark{30}}
 
\email{amc@ipac.caltech.edu}
\altaffiltext{1}{Spitzer Science Center, California Institute of Technology, 1200 E California Blvd., Pasadena, CA
91125, USA}
\altaffiltext{2}{LESIA, Observatoire de Paris-Meudon, 5 place Jules 
Janssen, 92195, Meudon, France}
\altaffiltext{3}{INAF - Osservatorio Astronomico di Palermo, Piazza
del Parlamento 1, 90134, Palermo, Italy}
\altaffiltext{4}{Centro de Astrobiolog\'ia, Dpto. de
Astrof\'isica, INTA-CSIC, PO BOX 78, E-28691, ESAC Campus, Villanueva de
la Ca\~nada, Madrid, Spain}
\altaffiltext{5}{Department of Astrophysics, Denys Wilkinson Building, University of Oxford, Oxford OX1 3RH, UK}
\altaffiltext{6}{UJF-Grenoble 1 / CNRS-INSU, Institut de Plan\'{e}tologie
et d'Astrophysique de Grenoble (IPAG) UMR 5274, Grenoble, F-38041, France}
\altaffiltext{7}{Astronomy Department, California Institute of
Technology, Pasadena, CA 91125, USA}
\altaffiltext{8}{Dept. of Astronomy, University of Massachusetts, Amherst, MA  01003}
\altaffiltext{9}{Department of Physics and Astronomy, The University
of Georgia, Athens, GA 30602-2451, USA}
\altaffiltext{10}{Jet Propulsion Laboratory, California Institute
of Technology, Pasadena, CA 91109, USA}
\altaffiltext{11}{Departamento de F\'{\i}sica -- ICEx -- UFMG, Av. Ant\^onio Carlos
6627, 30270-901, Belo Horizonte, MG, Brazil}
\altaffiltext{12}{Instituut voor Sterrenkunde, K. U. Leuven, Celestijnenlaan 200D, 3001 Leuven, Belgium}
\altaffiltext{13}{Infrared Processing and Analysis Center, California Institute of
Technology, Pasadena, CA 91125, USA}
\altaffiltext{14}{Department of Physics and Astronomy, 5151 State University
Drive, California State  University at Los Angeles, Los Angeles, CA 90032}
\altaffiltext{15}{Department of Astronomy, University of Michigan,
500 Church Street, Ann Arbor, MI 48105, USA}
\altaffiltext{16}{University of Vienna, 
Department of Astrophysics, T\"urkenschanzstr. 17, 1180 Vienna, Austria}
\altaffiltext{17}{U.S. Naval Observatory, Flagstaff Station, 10391
West Naval Observatory Road, Flagstaff, AZ 86001, USA}
\altaffiltext{18}{Harvard-Smithsonian Center for
Astrophysics, Cambridge, MA 02138, USA}
\altaffiltext{19}{Lowell Observatory, 1400 West Mars Hill
Road, Flagstaff, AZ 86001, USA}
\altaffiltext{20}{Astronomy Department, University of Wisconsin-
Madison, 475 N. Charter St., Madison, WI 53706, USA}
\altaffiltext{21}{Astronomy Department, Wesleyan University,
Middletown, CT 06459, USA}
\altaffiltext{22}{Department of Astronomy, New Mexico State University, Box 30001, Las Cruces, NM 88003, USA}
\altaffiltext{23}{Carl Sagan Center at the SETI Institute, 189 Bernardo Av., Mountain View, CA 94043, USA}
\altaffiltext{24}{School of Physics and Astronomy, University of St Andrews, North Haugh, St Andrews, Fife KY16 9AD, UK }
\altaffiltext{25}{Departamento de F\'{i}sica e Matem\'{a}tica - UFSJ - Rodovia MG 443 KM7 -36420-000 - Ouro Branco - MG - Brazil}
\altaffiltext{26}{School of Physics and Astronomy, Raymond and Beverly Sackler, Faculty of Exact Sciences, Tel Aviv University, 69978 Tel Aviv, 
Israel}
\altaffiltext{27}{Department of Astronomy, Boston University, 725 Commonwealth Avenue, Boston, MA 02215, USA}
\altaffiltext{28}{National Optical Astronomy Observatories, Tucson, AZ, USA}
\altaffiltext{29}{Centro de Radioastronomía y Astrofísica, UNAM, Apartado Postal 3-72 (Xangari), 58089 Morelia, Michoacán, Mexico}
\altaffiltext{30}{European Space Agency, 8-10 rue Mario Nikis, F-75738 Paris Cedex 15, France}

\begin{abstract}

We present the Coordinated Synoptic Investigation of NGC 2264, a continuous 30-day multi-wavelength photometric
monitoring campaign on more than 1000 young cluster members using 16 telescopes.  The
unprecedented combination of multi-wavelength, high-precision, high-cadence, and long-duration data opens a new
window into the time domain behavior of young stellar objects. Here we provide an overview of the observations, focusing on
results from {\em Spitzer} and {\em CoRoT}. The highlight of this work is detailed analysis of 162 classical T Tauri stars for
which we can probe optical and mid-infrared flux variations to 1\% amplitudes and sub-hour timescales.  We present a
morphological variability census and then use metrics of periodicity, stochasticity, and symmetry to statistically separate the
light curves into seven distinct classes, which we suggest represent different physical processes and geometric effects. We
provide distributions of the characteristic timescales and amplitudes, and assess the fractional representation within each
class.  The largest category ($>$20\%) are optical ``dippers'' having discrete fading events lasting $\sim$1--5 days.  The
degree of correlation between the optical and infrared light curves is positive but weak; notably, the independently assigned
optical and infrared morphology classes tend to be different for the same object.  Assessment of flux variation behavior
with respect to (circum)stellar properties reveals correlations of variability parameters with H$\alpha$ emission and with
effective temperature. Overall, our results point to multiple origins of young star variability, including circumstellar
obscuration events, hot spots on the star and/or disk, accretion bursts, and rapid structural changes in the inner disk.

\end{abstract}

\keywords{}

\section{Introduction}

Photometric variability on a variety of timescales is a long appreciated
characteristic of young stellar objects (``YSOs''). Since the initial
association of brightness fluctuations with emission line objects near
molecular clouds \citep{1949ApJ...110..424J}, it has been inferred that YSO variability
arises from a combination of physical processes operating at and near
the stellar surface. The weak-lined T Tauri stars (WTTSs), so
called for their lack of spectroscopic accretion signatures, tend to
display stable sinusoidal light curves attributed to cool magnetic
spots on the stellar surface \citep[e.g.,][]{1999AJ....117.2941S,2008A&A...479..827G,2009A&A...502..883R,2009A&A...508.1313F}. The classical T Tauri stars
(CTTSs), on the other hand, typically exhibit much more complex time
domain behavior, with light curves categorized as stochastic \citep[e.g.,][]{2008MNRAS.391.1913R,2011MNRAS.410.2725S},
intermittently fading \citep[e.g.,][]{2010ApJS..191..389C,2010A&A...519A..88A}, or semi-periodic \citep[e.g.,][]{1993AJ....106.1608V,1994AJ....108.1906H}. Most of the
photometric monitoring surveys conducted over the past few decades
have focused on optical or near-infrared variability on timescales of
days to years \citep[e.g.,][]{1993A&A...272..176B,1994AJ....108.1906H,1996AJ....112.2168S,1998A&AS..128..561B,
2000AJ....120..349H,2004AJ....127.1602C,2004AJ....127.2228M,2004A&A...424..857C,2007A&A...461..183G,2008A&A...485..155A,
2012MNRAS.420.1495S,2013ApJ...773..145W,2013arXiv1309.5300P}. While they showed that brightness fluctuations are common at the
1--10\% level, sparse or uneven time sampling often precluded full
assessment of variability, for CTTSs in particular.
Ultimately, a full understanding of the time-domain properties of
young stars is needed to inform models of their interaction with
surrounding disks, the accretion process, as well as structure and
geometry of star-disk systems.

Aperiodic or partially periodic variability in CTTSs has been attributed to a number of mechanisms, 
including obscuration by circumstellar material \citep[e.g.,][]{1994AJ....108.1906H,1999A&A...345L...9C,2010A&A...519A..88A}, instabilities in the 
accretion shock at the stellar surface \citep{2008MNRAS.388..357K}, unsteady accretion and hot spot 
evolution
\citep[e.g.,][]{1996A&A...310..143F,1999ApJ...510..892S,2002AJ....124.1001C,2009MNRAS.398..873S,2010A&A...517A..16V}, 
and instabilities in the accretion disk 
\citep[e.g.,][]{1988ApJ...330..350B,1989ARA&A..27..351B,2007A&A...463.1017B}. A number of parameters, 
including magnetic field strength and shape \citep{2012ApJ...756...68C}, disk structure 
\citep{2012ApJ...748...71F,2013ApJ...773..145W}, stellar mass \citep{1999AJ....118.1043H}, and rotation rate 
\citep{2007A&A...461..183G} also appear to influence variability properties. Given these complexities, few 
theoretical models offer detailed, verifiable predictions on the time domain behavior of young stars. Some 
attempts to match the optical and near-infrared time-domain properties of young stars to simple models 
\citep[e.g.,][]{1994AJ....108.1906H,2001AJ....121.3160C,2009MNRAS.398..873S,2012ApJ...755...65R,2012PASP..124.1137F} have noted photometric 
behavior that is largely consistent with variable accretion, hot spots, or obscuration. 
Nevertheless, these very different physical scenarios could not be distinguished unambiguously from the 
limited time and wavelength coverage of ground-based data.

Recently, it has become increasingly clear that YSOs are not only variable on timescales as short as hours 
\citep{2008MNRAS.391.1913R}, but these brightness changes also appear at a wide range of wavelengths in 
individual objects \citep[][and references therein]{2011ASPC..448....5R}.  Monitoring of disk-bearing stars by 
\citet{2002A&A...384.1038E} revealed optical and near-IR flux changes on 1--2 day timescales. While they speculated 
that changes in disk structure could produce disk emission or 
scattered light variations, the rapidity is difficult to explain. \citet{2013ApJ...773..145W} too found a variety 
of near-IR variability, but on much longer timescales of multiple months.
At longer wavelengths, 
\citet{2005ApJ...630..381B} found large-amplitude mid-infrared variability preferentially among disks at 
early evolutionary stages. Instruments aboard the {\em Spitzer Space Telescope} \citep{2004ApJS..154....1W} 
and {\em Herschel} \citep{2010A&A...518L...1P} have also enabled \citet{2009ApJ...704L..15M}, \citet{2011ApJ...728...49E},
\citet{2011ApJ...733...50M}, and \citet{2012ApJ...753L..35B}
to uncover mid- to far-infrared brightness fluctuations in disk-bearing young
stars. Complementary modeling efforts 
such as those by \citet{2003ApJ...594L..47D}, \citet{2010ApJ...719.1733F}, and \citet{2011MNRAS.416..416R} 
have begun to offer detailed descriptions of inner disk dynamics and star-disk interaction but nevertheless 
require more extensive input from observations on more varied timescales and wavelengths. Despite the 
headway in matching observed brightness fluctuations to physical models, the sparseness or non-simultaneity 
of data taken in different bands has hindered a full explanation of young star variability.

The ongoing Young Stellar Object Variability (YSOVAR) project \citep[][Rebull 2014]
{2011ApJ...733...50M} is exploring the variability properties of young stars in several young clusters at 
an unprecedented combination of cadence, photometric precision, and wavelength coverage, particularly in the
infrared. Variable accretion and 
extinction, as well as disk warps, shadowing, and magnetic instabilities have been cited as plausible mid-IR 
variability mechanisms, and YSOVAR offers the best opportunity to untangle them. On the heels of this project, 
we have performed photometric monitoring of young NGC~2264 cluster members using the {\em Spitzer} Infrared 
Array Camera \citep[IRAC;][]{2004ApJS..154...10F} and the Convection, Rotation and Planetary Transits satellite 
\citep[{\em CoRoT};][]{2006ESASP1306...33B} simultaneously. This campaign-- the Coordinated Synoptic 
Investigation of NGC~2264 (CSI 2264)-- comprises a unique cooperative effort including 15 ground and space-based 
telescopes, listed in Table 1. In this paper we focus exclusively on the results of optical and infrared 
photometric monitoring with {\em CoRoT} and {\em Spitzer} as part of CSI~2264; discussion of data acquired from other 
instruments is deferred to subsequent papers.

With sub-1\% photometry at cadences down to a few minutes, these two
space telescopes have provided the first set of simultaneous fully sampled light
curves in the optical and infrared. Whereas previous studies of YSOs in
NGC~2264 \citep[e.g.,][]{2004A&A...417..557L,2004AJ....127.2228M} and other regions had insufficient time sampling to 
identify more than generic aperiodic flux variations, we are able to resolve brightness fluctuations on all relevant
timescales expected from significant variability mechanisms. From the
exquisite {\em CoRoT} and {\em Spitzer} time series, we present here a
census of light curve morphologies based on an unbiased set of disk
bearing stars in NGC~2264. We first develop a visual classification scheme, which we
then confirm via quantitative metrics that can be determined on any
light curve without manual intervention. With distinct morphology
classes in hand, we proceed to provide quantitative measures of timescales, periodicities, amplitudes, and
correlations, as constraints on the theoretical models currently in
development. We use only measurements from {\em CoRoT} and {\em Spitzer} light curves to explore connections between
morphology and physical processes in this paper. In future work, we will incorporate
additional multiwavelength and spectroscopic data collected by the CSI 2264 project to 
futher constrain variability mechanisms.

The structure of this paper is as follows: In Section 2 we describe
the stellar sample; in Sections 3 and 4, the observations and data
reduction. In Section 5, we present our classification
of optical/infrared light curve morphologies, and in Section 6 we provide a
statistical characterization of the variables. In Section 7 we
investigate the correlation of optical and infrared light curve
morphologies, subsequently exploring the relationship between
variability properties and stellar and disk observables (Section 8).

\LongTables
\begin{deluxetable*}{ccccc}
\tabletypesize{\scriptsize}
\tablecolumns{10}
\tablewidth{0pt}
\tablecaption{\bf Coordinated Synoptic Investigation of NGC 2264: observations}
\tablehead{
\colhead{Telescope} & \colhead{Instrument} & \colhead{Dates} &
\colhead{Band(s)} &  \colhead{Time Sampling} 
}
\startdata
{\em Spitzer} & IRAC/mapping & Dec. 3, 2011--Jan. 1, 2012 & 3.6~$\mu$m,
4.5~$\mu$m & 101 min \\
{\em Spitzer} & IRAC/staring & Dec. 3; Dec. 5--6; Dec. 7--8;
Dec. 8--9, 2011 & 3.6~$\mu$m, 4.5~$\mu$m & 15~s \\
{\em CoRoT} & E2 CCD & Dec. 1, 2011-- Jan 3, 2012 & 3000-10000\AA  &
32~s (high cadence), 512~s\\
{\em MOST}  & Science CCD & Dec. 5, 2011--Jan. 14, 2012&
3500--7500\AA  &   24.1, 51.2~s$^1$ \\
{\em Chandra} & ACIS-I & Dec. 3, 2011--Dec. 9, 2011 & 0.5--8 keV &
$\sim$3.2~s$^2$ \\
VLT  &  Flames, UVES & Dec. 4, 2011--Feb. 29, 2012 & 4800--6800\AA  &  20--22 epochs  \\
CFHT  & MegaCam & Feb. 14, 2012--Feb. 28. 2012 &  $u$,$r$ & 30 epochs \\
PAIRITEL  &  2MASS camera & Dec. 5, 2011--Jan. 3, 2012  & $J$,$H$,$K$  & 1--12 epochs \\
USNO 40-inch telescope & CCD  & Nov. 22, 2011--Mar. 9, 2012 & $I$  & 912--1026 epochs \\
Super-LOTIS & CCD &  Nov. 11, 2011--Mar. 1, 2012  & $I$  & 495--522 epochs \\
NMSU 1m telescope & CCD & Oct. 12, 2011--Mar. 4, 2012 & $I$  & 47--54 epochs \\
Lowell 31-inch telescope & CCD & Oct. 12, 2011--Jan. 14, 2012  & $I$  & 44 epochs\\
OAN 1.5m telescope & CCD & Jan. 10, 2012--Feb. 15, 2012 & $V$, $I$ & 23--28 epochs \\
KPNO 2.1m telescope & FLAMINGOS & Dec. 16, 2011--Jan. 3, 2012 & $J$,$H$,$K_S$ & 40--52 epochs \\
FLWO 60-inch telescope & KeplerCam & Nov. 30, 2011--Jan. 26, 2012 & $U$  & 35--60 epochs \\
ESO 2.2m telescope  & WFI  &  Dec. 24, 2012--Dec. 29, 2011  &  $U$, $V$, $I$  & 25--45 epochs \\
CAHA 3.5m telescope & Omega 2000 & Dec. 5, 2011--Feb. 18, 2012   & $J$,$H$,$K$ & 35 epochs \\   
CAHA 3.5m telescope & LAICA      & Jan. 25-26, 2012 & $u$, $r$   & 20 epochs 
\enddata
\tablecomments{We provide the details of observing runs associated
  with the Coordinated Synoptic Investigation of NGC~2264. Some of the
  data were not used due to non-photometric conditions. We also note that the fields of view
were not the same for all instruments. In many cases we monitored a number of 5--20\arcmin\ regions
at slightly different cadences, depending on weather. In the time sampling column, we show
  either time between each datapoint, or the total number of
  datapoints per field for the unevenly sampled ground-based runs. $^1 $We note that
  MOST data for each of two fields were taken over only about half of
each 101-minute orbit. Observations with a number of epochs listed
were not taken at regular intervals, because of weather
interruptions, and many of them involved multiple fields. $^2$The cadence listed for {\em Chandra} 
observations is the temporal resolution of photon arrival times during the exposures.}
\end{deluxetable*}

\section{Sample selection}

\begin{figure}
\begin{center}
\includegraphics[scale=0.55]{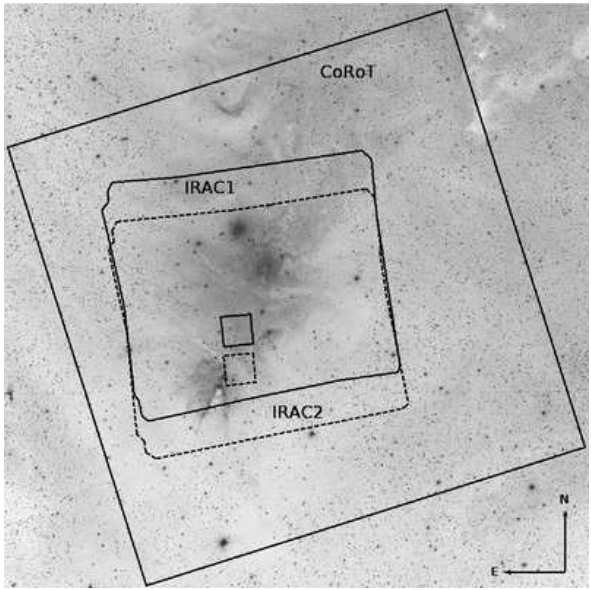}
\end{center}
\caption{\label{fov}Field of view for IRAC and {\em CoRoT} observations of NGC
  2264 carried out in December 2011. The underlying image is from the
  Digitized Sky Survey \citep{1990AJ.....99.2019L}. The smaller square regions
  represent the IRAC staring fields in channel 1 (solid line) and
  channel 2 (dashed line). The large region observed by {\em CoRoT} is
1.3\arcdeg$\times$1.3\arcdeg.}
\end{figure}

At $\sim$760~pc \citep{2000AJ....120..894P} and $\sim$1--5~Myr 
\citep{2002AJ....123.1528R,2008hsf1.book..966D}, NGC~2264 provides a rich selection of pre main sequence 
objects with masses from the substellar regime through $\sim$7~$M_\odot$. It includes areas of recent star 
formation where embedded stars are visible only at infrared and longer 
wavelengths \citep{2012A&A...540A..83T}, as well as lower extinction regions with one to two thousand 
optical sources.  The membership of this cluster is well characterized 
\citep{1997AJ....114.2644S,2002AJ....123.1528R,2004AJ....127.2659R,2005AJ....129..829D}, encompassing a large 
population for time series studies. The bulk of members are within a one square degree region of sky, comparable 
to the fields of view covered by {\em CoRoT} and {\em Spitzer}, as shown in Figure~\ref{fov}. Our goal in
studying variability among the YSOs in NGC~2264 is to
analyze multiwavelength flux behavior in a relatively unbiased sample
of disk-bearing members. Since {\em CoRoT} only observes using pre-selected pixel masks, we mined all of the available
published data in advance to identify highly probable cluster members with previously derived stellar
properties. A summary of our membership selection criteria is available in the
Appendix; these resulted in a list of over $\sim$1500 cluster members.  For convenience of identification,  we have 
assembled a comprehensive catalog of all cluster members, candidates, and field
stars in the NGC~2264 field of view (FOV), using
the id structure ``CSI Mon,'' or ``Mon'' in shorthand. The
full catalog of over 100,000 sources will be deferred to
a later publication, and we list only the objects considered in
this work by their Mon numbers.

We selected 489 high probability cluster members for observation by
{\em CoRoT} in 2011, along with a {\em Spitzer} FOV
encompassing 1266 members. Selection was done by priority, with
450 members and candidate members (i.e., meeting only one of two
membership criteria listed in the Appendix) previously observed by
{\em CoRoT} during its 2008 SRa01 run having highest priority. The 
second level of priority consisted of 322 CTTSs not previously
observed by {\em CoRoT}, and with $I<17$. Lower priority objects were
all WTTSs not previously observed by {\em CoRoT}. 
Previous variability was not a selection criterion, and
therefore we expect the CTTS sample analyzed in this paper to be nearly unbiased. The
break-down of CTTSs versus WTTSs, however, is probably not reflective
of the cluster distribution as a whole. To retain only those members
with circumstellar disks, we next apply additional criteria, as discussed
below. It is this subset of stars whose variability properties we
will study in Sections~5--8. For reference, we provide the numbers of
member and field objects in the {\em Spitzer} and {\em CoRoT}
samples in Table 2. 

\LongTables
\begin{deluxetable}{rccc}
\tabletypesize{\scriptsize}
\tablecolumns{10}
\tablewidth{0pt}
\tablecaption{\bf CSI 2264 sample subsets}
\tablehead{
\colhead{} & \colhead{{\em Spitzer}} & \colhead{{\em CoRoT}} &
\colhead{Both} 
}
\startdata
Total stars observed... & 19892 & 4235 & 1303 \\
  Field stars............. & 17043 & 2129 & 184 \\
  Candidates............ & 1583 & 1617 & 665 \\
  {\bf Members}............. & 1266 & 489 & 454 \\
    Class III............ & 574 & 305 & 288 \\
    Unknown SED.. & 59 & 8 & 3 \\
    {\bf Disk bearing}.. & 633 & 176 & 163* \\
      Class II/III.. & 90 & 24 & 23 \\
      Class II........ & 389 & 140 & 129 \\ 
      Flat SED..... & 65 & 9 & 8 \\
      Class I......... & 89 & 3 & 3 
\enddata
\tablecomments{The number of stars observed in each subset of the CSI 2264 sample. Members were selected according to the criteria in the Appendix,
and candidates are objects that satisfy only one of those criteria. SED classes are derived in Section 2.1. *Group discussed in this paper (i.e., disk bearing members in both {\em CoRoT}
and {\em Spitzer}); the actual sample size was 162, after removal of a low quality light curve with few data points.}
\end{deluxetable}

\begin{figure*}[!t]
\begin{center}
\includegraphics[scale=0.8]{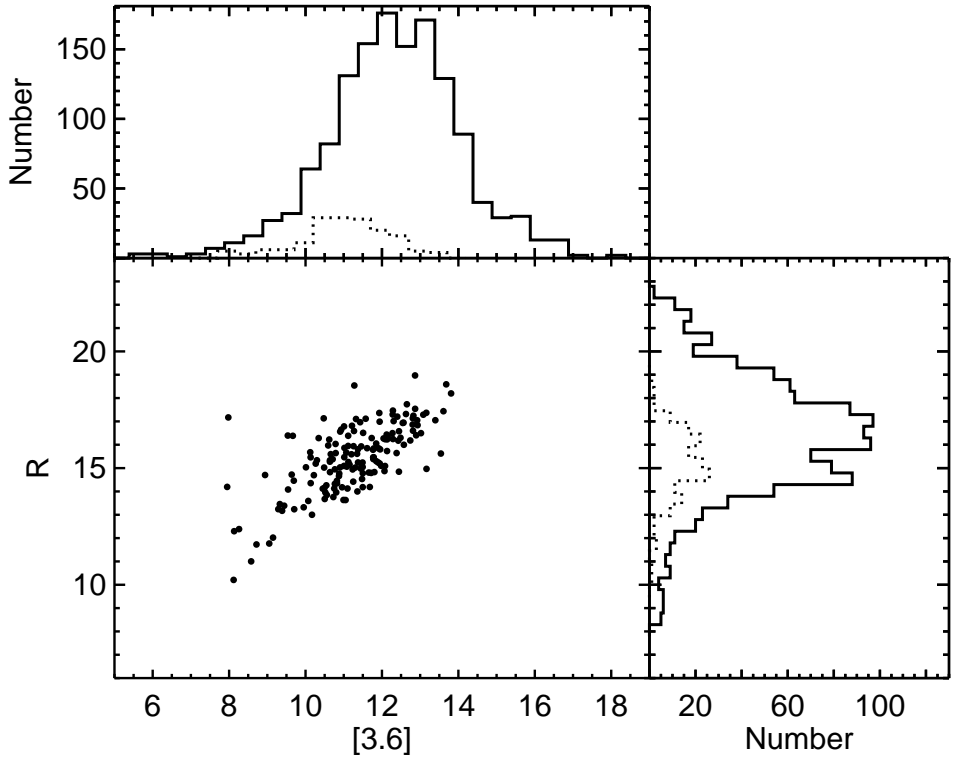}
\end{center}
\caption{\label{maghist}We present the distributions of 3.6~$\mu$m and
  $R$-band magnitudes characterizing our disk bearing sample, shown as dotted
  histograms. The solid histograms show the distributions of
  all $\sim$1500 confirmed cluster members in these bands.}
\end{figure*}

\subsection{Identification of disk-bearing objects}

We now narrow our focus to the NGC~2264 sample of stars with
infrared excesses indicative of circumstellar disks. Our main approach
is to measure the slope, $\alpha$, of the spectral energy distribution (SED) at
near and mid-infrared wavelengths, and to compare to the expectation
of a bare stellar photosphere. Several different disk classification schemes
exist in the literature. The most recent analysis of NGC~2264
members by \citet{2009AJ....138.1116S} involved the five classes I, II, III,
pre-transitional and transitional, based primarily on the loci of photometric
points on infrared color-color diagrams.  Since the definition of transitional disks has been a
subject of debate and the set of infrared photometry is not complete
for all objects, we prefer the SED slope definition \citep[see][]{2009arXiv0901.1691E}.
These are systems that have outer disk properties of normal T Tauri stars but with inner disk
holes or gaps.

Following \citet{2001ApJ...551..357W} with some guidance also from \citet{1984ApJ...287..610L}, 
\citet{1994ApJ...434..614G}, and \citet{1996ARA&A..34..111B}, we define $\alpha = d \log \lambda F_{\lambda}/d 
\log \lambda$ for flux $F_{\lambda}$ as a function of wavelength, $\lambda$. We classify $\alpha > 0.3$ for 
Class I, 0.3 to $-$0.3 for flat-spectrum sources, $-$0.3 to $-$1.6 for Class II, and $<-$1.6 for Class III. The 
notation II/III is reserved for transitional type disks, as described below.  For all objects in our {\em 
Spitzer} sample, we performed a least squares linear fit to all available photometry (not including upper or 
lower limits) as observed between 2 and 24~$\mu$m. We obtained data at 5.8~$\mu$m and longer wavelengths from 
cryogenic {\em Spitzer} observations. To maintain consistency with the approach taken by the YSOVAR project 
\citep{Reb14}, we obtained archival IRAC and Multiband Imaging Photometry for Spitzer (MIPS) data 
(program IDs 3441, 3469, 50773, and 37) and performed our own photometric reduction with Cluster Grinder 
\citep{2009ApJS..184...18G}. We produced photometry for all sources in the NGC~2264 region with a signal to 
noise ratio of at least five. For objects in the \citet{2009AJ....138.1116S} sample, our absolute photometry 
agrees with theirs to within 1\%. We note that formal errors on the infrared points are so small as to not 
affect the fitted SED slope. The fit was performed on the observed SED, with no reddening corrections. 
Accounting for reddening is unlikely to change the SED class, since all of our targets with {\em CoRoT} data are in 
low extinction regions of the cluster.  Variability could also alter the shape of the SEDs slightly, since
not all of the photometry was obtained simultaneously. However, we find that an allowance of 15\% uncertainty 
does not change the classification appreciably. We provide this slope assessment, as opposed to that of 
\citet{2009AJ....138.1116S} to ensure internal consistency and enable comparison with other clusters in the 
YSOVAR sample (see Rebull 2014).

For objects with MIPS data (131 in the combined {\em CoRoT}/{\em
Spitzer} sample), we compared the $\alpha$ value and resulting disk class derived with and without the
24~$\mu$m point. Disagreement occurred for around one quarter of these,
primarily due to disjoint 24~$\mu$m
photometry compared to the shorter wavelength SED. In the majority of these cases, the disk is transitional. However, in others,
it is possible that nebulosity is causing a false excess at long
wavelengths, and we have adopted the $\alpha$ value from 2--8~$\mu$m data.
For all disagreements, we visually inspected the SEDs to determine the
most appropriate wavelength range for slope determination. We
identified a total of 140 disk-bearing objects among NGC~2264 members targeted
for observation with both {\em Spitzer}/IRAC and {\em CoRoT}.

There are additional cases where misclassifications may have occurred, such that disk-bearing stars are 
labeled as diskless. Several objects display high-amplitude variability but their mid-infrared SED slopes 
of less than -1.6 result in class III status. To assess whether they might have weak disks, we examined 
their [3.6]-[8.0] IRAC colors. Analysis of infrared photometry in
other clusters \citep[e.g.,][]{2006ApJ...649..862C, 2010ApJS..191..389C} 
has shown that the requirement [3.6]-[8.0]$>$0.7 is a fairly robust disk 
selection criterion, whereas [3.6]-[8.0]$<$0.4 selects diskless stars with high accuracy. Using the 0.7  
cut-off, we identified 23 additional stars in the {\em Spitzer} and
{\em CoRoT} fields with weak or transitional disks, which are confirmed by visual 
inspection of the SEDs, revealing infrared excess above the expected photospheric flux level predicted from 
the stellar spectral type. We adopt the disk class ``II/III'' for these objects and present them along with 
the rest of the disk-bearing sample in common with {\em CoRoT} observations in Table 3. We also list the 2MASS and
{\em CoRoT} cross matches here.

\subsection{Overall sample properties}

\begin{figure}
\begin{center}
\includegraphics[scale=0.45]{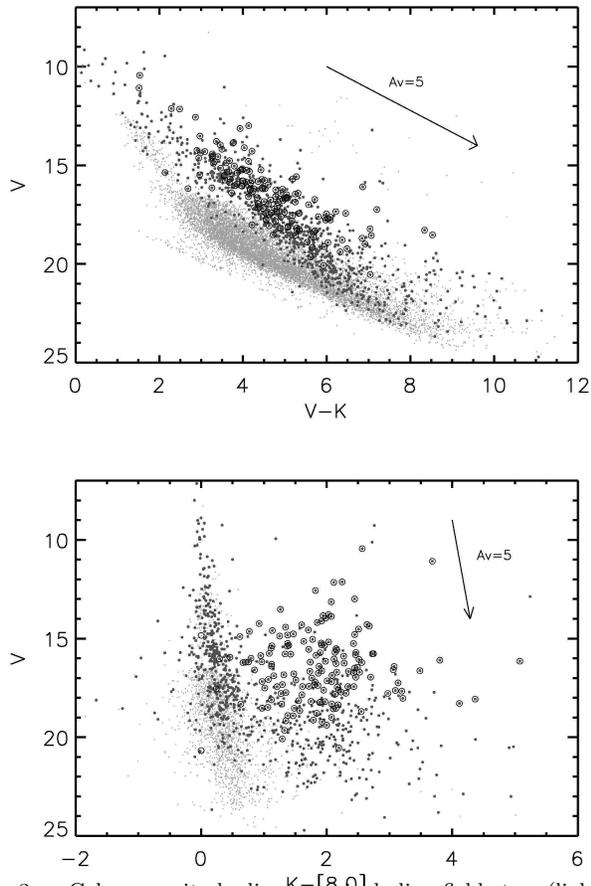}
\end{center}
\vspace{-1.cm}
\caption{\label{cmd}Color-magnitude diagrams, including field stars
  (light grey points), NGC 2264 members (dark grey dots), and objects
  in the 162-member disk bearing dataset highlighted in this paper
  (black circles). The optical/near-infrared diagram at top indicates
  that the sample considered here is consistent with a
  pre-main-sequence track, while the near-infrared/mid-infrared
  diagram at bottom confirms that the SED slope selection has resulted
  in a robust set of infrared excess objects. The infrared extinction value
  at 8.0~$\mu$m is taken from \citet{2005ApJ...619..931I}.}
\end{figure}

The full sample of $\sim$1500 cluster members in NGC~2264 includes spectral types ranging from M5 to A7
\citep{1956ApJS....2..365W,2004AJ....127.2228M,2005AJ....129..829D}, corresponding to masses of
$\sim$0.1~$M_\odot$ to several $M_\odot$. Restricting this to the 162
disk-bearing objects among the IRAC and {\em CoRoT} targets, the spectral
types run from M5 to G0. We present the distributions of brightness
in {\em CoRoT} and IRAC bands for this disk-bearing sample in Figure~\ref{maghist}. Both the spectral type and
$R$ magnitude distributions are representative of the cluster sample
as a whole. However, the collection of 3.6~$\mu$m points is skewed
considerably toward brighter values than compared to the available infrared
dataset. This is because the requirement of {\em CoRoT} observations
eliminates most faint and embedded objects from the set. 
We have also produced color-magnitude diagrams to compare the
162-object sample with the distributions of field stars and other
cluster members; these are presented in Figure~\ref{cmd}.

The pixel sizes for {\em Spitzer}/IRAC
and {\em CoRoT} are 1.2\arcsec\ and 2.3\arcsec\, respectively. We are therefore unable to separate
some visual binaries. However, thanks to the higher spatial resolution of
previous ground-based datasets \citep[i.e.,][]{2008AJ....135..441S}, we have a fairly
complete list in Table~3 of blended objects. The effect of binarity on variability will need to be assessed once a more complete 
multiplicity census is available.

\section{IRAC data}
\subsection{Observations}

{\em Spitzer} has been operating in its Warm Mission mode since the exhaustion of cryogen in 
mid-2009, and it now observes exclusively in the IRAC 3.6~$\mu$m and 4.5~$\mu$m channels. 
Observations of a $\sim$0.8\arcdeg$\times$0.8\arcdeg\ region of NGC~2264, including the more 
embedded Cone Nebula and Christmas Tree Cluster regions, were carried out from 2011 December 3 to 
2012 January 1 with Warm {\em Spitzer}/IRAC, under program 80040 (P.I.: J. Stauffer). The field center was 
R.A.=06$^{\rm h}$40$^{\rm m}$45.0$^{\rm s}$, decl.=+09$\arcdeg$40$\arcmin$40\arcsec. Since the 
cluster is considerably larger than the camera's 5.2\arcmin$\times$5.2\arcmin\ FOV 
for each channel, we visited targets $\sim$12 times a day with a series of mapping fields.  For 
each such Astronomical Observation Request (AOR), we executed four pointings. Starting at the first 
position, two exposures were acquired with a 8\arcsec\ dither to mitigate any detector artifacts 
or cosmic ray hits. Total frame time for each was 12 seconds. Data were taken simultaneously 
in the two IRAC bands, but due to the non-overlapping IRAC FOVs, individual objects were monitored 
in one channel at a time. A spatial shift of just over half the field size was then performed, providing a 
second pair of dithers for half of the targets and a first dither pair
for an equivalently sized set of targets in the newly observed
section of the FOV. After three telescope offsets, the FOV has
advanced by 7.2\arcmin, such that targets in the newly observed
section of the ``trailing'' IRAC dectector's FOV will have been
observed by the ``leading'' IRAC detector less than a minute earlier. 
Accordingly, most of our targets have two-color IRAC data, but for the approximately 40\% of 
objects near the edges of the mapping region, there is only single-channel coverage.

Since stars brighter than magnitude 9.5 saturate the IRAC detector in
12~s exposures, we also acquired 0.6~s integrations of each field. 
This high dynamic range (HDR) mode was employed during every sixth
round of mapping. The resulting observing cadences were every $\sim$101 minutes for mapping
data, and $\sim$12 hours for the 0.6~s HDR mode data.

To capture light curve behavior on even shorter timescales, we
operated IRAC in staring mode for four blocks of 20, 26, 16, and 19 hours, respectively, toward
the beginning of the run (see Table~1 for dates). The channel 1 and 2
FOVs were fixed on a region near the center of NGC~2264, and exposures with
6s integration times were acquired repeatedly, with no dithering or HDR mode images.
After taking readout and data volume restrictions into account, the
cadence for staring mode data was $\sim$15 seconds. A total of 549 stars were observed 
in the staring fields. We display these regions, along with the full 3.6 and 4.5~$\mu$m FOVs in Figure 1.

\subsection{Light curve production}

For both the mapping mode and the staring mode data, we used the
Interactive Data Language (IDL) package Cluster Grinder
\citep{2009ApJS..184...18G} to generate light curves from the basic calibrated data (BCD) images released by the
Spitzer Science Center (SSC) pipeline, version S19.1.  Each
flux-calibrated BCD was processed for standard bright
source artifacts. Cluster Grinder delivers mosaics for each AOR,
point source locations, and photometric measurements from the mosaics (``by-mosaic''
lightcurves).  We then re-performed photometry on the BCD images using
the Cluster Grinder source list, with a 5--$\sigma$ source detection threshold.
We applied array position dependent systematic corrections for residual
gain and pixel phase effects (although the treatment of this was
modified for staring data; see the Appendix). The pipeline also
computes a variety of flagging information, including maximum pixel
value for saturation detection, spatial coverage, and outlier rejection.

Aperture photometry was obtained from both the BCDs (``by-BCD'' photometry) and the
mosaics (``by-mosaic'' photometry) using an aperture
radius of 2.4$\arcsec$ (2 pixels), and a sky 
annulus from 2.4$\arcsec$ to 7.2$\arcsec$ (2--6 pixels). 
We also combined the resultant photometric by-BCD measurements into ``by-AOR'' photometric
products, mimicking the cadence of the by-mosaic photometry.  
We have found that the photometric precision achieved with this
approach is higher than that of the by-mosaic data, presumably because
the latter does not allow for easy treatment of
array-position-dependent systematics, such as residual gain and pixel phase
effects. Hence we present only the by-BCD light curves (for staring
data) and by-AOR light curves (for mapping data) here. 

We flagged by-BCD photometry for saturation, low signal-to-noise
ratio (SNR; $<$5), and outlier
status (5--$\sigma$ above the median flux of other points in the same AOR). 
Affected points were omitted from by-AOR photometry, and unaffected
points were combined via an unweighted mean. By-AOR photometry values
resulting from fewer than three by-BCD measurements were flagged as well.
Following these procedures, the majority of mapping mode light curves
contain 300-320 valid datapoints.

For pairs of 0.6~s and 12~s exposures taken during the HDR mode, we
selected the latter only if it did not exceed a count level of 20,800 DN (Warm
IRAC saturates at $\sim$30,000 DN); otherwise the measurements from
the 0.6~s exposures were swapped in. For saturated objects brighter than a magnitude of
$\sim$9.5, we retained only the twice daily short HDR exposures, resulting in by-AOR
light curves with $\sim$50 points. 

We adopted the standard zero points derived from the official
Warm Mission FLUXCONV header values: 19.30 and 18.67 for 1
DN/s total flux at 3.6 and 4.5 microns, respectively. 
These values include standard corrections for our chosen aperture and sky annulus
sizes (Reach et al.\ 2005). 

Uncertainties for the by-BCD photometry were derived by combining
three terms in quadrature: shot noise in the aperture, shot noise in
the mean background flux per pixel integrated over the aperture, and
the standard deviation of the sky annulus pixels (to account for the
influence of non-uniform nebulous background).  Finally, the BCD-level photometric
uncertainties were combined in quadrature to yield uncertainties for the by-AOR
photometry reported here.

To weed out extremely faint sources and artifacts, we required that each light curve contain at least five 
unflagged datapoints and have a mean photometric uncertainty of less than 10\%. Applying these cuts, we 
generated a total of 13,856 mapping mode light curves in the 3.6~$\mu$m band and 12,186 light curves in the 
4.5~$\mu$m band, 9541 of which include data in both bands. For our eventual comparison of behavior in the optical
and infrared (e.g., Section 7), we performed further cuts and retained only IRAC light curves with at least 30
unsaturated points covering a minimum timespan of 20 days.
For the subset of observations executed in 
staring mode, we have additional high cadence light curves for 290 objects with 3.6~$\mu$m band data, and 
259 objects with 4.5~$\mu$m band data. We have also produced binned staring light curves with smaller 
errorbars at 2.5-minute cadence, which we describe below. Since the staring fields were observed 
simultaneously, there is no overlap between the 3.6 and 4.5~$\mu$m sets.

\begin{figure}
\begin{center}
\includegraphics[scale=0.45]{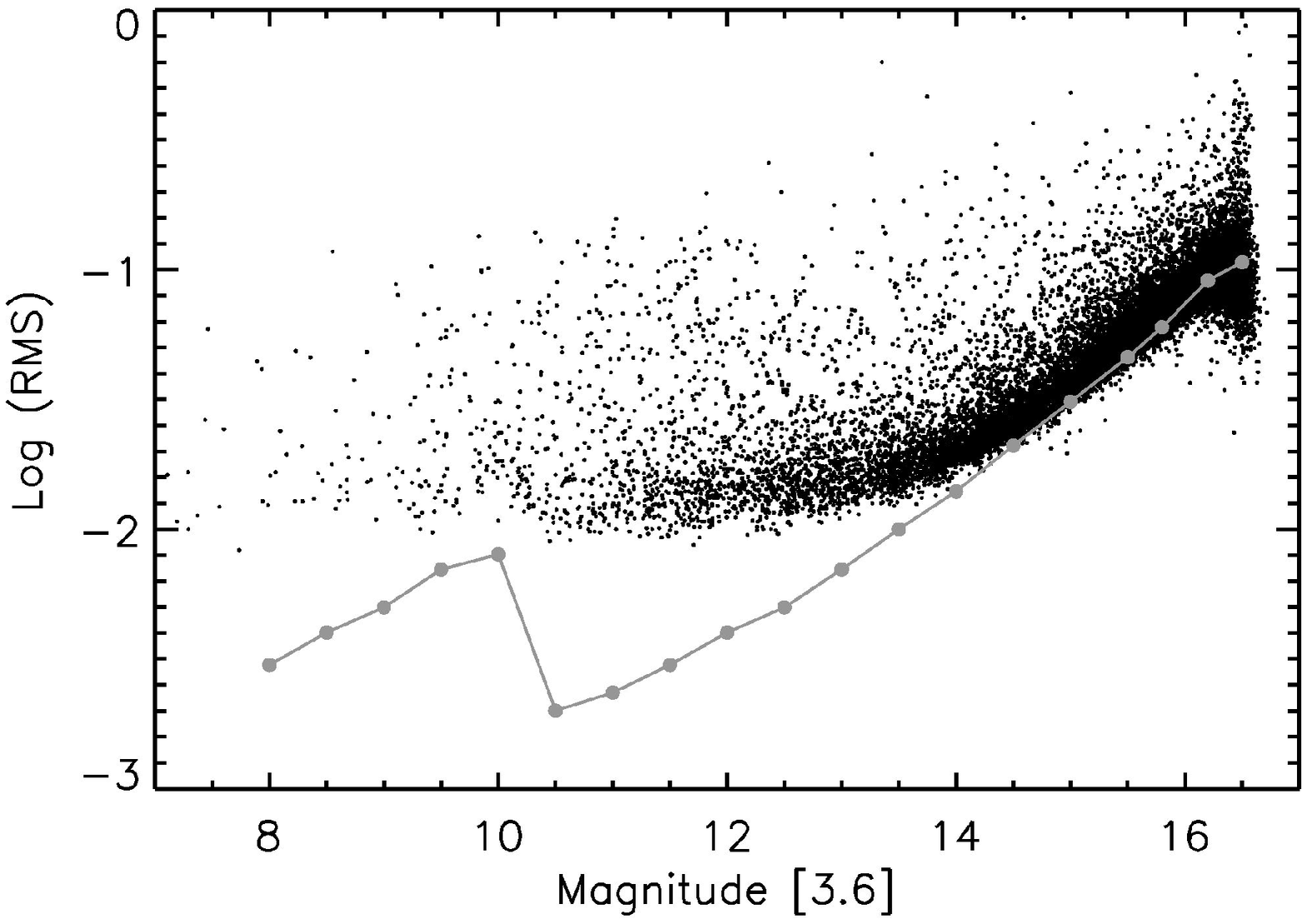}
\includegraphics[scale=0.45]{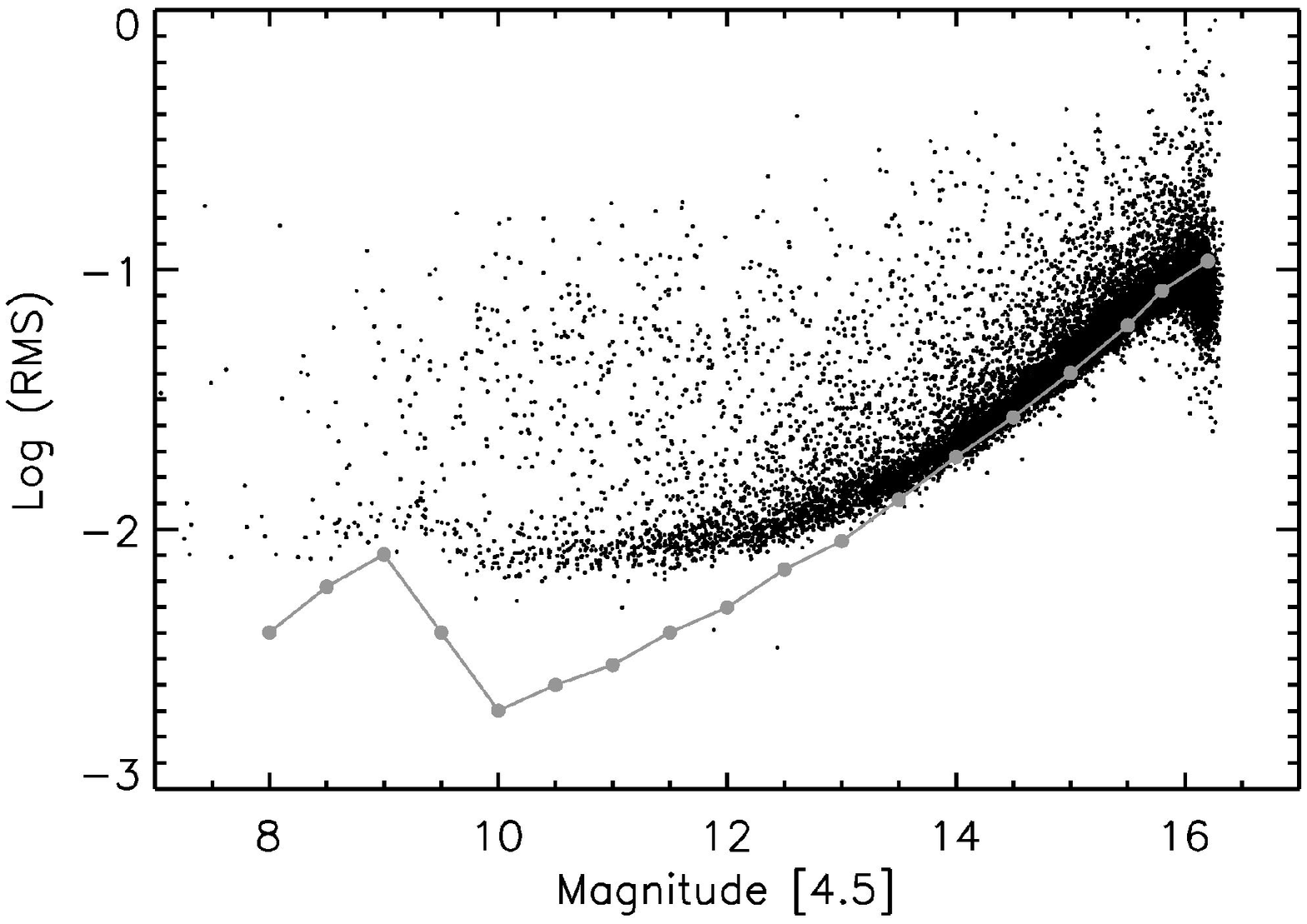}
\end{center}
\vspace{-0.5cm}
\caption{\label{mappingrms}Expected and observed standard deviations of IRAC mapping light curves in channel 1 (top) and
channel 2 (bottom), as a function of magnitude. Black points are the
measured RMS values, while the gray curve marks the estimated 1--$\sigma$ white
noise level in 0.5-magnitude bins (dots), as predicted by Poisson noise and sky background. The
jump near magnitude 10 is due to the shift to shorter exposure times
for bright stars observed with short HDR exposures. There is an empirical noise floor that is limited to
just under 0.01 magnitudes for the brightest objects. Points with large
RMS values above the empirical noise floor at each magnitude are predominantly variable cluster members.}
\end{figure}

\subsection{Mapping data quality}

Comparison of the root mean square (RMS) values for all mapping light curves with their mean predicted
uncertainties revealed additional systematic errors that limited the
photometric accuracy to 1\%, as seen in Figure~\ref{mappingrms}. In order to
identify variable stars, we need to characterize the systematic
effects as a function of magnitude in detail. Because of the transition to shorter
exposure times toward the bright end of the sample, there is an upward bump in
the RMS distribution near 10th magnitude, and it is difficult to assess the systematic
contribution here. We therefore computed a different measure of the systematic, ``$S$,'' which should be
independent of exposure time:
\begin{equation}
S=\sqrt{{\rm RMS}^2-\sigma^2},
\end{equation} 
where $\sigma$ is the uncertainty estimate described above, without systematics, and RMS is the root-mean-square deviation
of each light curve omitting flagged points.
Plotting $S$ against magnitude, we found that the nonvariable stars
traced out an exponential trend, with a larger systematic contribution for fainter
objects. To omit as many variable objects as possible while characterizing this effect, we
considered only stars lacking firm NGC~2264 membership status (see
the Appendix for membership evaluation criteria), and computed the median of the distribution of $S$ values 
in 0.5-magnitude bins. We then fit an exponential curve of form
\begin{equation}
S(m)=e^{(m-m_0)}+S_0,
\end{equation}
where $m$ is magnitude and $m_0$ and $S_0$ are free parameters in units of magnitude.

\begin{figure}
\begin{center}
\includegraphics[scale=0.45]{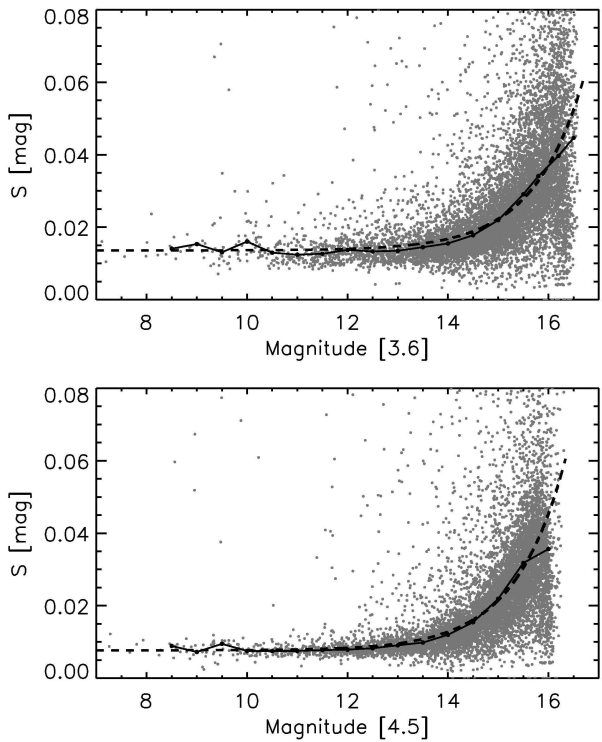}
\end{center}
\vspace{-0.3cm}
\caption{\label{syserr}Systematic error (``$S$'') as a function of IRAC magnitude, for
channel 1 (top) and channel 2 (bottom). Gray points are field stars or
objects with uncertain membership, culled by a lack of photometric or spectroscopic
signatures of youth. For more information on NGC~2264 membership criteria, see the appendix.
Black circles mark the median trend of S, in 0.5 magnitude bins. The
dashed curve represents the best fit exponential trend.}
\end{figure}

The distribution of $S$ values, as well as the fits for each IRAC band, are shown in Figure~\ref{syserr}.
We obtained best fitting values of $S_0$=0.014, $m_0$=19.75 in the 3.6~$\mu$m band
and $S_0$=0.008,  $m_0$=19.28 in the 4.5~$\mu$m band, indicating that the
photometry includes systematics at the 1\% level. Indeed, this is
expected from intrapixel sensitivity variations and detector
gain variations, both of which are known to result in
position-dependent flux measurements (see the Appendix).
As shown in Figure~\ref{syserr}, $S$ increases exponentially from
$\sim$0.01~mag for 13$^{\rm th}$ magnitude sources, to 0.04~mag for 16$^{\rm th}$
magnitude sources. Figure~\ref{mappingrms} demonstrates that the
expected Poisson and sky noise increase from $\sim$0.01~mag to
$\sim$0.1~mag over the same range of magnitudes. Thus, while both
random and systematic uncertainties increase toward faint magnitudes,
the systematic effects increase more slowly towards the faint end,
such that they dominate the error budget for sources brighter than
$\sim$14$^{\rm th}$ magnitude, with random uncertainties becoming increasingly
dominant for fainter sources.

To fully account for the errors, we added the fitted
magnitude-dependent systematic offsets in quadrature to the derived uncertainties and adopted
the result for the selection of variables and production of 
variability statistics, as will be discussed in Section~6. In addition, we have performed similar
assessments of the staring data. Since those light curves are not
presented extensively here, we refer the reader to the appendix for a
discussion of their quality and the mapping/staring data merging process.

\section{{\em CoRoT} data}

The {\em CoRoT} satellite \citep{2006ESASP1306...33B} was launched in 2006 and observed until November 2012. Consisting of a 27-inch 
diameter telescope with an external baffle to suppress scattered light from Earthshine, it performs 
broadband photometric monitoring ($\sim$ 3700 to 10,000\AA) of stars in two 10\arcdeg\ diameter regions 
near the galactic center and anti-center, known as the {\em CoRoT} ``eyes.'' Observations were originally 
carried out with two exoplanet-dedicated and two seismology-dedicated CCDs, but by the time of our 
observations, only one of each was in operation. Data from the exoplanet CCD is passed through 
low-resolution spectral dispersing prism to produce three-color photometry in approximately ``red,'' 
``green'' and ``blue.'' For most targets, data from these three bands is coadded to generate a white light 
curve. For 910 targets (including 262 with {\em Spitzer} observations), the full chromatic dataset was 
retained. However, we did not use these data at this time as they showed signs of strong systematics 
as a function of both band and time.  In addition to the 2011 obserations described 
here, a prior {\em CoRoT} short run including many of the same NGC~2264 targets in 2008 provides
long time baseline information on optical variability phenomena, including rotation \citep{2013MNRAS.430.1433A} 
and pulsation properties \citep{2011ApJ...729...20Z}.

\subsection{Observations}

NGC~2264 is the only rich 1--5~Myr cluster situated within one of the {\em CoRoT} eyes, making it the 
primary target for young star variability with this telescope. For the CSI~2264 campaign, we monitored a 
$\sim$1.3\arcdeg$\times$1.3\arcdeg\ field in NGC~2264 centered at R.A.=06$^{\rm h}$40$^{\rm m}$18.0$^{\rm 
s}$, decl.=+09$\arcdeg$41$\arcmin$46.24$\arcsec$ (see Figure 1) from 2011 December 1 to 2012 January 9. 
These observations comprised the fifth {\em CoRoT} short run (``SRa05''). All stars were placed on the second exoplanet 
channel CCD (E2), as the first channel ceased to function in early 2009. NGC~2264 was previously observed 
during the SRa01 short run in 2008, using the first exofield CCD, E1. In each case, targets 
consisted of confirmed and candidate NGC~2264 members as well as field objects selected for {\em CoRoT}'s 
transiting planet search program. Only data for pre-selected targets are downloaded from the satellite, 
with photometry consisting of flux values within a pre-defined aperture mask for each star. For SRa05, we 
monitored 489 confirmed and 1617 candidate NGC~2264 members, along with an additional 2129 field stars.

Light curves are produced by the {\em CoRoT} pipeline \citep{2006ESASP1306..317S}, which performs corrections for gain 
and zero offset, jitter, and electromagnetic interference, as well as background subtraction. We obtained level N2 
reduced data from the {\em CoRoT} archive, which consists of fully processed light curves. This includes fluxes and 
background levels, flagged for outliers and hot pixels. Typically $\sim$10--15\% of these were flagged and omitted from 
the light curves.

Exposure times were 512~s, resulting in up to $\sim$6300 datapoints per light curve.
For a subset of stars with signs of eclipses or transits, a
high-cadence mode was triggered with 32-second exposure times. This
mode generated light curves with up to 100,850 points and was mainly used for stars
brighter than 14th magnitude in $R$ band. 

The full range of magnitudes for observed stars was $R\sim$11--17.
Light curve RMS values ranged from 0.00055 for the brightest stars to 0.01--0.1 for the faintest
objects. There is a substantial spread in RMS as a
function of magnitude, because of strong systematics in a subset of
the light curves, which we address below.

We produced light curves from the {\em CoRoT} data by converting raw fluxes to the magnitude scale 
and subtracting the median after removal of outliers automatically
flagged by the {\em CoRoT} pipeline. The zero point for {\em CoRoT} photometry is not well determined and may
vary slightly between runs. To estimate a calibration for the SRa05
dataset, we compared the logarithmic mean flux
with $R$-band photometry reported in the literature by
\citet{2002AJ....123.1528R}, \citet{2004A&A...417..557L}, and \citet{2008AJ....135..441S}. A
fit of the intercept results in an $R$-band zero point of 26.74.
 
The full 2011 {\em CoRoT} dataset on NGC~2264 contains light curves
for a total of 4235 objects, just over half of which
are field stars, based on the membership criteria outlined in the Appendix. These were included for the transit monitoring portion of the
campaign. 2500 of the {\em CoRoT} targets were previously observed
during the SRa01 run in 2008. In total, 1303 stars were
observed by both {\em CoRoT} and {\em Spitzer}, about 65\% of which are possible or likely field
stars. All {\em CoRoT} light curves encompassed the full
39-day campaign. 

In general, we performed source cross matching
with a 1\arcsec\ radius. For targets with source confusion within the
{\em CoRoT} mask, we first identified the object in the optical
by requiring its {\em CoRoT} calibrated $R$ magnitude to match
photometry of a known source to within 0.5 magnitudes. We then
identified the 2MASS counterpart and used those coordinates to select
the appropriate {\em Spitzer} source. 

\begin{figure}
\begin{center}
\includegraphics[scale=0.44]{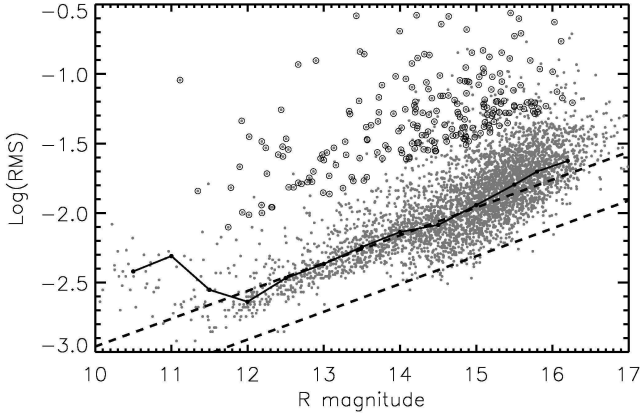}
\end{center}
\vspace{-0.5cm}
\caption{\label{corotrms} RMS values of {\em CoRoT} light curves, as a function of
magnitude. The black points and curve track out the median from 0.5
magnitude bins. A transition in RMS behavior is seen near $R$=14, as this
was the cut-off used to select exposure times. Most objects brighter than $R$=14
were monitored at high cadence (32s exposures), while fainter objects were
monitored at low cadence (512s exposures). The bottom dashed line illustrates the estimated
Poisson error. We have approximated the median error, including
systematics, by shifting this upward by 0.35 in log(RMS), as shown
with the middle dashed line. We shift this another $\sim$0.5 dex to
select variables (shown as black circles) at the 3--$\sigma$ level.}
\end{figure}

\subsection{Correction of light curve systematics}

{\em CoRoT} light curves include a number of well known systematic effects,
including outliers and flux discontinuities due to radiation events or changes in
detector temperature \citep{2009A&A...506..411A}. Isolated outliers
are automatically detected by the {\em CoRoT} pipeline and flagged in the
resulting photometry. Other types of systematics can be removed
if the light curves are well behaved or if color data is available
\citep[e.g.,][]{2010A&A...522A..86M}. However, correction becomes more difficult in
a highly variable, mostly monochromatic dataset such 
as ours. 

We found that $\sim$10\% of light curves contained abrupt jumps in flux not
attributable to stellar variation. Two prominent discontinuities
appeared at the same time stamps in many light curves. 
These are due to detector temperature jumps, the times of which were
provided by the {\em CoRoT} team. We searched for and corrected
discontinuities at these locations by computing a weighted difference
between the flux difference on either side of each jump. 

Background correction has also introduced time-dependent systematics
into the light curves. Background levels are measured as a median
across 400 10$\times$10 pixel windows placed in star-free regions of
the FOV. As the E2 CCD has aged, the level of dark current, as well as its
gradient across the detector, has increased. Further complicating
background measurements is the intrinsic peak in background
due to nebulosity near the NGC~2264 cluster center. As a result, the median background value
subtracted from stellar fluxes can be an underestimate for many
sources. The combination of these effects, along with the different CCD
used to observe NGC~2264 during 2008 has resulted in artificial
variability amplitude changes for stars observed in both the SRa01 and
SRa05 runs. For well characterized variables such as eclipsing binaries,
we note amplitude differences of up to 10\%.

To omit problematic data from variability statistics, we visually inspected
all 2011 {\em CoRoT} light curves and flagged those with
discontinuities that were at least a factor of seven times the RMS
(and often much more). About 10\% of light curves received flags. 

To assess the noise levels and fit the RMS distribution as a function
of magnitude, we have plotted the RMS values of the entire $\sim$5000-object
{\em CoRoT} dataset versus the magnitude values, using the zero point of
26.74; the result is displayed in Figure \ref{corotrms}. For bright objects ($R<14$), the trend of RMS
values follows a slope consistent with Poisson noise plus a
constant systematic. The distribution expands to lower values
for objects beyond $R$=14, with some RMS values reaching the Poisson limit. This structure in the diagram is
related to the change from high to low cadence near $R$=14 and is not explained by the flagged objects, as these 
have a large range of RMS values at all magnitudes. As seen in the figure, the RMS values of the final light curves ranged from 0.001 mag (at $R\sim 11.5$)
to 0.04 (at $R\sim 17$), with larger amplitudes for variable objects.

\section{Light curve morphology classes}

\begin{figure}
\begin{center}
\includegraphics[scale=0.45]{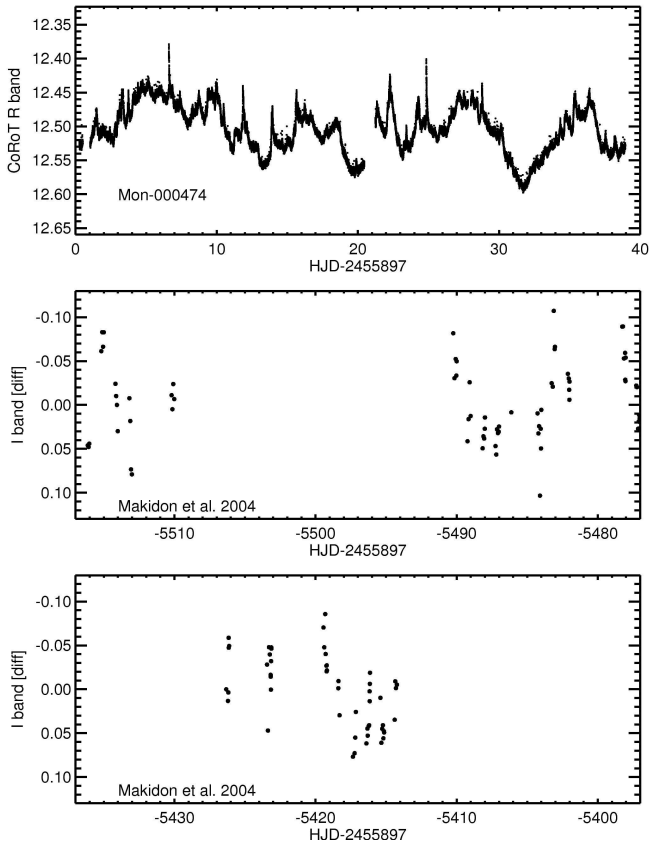}
\end{center}
\vspace{-0.3cm}
\caption{\label{makcompare}Light curves for Mon-000474, as
  observed by {\em CoRoT} from space (top) and
  \citet{2004AJ....127.2228M} from the ground (middle, bottom). Short duration flux increases appear in the high
  cadence data but are difficult to discern in ground-based time series.}
\end{figure}

The high cadence and precision of our {\em CoRoT} and {\em Spitzer}
time series data enable an unprecedented window into the diversity of light
curve morphologies. We are able to analyze in detail the patterns and timescales of
brightness fluctuations, and in some cases to separate them into
multiple components. In comparison with ground-based monitoring results,
finer morphological divisions appear in our dataset. We provide an
example comparison of one of our {\em CoRoT} light curves with typical
ground-based data from \citet{2004AJ....127.2228M} in
Figure~\ref{makcompare}.

\begin{figure}
\begin{center}
\includegraphics[scale=0.48]{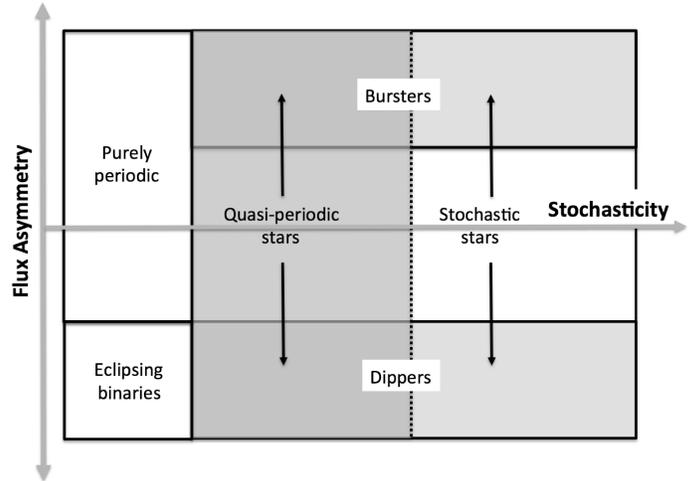}
\end{center}
\vspace{-0.5cm}
\caption{\label{vartypes} Schematic of the classification system we
  have applied for optical and infrared variability in the CSI 2264
  dataset. At the top end of
the schematic are light curves preferentially populated with
brightening events on top of a continuum, while at the bottom of the
schematic are light cruves with fading events. The middle is occupied
by variability that is more or less symmetric in flux. Moving from right to left are variables with increasingly stable repeating patterns. }
\end{figure}

Our own approach to classifying variability in the 162 member disk bearing {\em CoRoT}/{\em Spitzer}
dataset involves an initial visual classification, followed by statistical confirmation and analysis.
As shown in Figure~\ref{vartypes}, we select variable
types based on two parameters: stochasticity (or alternatively,
periodicity), and degree of asymmetry in the flux. 

For many types of variability mechanisms, we expect morphological properties to correlate with
physical processes or geometry. For example, we expect the fraction of extinction dominated 
variables to relate to disk scale height; this was estimated by \citet{2000A&A...363..984B} to be $\sim$15\%. 
In addition, timescales inherent in variability point
to mechanisms associated with star or disk rotation, or disk and accretion instabilities. Multiwavelength
properties of morphology should also confirm which processes dominate flux changes.
For global changes in luminosity (e.g., from accretion), we might expect optical and infrared variations to
be correlated; starspots, on the other hand, could induce phase lags at these two wavelengths.
In summary, a careful accounting of morphological types can help to verify these assumptions.

In this section, we begin by highlighting progress made by previous campaigns, and then
present the categories of variability illustrated in Figure~\ref{vartypes}, followed by speculation on the physical processes behind
them. 

\subsection{Summary of previous classification schemes}

Prior optical and near-infrared YSO variability studies have proposed several different 
classes of T Tauri star behavior, depending on light curve morphology, accretion 
status, and stellar mass. The first general survey of YSO variability was performed by
\citet{1954TrSht..25....1P}, who classified T Tauri stars based on histograms of their
brightness levels. Objects that were more often bright than faint were referred to as
Class I, objects with typical brightness similar to the mean were Class II, and stars
that were more often faint than bright were called Class III. Following up on this, 
\citet{1980MNRAS.191..321W} explored the relationship between light curve class and emission 
line strength, finding that Class I objects were associated with preferentially weak strengths. 
Later, \citet{1994AJ....108.1906H} presented a variability classification scheme based on 15 years 
of photometric monitoring of 80 young stars. They divided light curve behavior into four 
types, including periodic variations caused by cool spots in WTTSs (``type 
I''), irregular and periodic variations in CTTSs (types ``II'' and ``IIp'') attributed to 
variable accretion luminosity and rotational modulation of hot spots, respectively. In 
addition, they identified flux changes of 0.8--3 magnitudes in the $V$ band among T Tauri stars 
earlier than spectral type K1 (``type III''). The mechanism for this behavior was proposed to 
be obscuration by circumstellar material, since it was not associated with veiling changes. It 
frequently involves pronounced flux decreases, interpreted as extinction events as in the 
prototype UX Ori. 

A similar scheme was suggested by \citet{2010A&A...519A..88A} based on higher cadence
light curves of T~Tauri stars from the first {\em CoRoT} short run on NGC~2264 in 2008 (``SRa01''). They
divided the variability sample into periodic sinusoidal (spot-like), periodic flat topped
(AA Tau-like), and aperiodic (irregular). These correspond roughly to the \citet{1994AJ....108.1906H}
classes, except that the AA-Tau type objects have much short timescales (a few to 10 days) than
the higher mass UX~Ori stars.

Further variability studies
\citep[e.g.,][]{2001AJ....121.3160C,2004A&A...417..557L,2007A&A...461..183G,2008A&A...479..827G,2009MNRAS.398..873S},
continued the strategy of dividing variability 
into periodic and aperiodic, and assessing physical mechanisms by analyzing color changes as a 
function of magnitude. These photometric campaigns typically involved monitoring at lower 
cadences (days to years) or with many more gaps in time sampling than
our own dataset. In the case of NGC~2264, \citet{2004A&A...417..557L}
classified stars only as periodic or irregular, based on periodograms and $\chi^2$ tests. They concluded
that variability was likely due to rotational modulation of cool spots (if periodic), or hot spots
generated by accretion flows onto the star (if irregular).

\subsection{``Dipper'' light curves}

One of the most common phenomena in our time series is transient optical fading events.
These have been noted previously in YSOs, from the
early type HAeBe stars \citep[e.g.,][]{1999AJ....118.1043H} to lower mass AA Tau stars
\citep[e.g.][]{1999A&A...349..619B}.  Based on color trends
and an inconsistent relationship of photometric behavior with veiling in accompanying spectra, these episodes have
been attributed primarily to extinction by circumstellar material
\citep[e.g.][]{1999A&A...349..619B,2010A&A...519A..88A}. In some cases, the events recur
periodically, although their depth changes. \citet{2011ApJ...733...50M} identified similar behavior in infrared
and near-infrared time series of YSOs in Orion, likewise attributing ``dipper'' events to stellar
occultations by regions of enhanced optical depth in a surrounding dusty
disk. In several cases, periodic fading on longer timescales (months
to years) has been traced to obscuration by a circumbinary disk
precessing about a double or triple star system \citep[e.g.,][]{2003ApJ...596L.243C,2008ApJ...684L..37P,
2010AJ....140.2025H,2013A&A...554A.110P}. If the dips are caused by
varying dust extinction along the line of sight, then the light curves
should become redder as they grow fainter. \citet{2003A&A...409..169B}
indeed found this behavior in the prototype AA~Tau itself, noting color
slopes similar to the values expected from an ISM extinction curve. 

\begin{figure*}
\begin{center}
\includegraphics[scale=0.7]{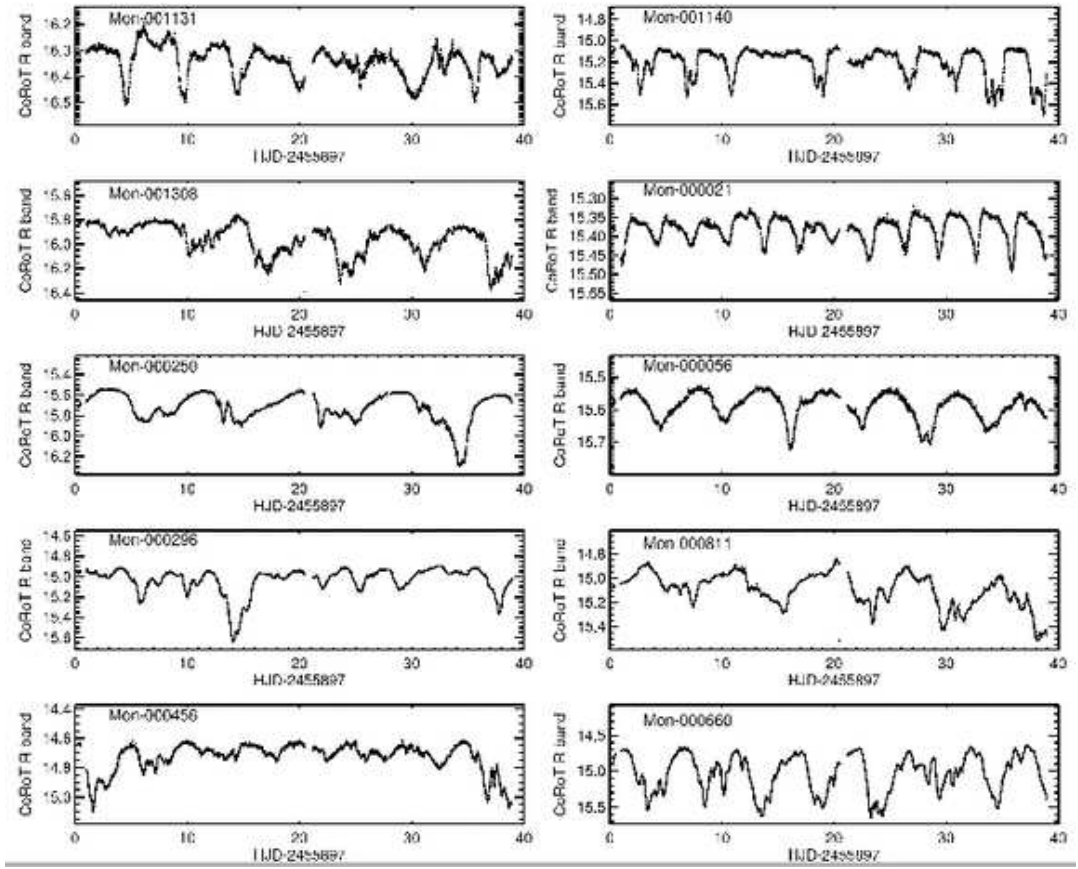}
\end{center}
\vspace{-0.2cm}
\caption{\label{perdippers} Prototypical quasi-periodic optical dippers
  observed with {\em CoRoT}. These light curves preferentially display fading events
  that repeat regularly, albeit with different amplitudes.}
\end{figure*}

\begin{figure*}[!b]
\begin{center}
\includegraphics[scale=0.7]{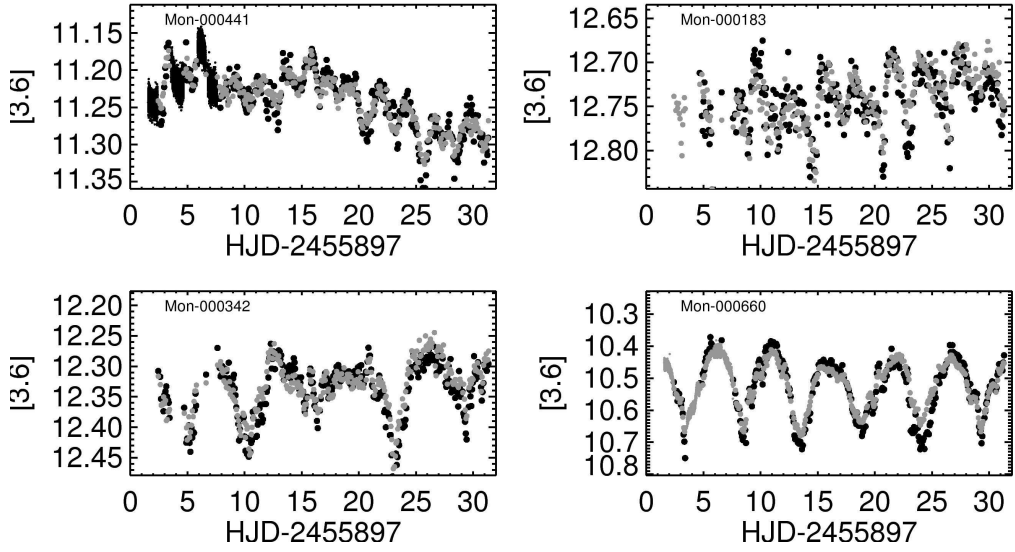}
\end{center}
\vspace{-0.2cm}
\caption{\label{perdippersIR} Quasi-periodic infrared dippers among
  NGC~2264 IRAC sources. Black points are 3.6~$\mu$m data, and grey
  points are 4.5~$\mu$m data. Only two of these were also identified as
  periodic dippers in the optical; the top two objects show
  small decrements in the {\em CoRoT} light curves, but the amplitudes
were too low to independently confirm dipper status. The lack of
larger amplitude behavior in the optical suggests that starspots are
not a good explanation for the variability.}
\end{figure*}

\begin{figure*}[!t]
\begin{center}
\includegraphics[scale=0.7]{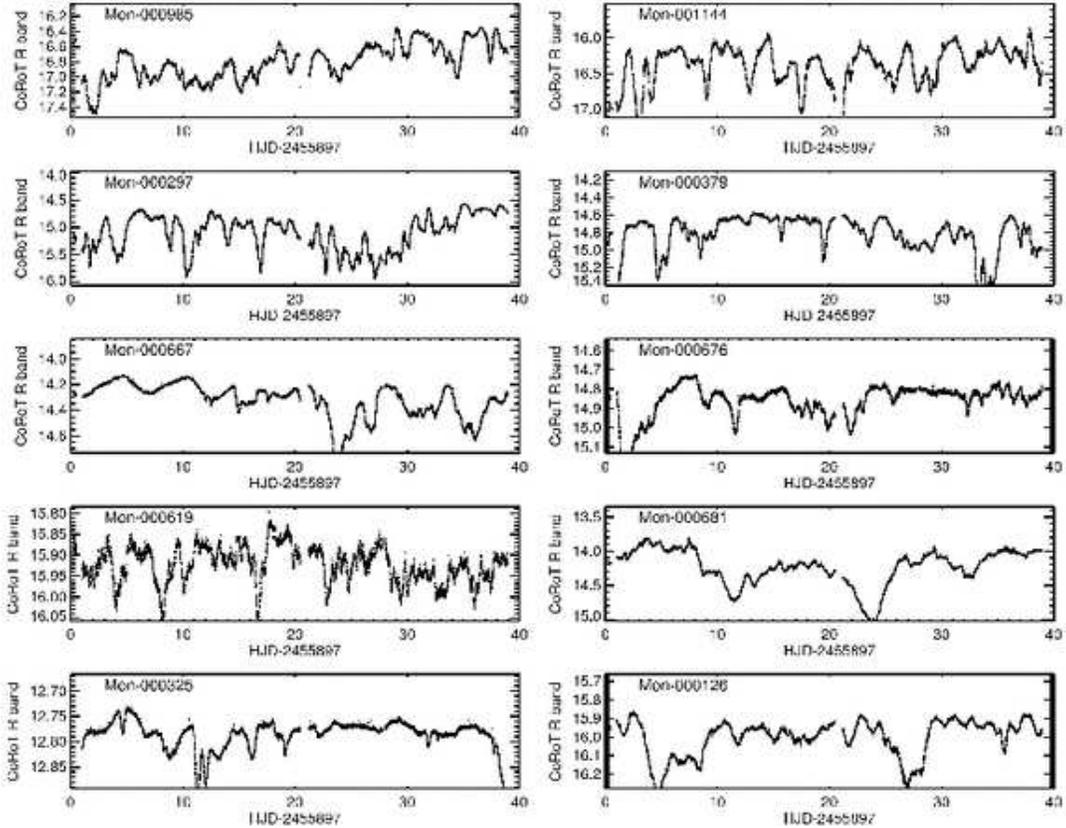}
\end{center}
\caption{\label{aperdippers} Prototypical aperiodic optical dippers
  observed with {\em CoRoT}. These light curves display prominent fading events
  with no detectable periodicity.}
\end{figure*}

\begin{figure}[!t]
\begin{center}
\includegraphics[scale=0.65]{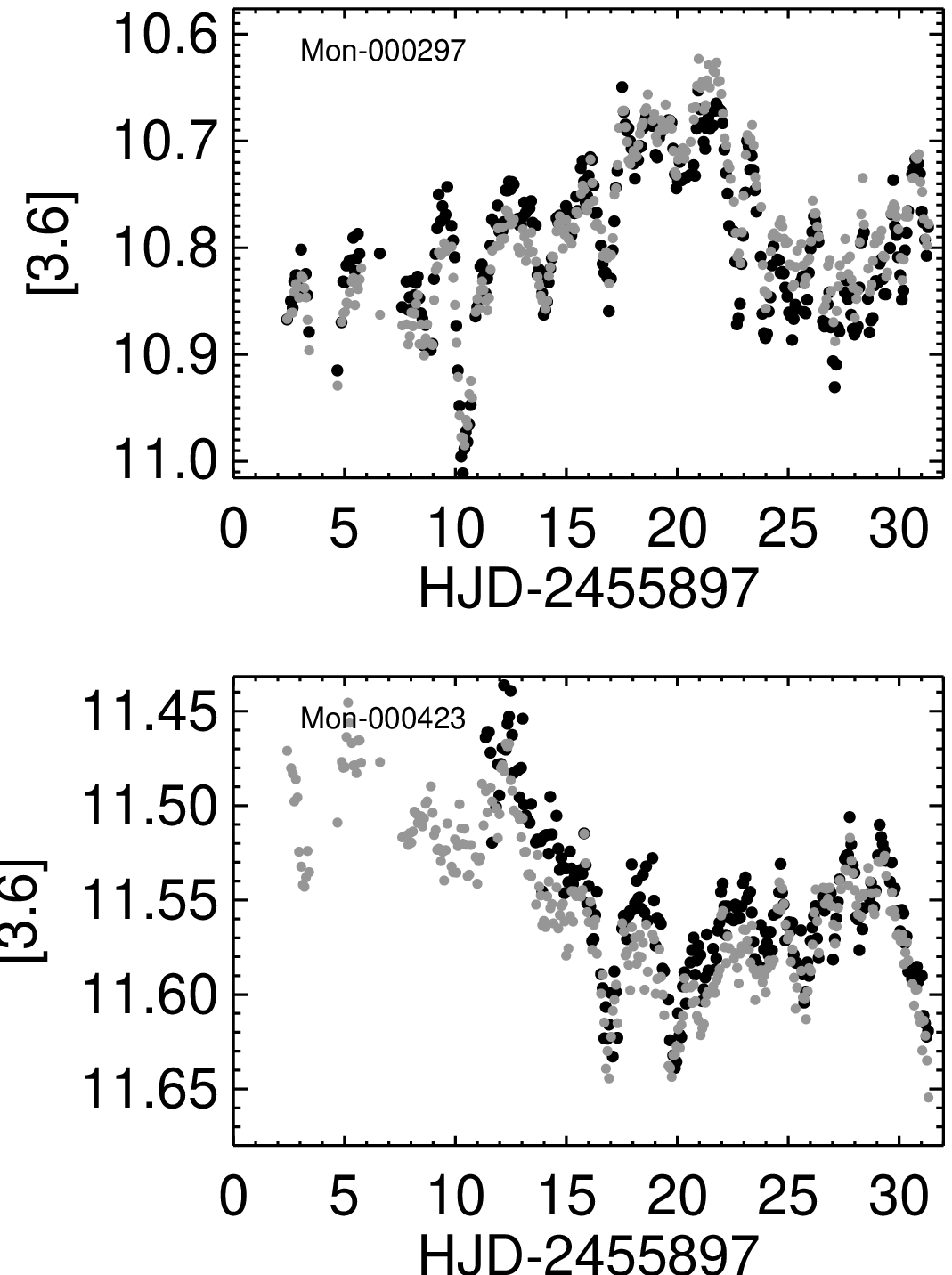}
\end{center}
\caption{\label{aperdippersIR} Aperiodic infrared dippers
  observed with IRAC; only two such objects were identified.}
\end{figure}

Unsurprisingly, prominent fading events appear in our {\em CoRoT} and {\em Spitzer}
light curves. These are distinguished from other types of variability by
both the sharpness and rapidity of the brightness troughs. Outside of
these events, the light curves display a ``continuum'' flux level with
less variability. To quantitatively identify and
compare dipper behavior in the different bands, we have selected by eye
objects that are clearly asymmetric, in that their light curves
appear different when flipped upside-down.

We have identified 35 optical and seven infrared dipper stars in the joint
{\em CoRoT}/{\em Spitzer} disk bearing sample. Two (Mon-000183 and Mon-000566) exhibit clear dips only
in the infrared, although with knowledge of the {\em Spitzer} data, one can see small dips (2-3\% depth) in the
{\em CoRoT} light curves at the same times. Many of the {\em optical} dippers, however, are accompanied
by entirely different infrared light curves, with high amplitude long timescale variability and little
to no sign of dips.

Prominent example light curves are shown in
Figures~\ref{perdippers}--\ref{aperdippersIR}. In Section~6.2, we introduce criteria to distinguish
between periodic, quasi-periodic, and aperiodic light curves. The dipper sample here is split
nearly evenly among quasi-periodic and aperiodic behavior, although it is not clear how the
physics variability mechanisms differ between the two groups. We find that
seventeen of the optical dippers are quasi-periodic, compared to four in the infrared; we label them
``QPD'' in Table~4, and these are synonymous with the periodic AA~Tau variables
described in \citet{2010A&A...519A..88A}. While these may be explained by a warped disk
or orbiting companion, it is more difficult to envision a mechanism for the {\em aperiodic} dippers.
A possible origin is that magnetic turbulence induces scale height changes in the inner disk, as
envisioned by \citet{2013AAS...22120505T}.

Also problematic is the total fraction of our dataset displaying dipping behavior. With more than
20\% of the sample in this category, the assumption that dust obscuration is happening in randomly
aligned disks implies that the occulting material lies at scale heights of $\sim$0.2, in units
of distance from the central star. The presence of material this high up in a hydrostatic disk 
might require large surface densities at this radius. 

In several cases, dipper behavior does not persist over the entirety
of the time series. The fact that we detect fewer
dips in the infrared is not simply a selection bias due to the lower
cadence of the IRAC time series; a number of objects display prominent
dips only in the optical, whereas any corresponding dipping behavior at
longer wavelengths is hidden by high-amplitude, long-timescale
behavior originating in the disk. 

In general, dipper light curves display infrared amplitudes that are
less than the optical amplitudes; this supports the idea that the
fading events are caused by enhanced dust extinction. 
We explore the degree of correlation between optical and infrared flux
in dipper light curves, as well as implications for the underlying physics, in Section 7.

\subsection{Short-duration ``bursting'' light curves}

\begin{figure*}
\begin{center}
\includegraphics[scale=0.7]{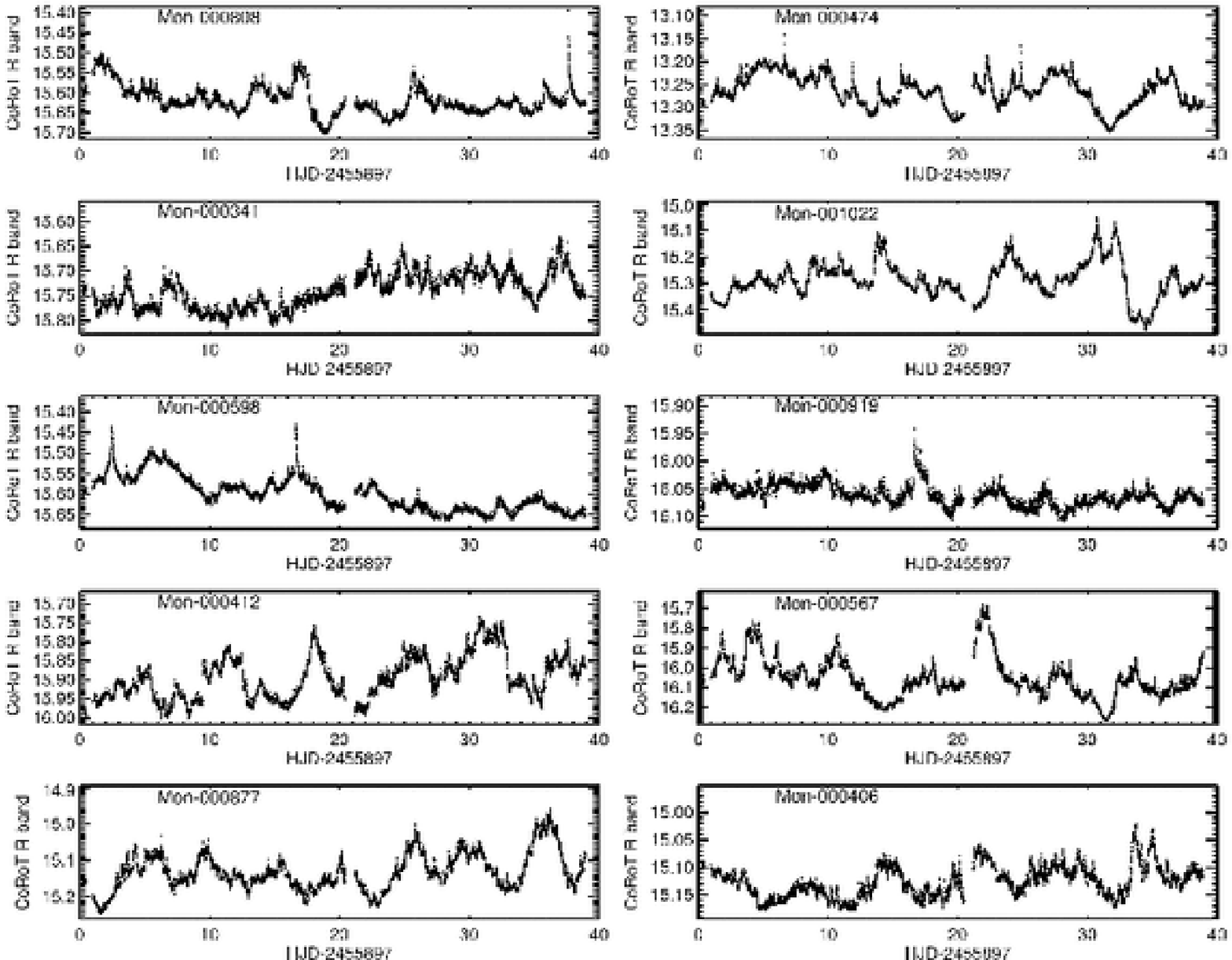}
\end{center}
\caption{\label{bursters} Prototypical bursters: short-duration flux increases
in the optical. The events seen here may represent accretion bursts; some repeat regularly, while others are aperiodic.}
\end{figure*}

\begin{figure*}
\begin{center}
\vspace{-0.2cm}
\includegraphics[scale=0.75]{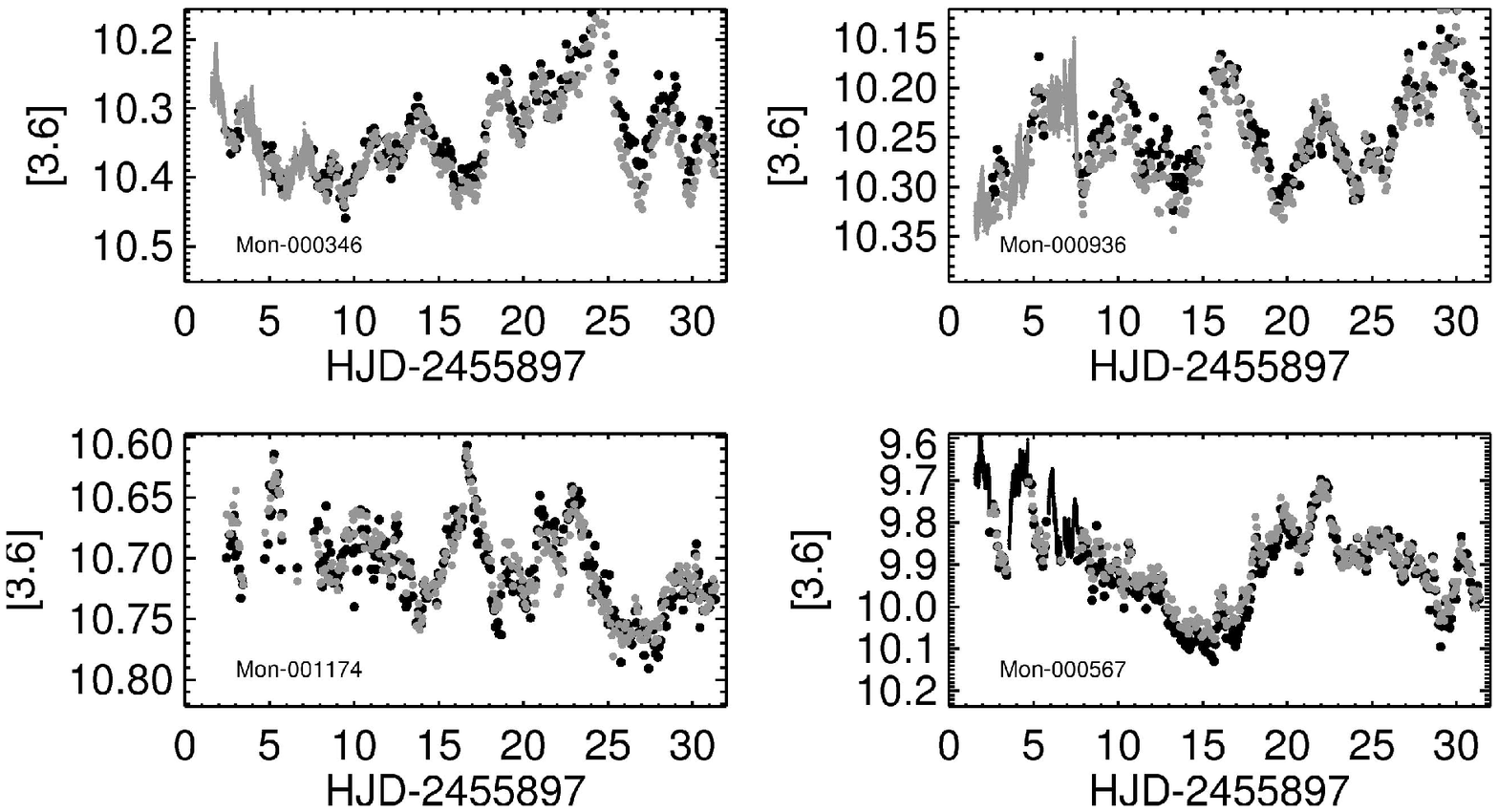}
\end{center}
\caption{\label{IRbursters} Infrared variables detected with bursting
  events in the IRAC staring and mapping time series. Black points are 3.6~$\mu$m data, and grey
  points are 4.5~$\mu$m data.  Densely sampled portions at the
  beginning of the Mon-000346, Mon-000936, and Mon-000567 light curves are staring data.}
\end{figure*}

Dipper light curves are not the only ones displaying asymmetric behavior with respect to flux. We have 
identified a separate set of variables in the {\em CoRoT} dataset that exhibit abrupt (i.e., 0.1--1 day) 
increases in flux, followed by decreases in flux on similar timescales. Examples are shown in 
Figures~\ref{bursters} and \ref{IRbursters}. We refer to these events as short-duration bursts (hereafter 
``bursts'' or ``B'' in Table 4). We wish to distinguish them from more powerful outbursts such as FU Ori events, 
as well as the longer bursts identified in young stars by \citet{2013ApJ...768...93F}. The bursts in our dataset 
can be differentiated from coronal flares by their symmetric shapes (with respect to time), generally much 
longer durations, and sequential nature (i.e., many bursts occur in a row; they are by no means isolated 
events). Our examination of the light curves of weak-lined T Tauri stars in the {\em CoRoT} sample revealed at 
most a few coronal flares per star over the 40-day observing window. Thus while some of our bursting events
may be due to stellar activity, most require a different explanation.

Some YSOs exhibit isolated bursts above a ``continuum'' flux level, while others display similar behavior 
superimposed on a longer term trend. The baseline flux level in bursting light curves of some members of this 
class even displays quasi-periodic variability (i.e., ``QPB'' in Table~4). Since the burst durations are 
typically less than one day, we have in some cases subtracted out underlying trends to see the bursts more 
clearly. It should be noted that we are consequently insensitive to bursting behavior in timescales of $\sim$10 
days or longer, unless there is no long-term trend.

We detect a total of 20 cases of short-duration optical bursting
behavior in our 162 disk bearing {\em CoRoT}/{\em Spitzer} dataset, four of which also occur in the infrared (see Table 4). In
addition, we find three examples (Mon-000119, Mon-000346, Mon-000185) of bursting in the IRAC dataset that
appear stochastic in the corresponding {\em CoRoT} light curves (as discussed in Section~5.5), as
well as one case where the object does not display prominent
variability at all in the optical (Mon-000273).

Bursting behavior in the {\em Spitzer} dataset is much more difficult
to identify with the sparser time sampling of $\sim$12 points per
day in the mapping observations. We only identify a few examples {\em exclusively} in the IRAC time
series but note that in about 50\% of cases, there are increases in
infrared flux coincident with strong bursts in the optical. 
It therefore appears that a combination of short cadences and
high precision ($\sim$1\% or better) is crucial to detecting the full range of timescales involved in bursting, at
least as it appears in our dataset. Longer timescale analogs may be
present in multi-year datasets, such as the one presented by \citet{2013ApJ...768...93F}.  

We believe the bursts are caused by accretion instabilities,
based on their resemblance to the predictions of
\citet{2008ApJ...673L.171R} and strong ultraviolet (UV) excesses
evident for many of these stars. To our knowledge, these are the first
observations of an entire class of objects with such behavior. As such,
a more detailed exploration of the bursters, including average
durations and strengths, is presented in a companion paper \citep{Stauff13}.

We wish to note that while this paper and \citet{Stauff13} are closely linked, they were written in parallel and
evolved somewhat independently.   The two papers used slightly
different sets of data  - most importantly, \citet{Stauff13} included
stars with only 2008 {\em CoRoT} light curves whereas such stars were
excluded here - and slightly different criteria for defining
light curve classes.  This resulted in slightly
different sets of stars belonging to each class.  In
particular, the burst-dominated class in Table 1 of \citet{Stauff13}
included 23 stars; 19 of those are also listed as burst-dominated
in Table 4 of this paper.   Of the remaining four, one (Mon-000185) was
included in \citet{Stauff13} based on its 2008 {\em CoRoT} light curve and
hence was not in the parent sample for this paper.  The other three
were all classified as stochastic in our Table 4.  Inclusion or
exclusion of these stars from the burst-dominated class would not
appreciably change the conclusions in either paper.

\subsection{Quasi-periodic symmetric behavior}

Our formal period search (Section 6.2) uncovered many quasi-periodic light curves in both the {\em 
CoRoT} and {\em Spitzer} datasets. Only five optical and four infrared light curves in our infrared-excess 
restricted dataset displayed strictly periodic behavior typical of the spotted WTTSs in the current and previous 
{\em CoRoT} run on NGC~2264 \citep{2013MNRAS.430.1433A}. This set includes only two stars that are periodic in 
both bands (Mon-001085 and Mon-001205; see Table 4). This is not entirely surprising considering that our stars 
were selected specifically to have disks based on their SEDS, and many are actively accreting based on 
spectroscopy.

\begin{figure*}
\begin{center}
\includegraphics[scale=0.7]{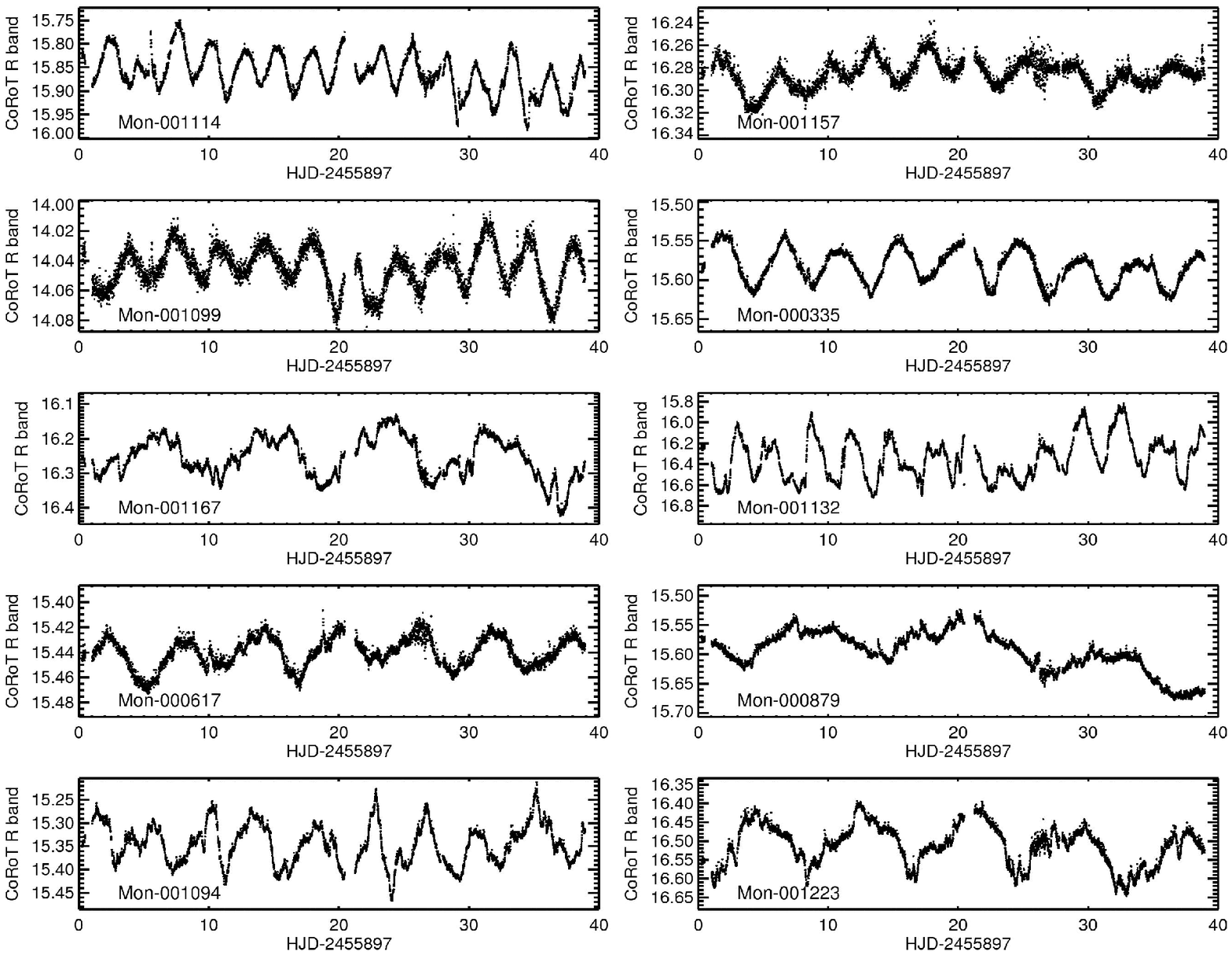}
\end{center}
\caption{\label{ppd} Quasi-periodic optical variability in
  NGC~2264 {\em CoRoT} sources. The patterns in these light curves show regular repetition,
but amplitudes and shapes change from one cycle to the next.}
\end{figure*}

\begin{figure*}
\begin{center}
\includegraphics[scale=0.7]{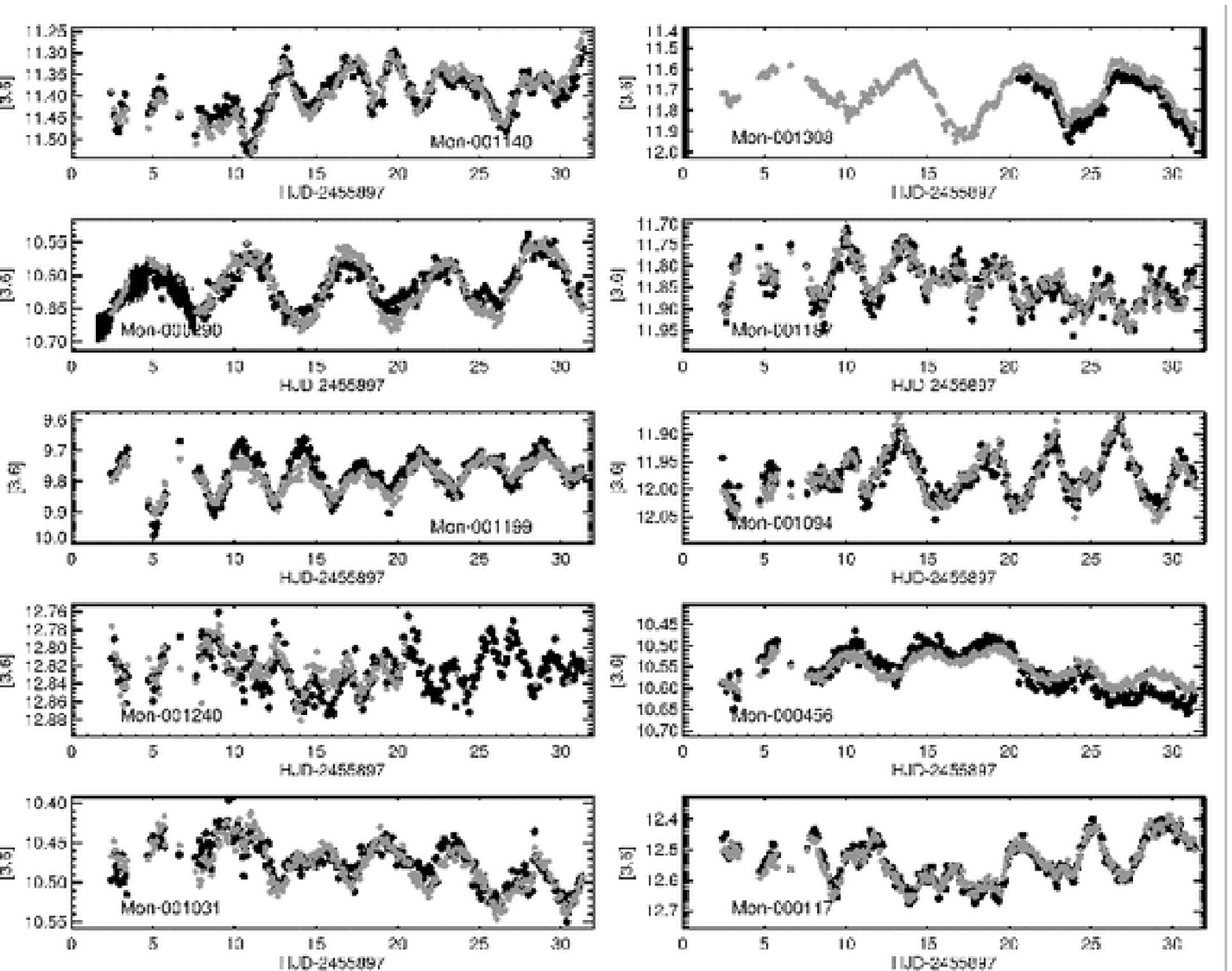}
\end{center}
\caption{\label{ppdIR} Quasi-periodic infrared variability in
  NGC~2264 IRAC sources. Black points are 3.6~$\mu$m data, and grey
  points are 4.5~$\mu$m data.}
\end{figure*}

The remaining quasi-periodic disk-bearing stars exhibit repeating patterns that
either change in shape from cycle to cycle or include lower amplitude
stochastic behavior superimposed on a periodicity.  We refer to these 
as quasi-periodic symmetric, or ``QPS'' in Table 4.
Excluding the quasi-periodic dipper class as well as the handful of quasi-periodic
bursters already accounted for above, we detect 27
quasi-periodic symmetric cases in the optical and 22 in the infrared, with 6 appearing in
both bands. Examples are shown in Figure~\ref{ppd} and \ref{ppdIR}.
Unlike the bursters and dippers, all of these light
curves are relatively symmetric with respect to
an upside-down flip. We consider here only variables with periods less
than half the total observation baseline; longer timescale behavior is
indistinguishable from the ``stochastic'' class, introduced below.

The infrared variables that show repeating patterns do tend to be partially correlated
with the optical behavior, even in cases for which we have assigned different
morphology classes for the two wavelength bands. This may reflect multiple origins for the infrared flux variations: if
a light curve is a superposition of periodic behavior and higher
amplitude aperiodic changes, then we may detect the former but assign
a different overall classification.

We propose two possible origins for this quasi-periodic symmetric
variability. First, it may be a combination of purely periodic
variation (such as cool starspot modulation) with longer timescale aperiodic
changes (e.g., from accretion). Second, it could reflect a single variability process that is
not entirely stable from cycle to cycle. Examples include stellar hotspots
for which brightness evolves as a function of stochastic accretion
flow (e.g., the type IIp variability highlighted by \citet{1994AJ....108.1906H}),
or structural inhomogeneities in the surrounding disk periodically
occulting the central star, but with geometries different from the standard dipper
orientation. The latter is indicated in cases where a $v$sin$i$ value suggests
that the measured period is too long to originate at the stellar
photosphere \citep[e.g.,][]{2012AstL...38..783A,2013AJ....145...79C}.
In addition, we cannot rule out that we are seeing cool spots which evolve
much faster on disk-bearing stars than on their weak-lined counterparts.

\subsection{``Stochastic'' Stars}

\begin{figure*}
\begin{center}
\includegraphics[scale=0.7]{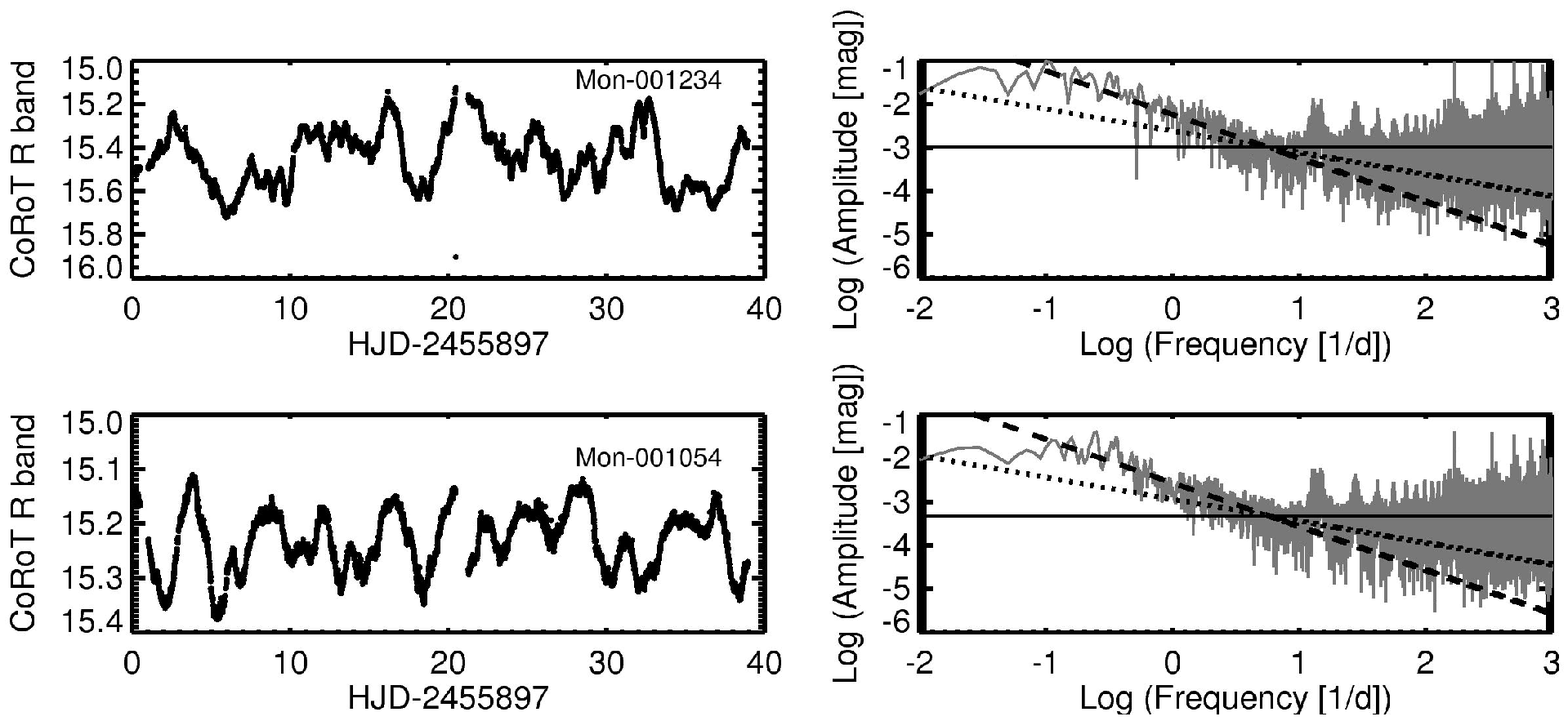}
\end{center}
\caption{\label{flicker}Light curves for two prototypical stochastically behaving stars, along with their Fourier
transform periodograms, shown here in log space. Each Fourier transform consists of three parts: a relatively 
flat low-frequency regime, a steeper red noise trend from log$(f)\sim -1$ to 1, and a flat white noise dominated
regime for log frequencies beyond 1.0. We have fitted the white noise with a single value, shown as a black line.
The red noise is modeled by a $1/\sqrt(f)$ trend in amplitude (i.e.,
1/f in power; dashed line) and a $1/f$ trend in amplitude (dotted line), the latter of
which fits the data much better.}
\end{figure*}

The majority of remaining variables show no preference for fading or
brightening but nevertheless exhibit prominent brightness changes on a
variety of timescales. Fourier transform periodograms (see Cody \& Hillenbrand 2010) for these objects display amplitudes devoid of
dominant periodicities and consistent with a ``1/$f$'' trend in amplitude; two examples are provided in Figure~\ref{flicker}. 
This is in contrast to typical red or
``flicker'' noise, which follows a 1/$\sqrt{f}$ trend in amplitude and 1/$f$ in power \citep{1978ComAp...7..103P}. 
Thus our stochastic objects display more power at low frequency than expected for standard flickering. 
Similar behavior was detected in a Herbig~Ae star by \citet{2010A&A...522A.113R}. 
The combination of lack of periodicity and coherence of the light curves suggests that these objects may 
represent a superposition of variable extinction (as in the dipper stars) and stochastic accretion
(as in the bursting stars). 

\begin{figure*}
\begin{center}
\includegraphics[scale=0.7]{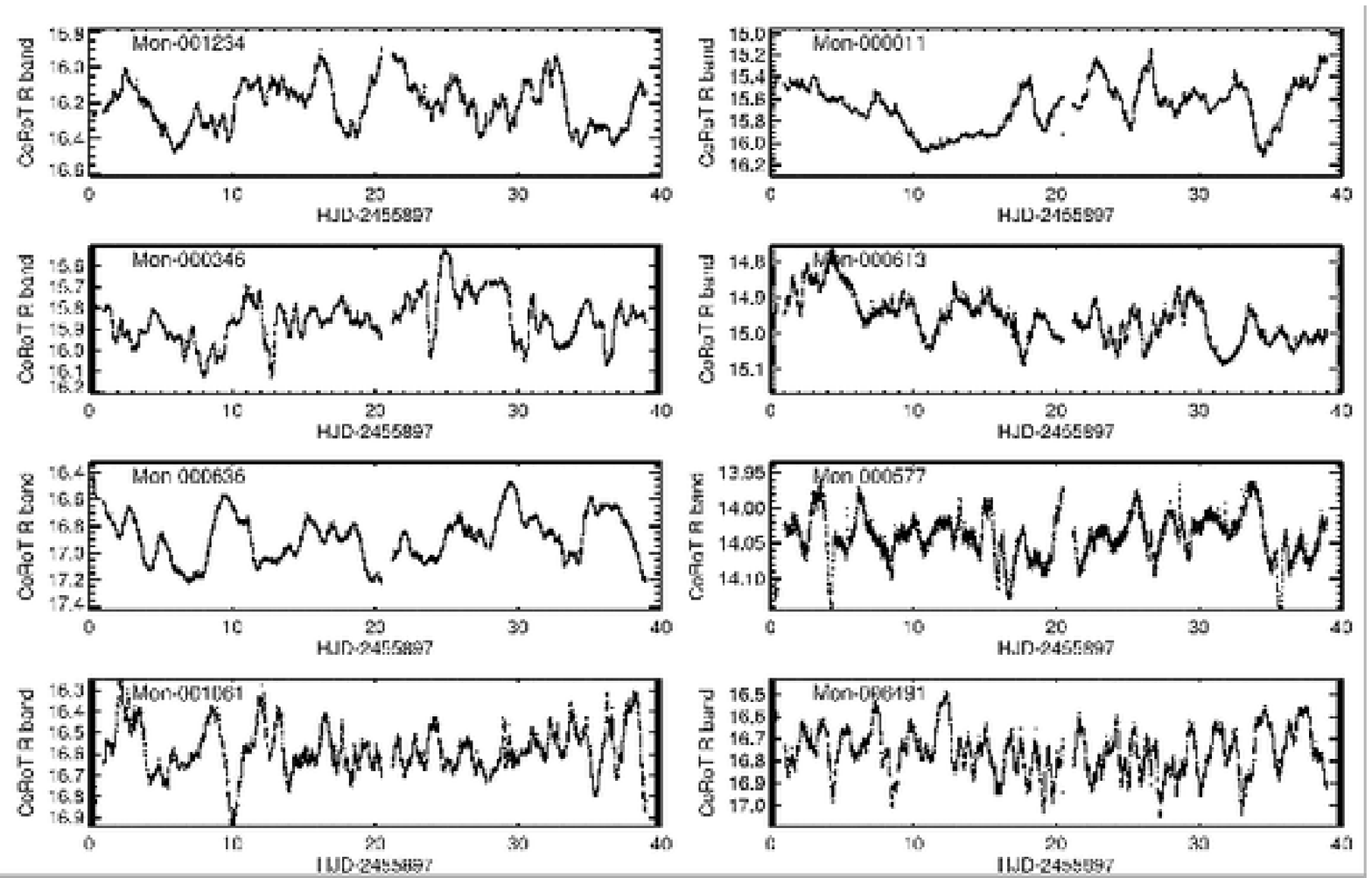}
\end{center}
\caption{\label{optstochastic} Stochastic optical light
  curves in the {\em CoRoT} sample. These variables show no detectable periodicity, nor any preference
for fading or brightening events.}
\end{figure*}

Stochastic behavior comprises one of the largest classes of variability in the
{\em CoRoT} dataset. Figure \ref{optstochastic} illustrates prototypical examples of optically stochastic
behavior. We focus specifically on short timescale stochastic YSOs, 
which have peak-to-peak timescales much less than the
40-day duration of the light curve. There are in fact only two longer duration
variables among the optical sample (see Section~6.5); thus there is a true dearth of optical behavior on 
timescales longer than a week.


Most {\em infrared} light curves are aperiodic, with
fewer examples of quasi-periodic or asymmetric flux behavior (as in
dippers or bursters) than in the optical. The timescales evident in
the mapping data are also somewhat longer.
Since the flux is generally disk dominated (although not by a large factor) at 3.6 and
4.5~$\mu$m, stochastic infrared variability could have a different
origin from that in the optical, although it may simply reflect reprocessed variable
starlight. This latter possibility is supported by the fact that we find stochastic behavior 
to be more than twice as common in the optical band than in the infrared. For
long timescale trends (see Section 5.6), the reverse is true: the behavior is much
more common in the infrared. Orbital timescales in the disk emission region range from a
couple of days to several weeks, depending on the mass and temperature of the star. It is
therefore difficult to produce stochastic infrared behavior without invoking reprocessed
optical variability or long timescales, which we consider separately below. 

\begin{figure*}
\begin{center}
\includegraphics[scale=0.65]{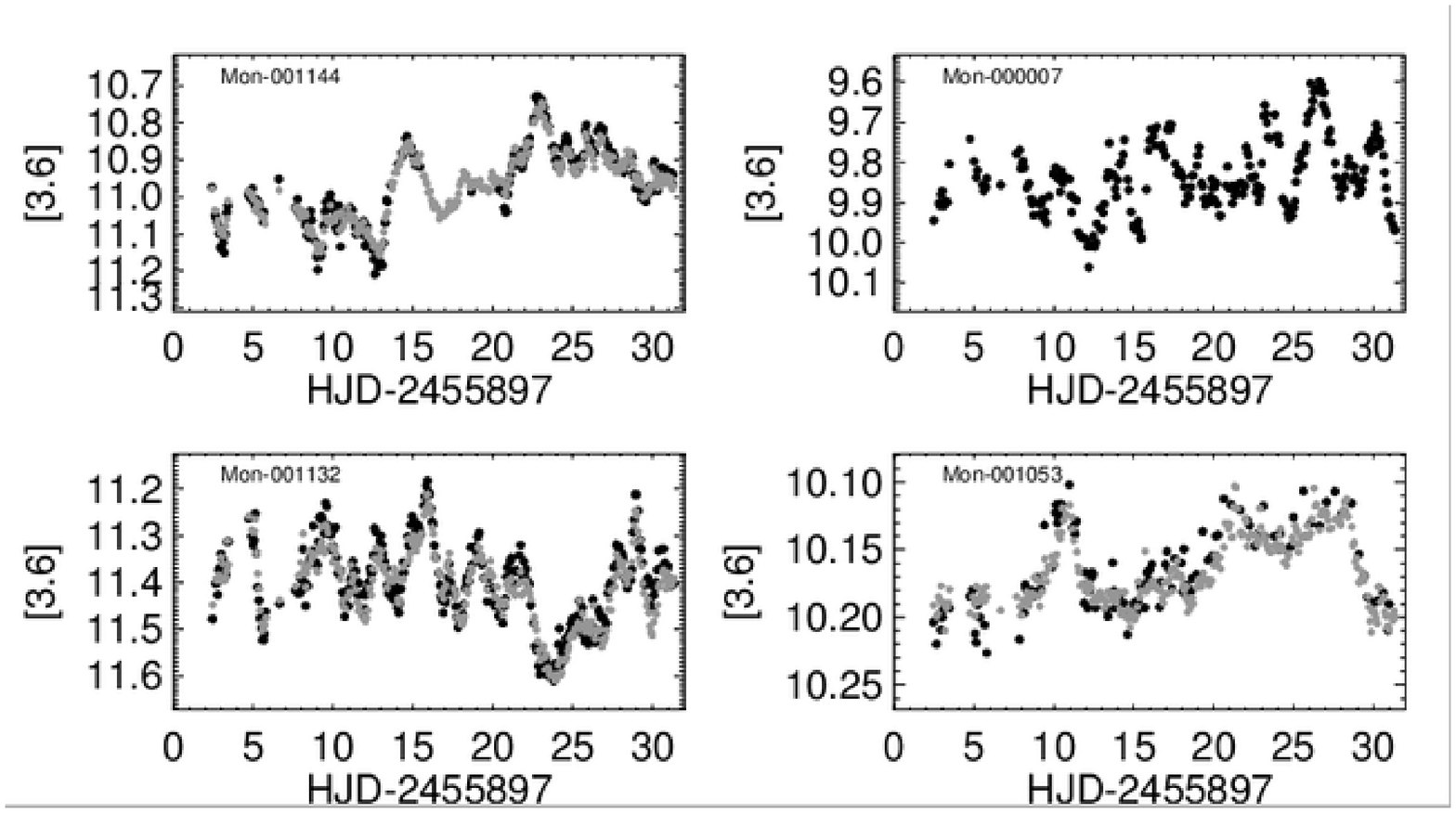}
\end{center}
\caption{\label{IRstochastic} Stochastic infrared light curves
  detected in IRAC time series. Black points are 3.6~$\mu$m data, and grey
  points are 4.5~$\mu$m data.}
\end{figure*}

We present a selection of stochastic infrared light curves in Figure \ref{IRstochastic}.

\subsection{Long timescale behavior}

\begin{figure*}
\begin{center}
\includegraphics[scale=0.7]{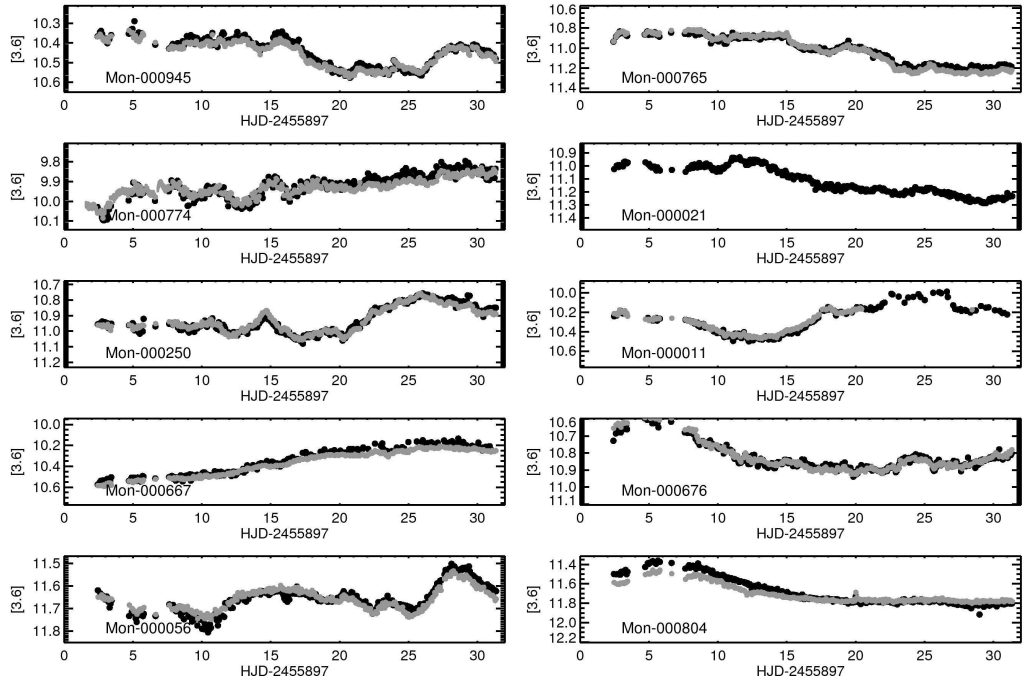}
\end{center}
\caption{\label{IRlong} Infrared light curves with long timescale
variability. Black points are 3.6~$\mu$m data, and grey
points are 4.5~$\mu$m data. The peak-to-peak amplitude range for these
objects is 0.18 to 0.42 magnitudes.}
\end{figure*}

A subset of our disk bearing stars displays variability that grows or declines all the way out to the
longest timescale of observation. We select as ``long timescale'' those objects for which the largest
peak-to-peak amplitudes occur on timescales of 15 days or longer.
 This behavior is much more common among the infrared light curves than
those in the optical. Because of the 30--40 day duration of our time series, we are unable
to classify the trends morphologically and simply refer to them as
``long timescale'' variables. The long duration of flux changes likely
reflects disk dynamics beyond the inner edge, where Keplerian
timescales are longer than one week. It also involves fairly large
amplitude changes, from 0.08 to 0.6 magnitudes among the sample
identified here. We present prototypical examples
of long-timescale infrared behavior in Figure~\ref{IRlong}.

Notably, the optical behavior of these objects involves much shorter timescales.
Only two {\em CoRoT} light curves contained trends gradual enough to be
classified as ``long timescale.'' The rest fall into a variety of classes,
from quasi-periodic dippers to stochastic. 

The distribution of stochastic and long timescale stars in each band suggests that
while some infrared variability reflects reprocessed starlight, this is a minor
contribution to the flux. Some other process must dominate the
long wavelength variability at long timescales, and this could involve changes in
the disk luminosity or shape.

\subsection{Unusual variability types in disk-bearing YSOs}

There are several notable cases of rare light curve types that
merit discussion, since they may signify unusual geometry, disk
evolutionary states, or variability processes.

\begin{figure}
\begin{center}
\includegraphics[scale=0.4]{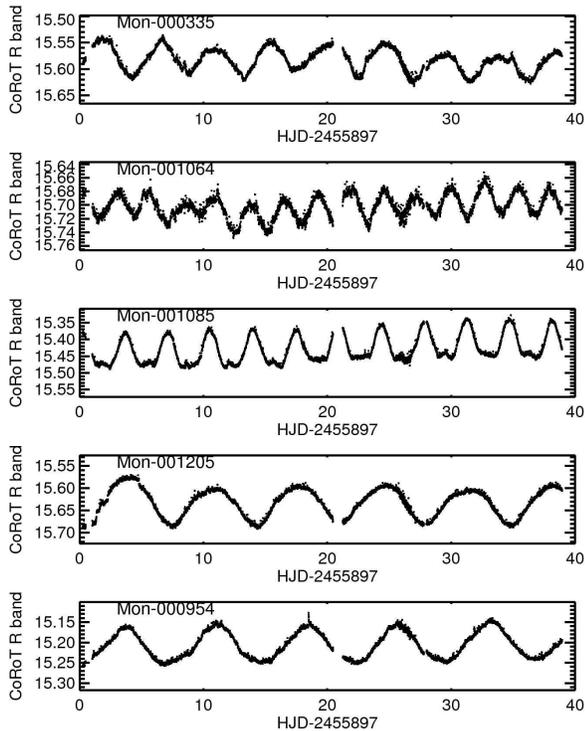}
\end{center}
\caption{\label{pureper}Periodic (as opposed to quasi-periodic)
  optical light curves. These are the most regular light curves in the entire disk bearing dataset. 
The weak infrared excesses, H$\alpha$, and low
UV excesses of these objects suggest that they host cool magnetic
spots, which appear in the light curves via rotational modulation.}
\end{figure}

We detect purely periodic optical behavior in only five stars (Mon-000335, Mon-001064, Mon-001085, 
Mon-001205, and Mon-000954), which we show in Figure~\ref{pureper}. The latter
two are nearly sinusoidal.  These light
curves are not easily distinguishable from the magnetically active
WTTSs, and in fact all but two of them have weak or transitional disk
SED classifications (``II/III''), along with weak reported H$\alpha$
emission equivalent widths and low UV excesses. By some schemes, these sources could have
been included with the WTTS set. We therefore believe the variability here reflects cold spots on the stellar
surface in systems where circumstellar dust obscuration and variable accretion 
are minimal. Nevertheless, it could be generated by very stable accretion
flow onto a hotspot in the two objects that have significant infrared
excesses (Mon-000335 and Mon-001064). 

\begin{figure}
\begin{center}
\includegraphics[scale=0.55]{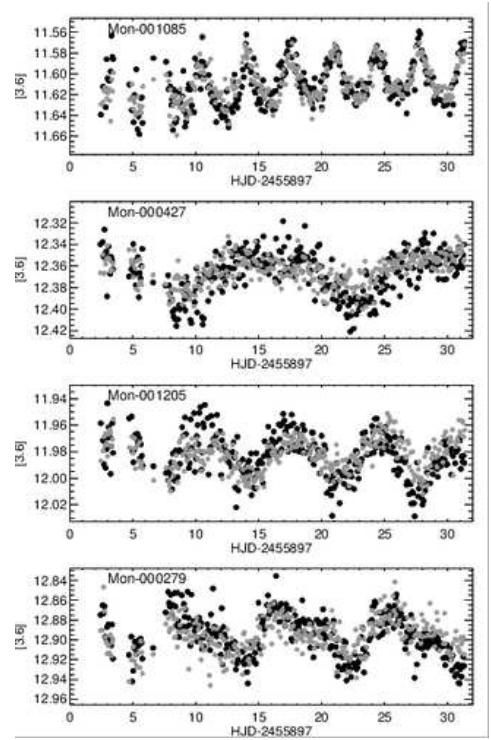}
\end{center}
\caption{\label{pureperir}Periodic infrared light curves. Black points are IRAC 3.6~$\mu$m data,
and grey points are 4.5~$\mu$m.}
\end{figure}

Similar to the optical light curves, we detect only four instances of periodic
infrared behavior, shown in Figure~\ref{pureperir}: Mon-000279, Mon-000427, Mon-001205,
and Mon-001085. The latter two of these are also periodic in the optical, likely reflecting the
variability induced by rotational modulation of starspots.

\begin{figure}
\begin{center}
\includegraphics[scale=0.45]{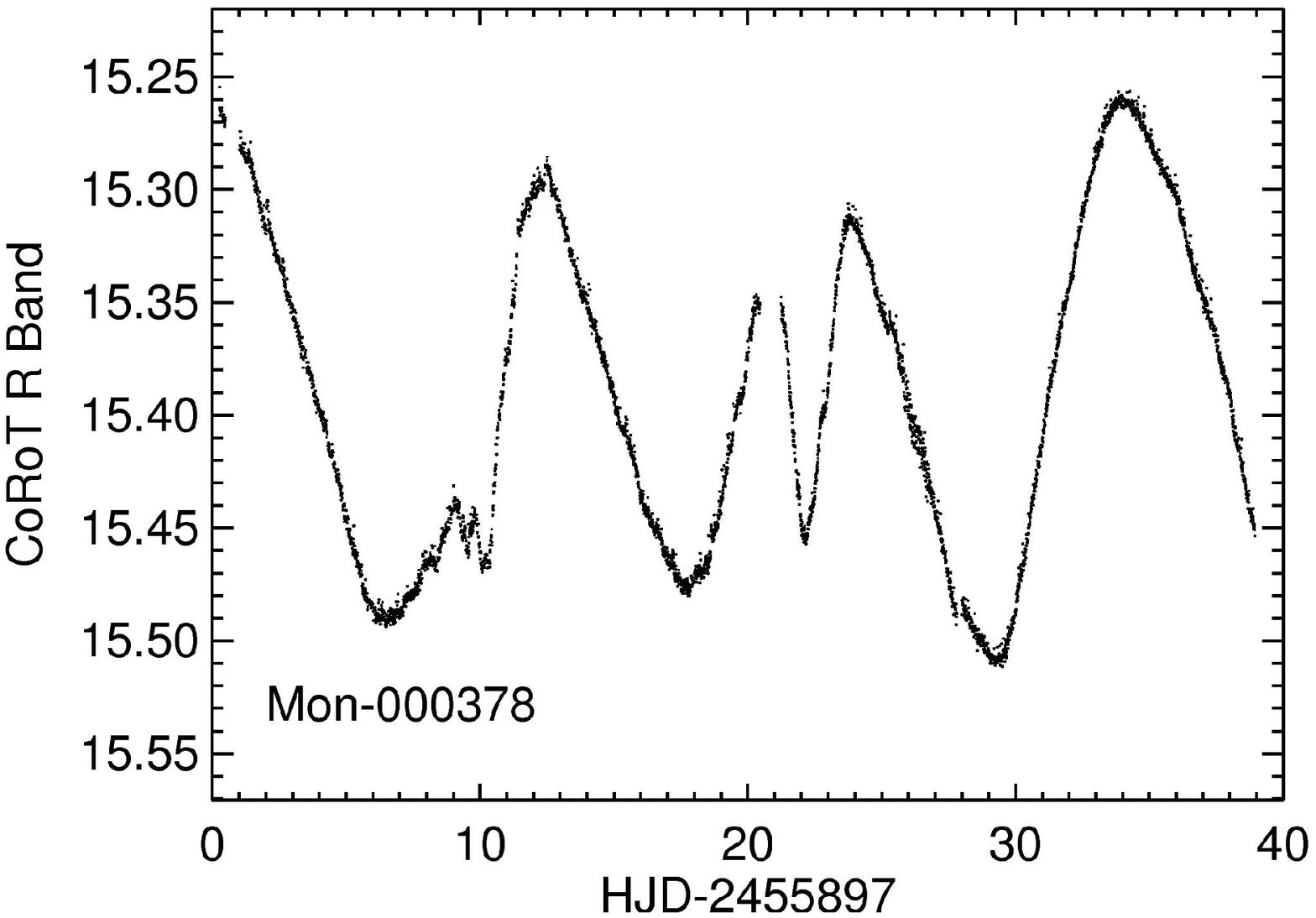}
\end{center}
\caption{\label{Mon378} The unusual optical light curve of
  Mon-000378. This may be a combination of periodic behavior and
 circumstellar obscuration.}
\end{figure}

We also highlight an unusual optical light curve consisting of a periodic 
morphology interrupted by fading episodes, as shown in Figure~\ref{Mon378}. The
origin of the variability in this K5.5 star, Mon-000378, star may be a combination
of spots interspersed by circumstellar extinction dip events; this is
difficult to verify since the infrared behavior does not correlate
with the optical in this case. 

\subsection{Statistical division of light curve morphologies}

\begin{figure}
\begin{center}
\includegraphics[scale=0.6]{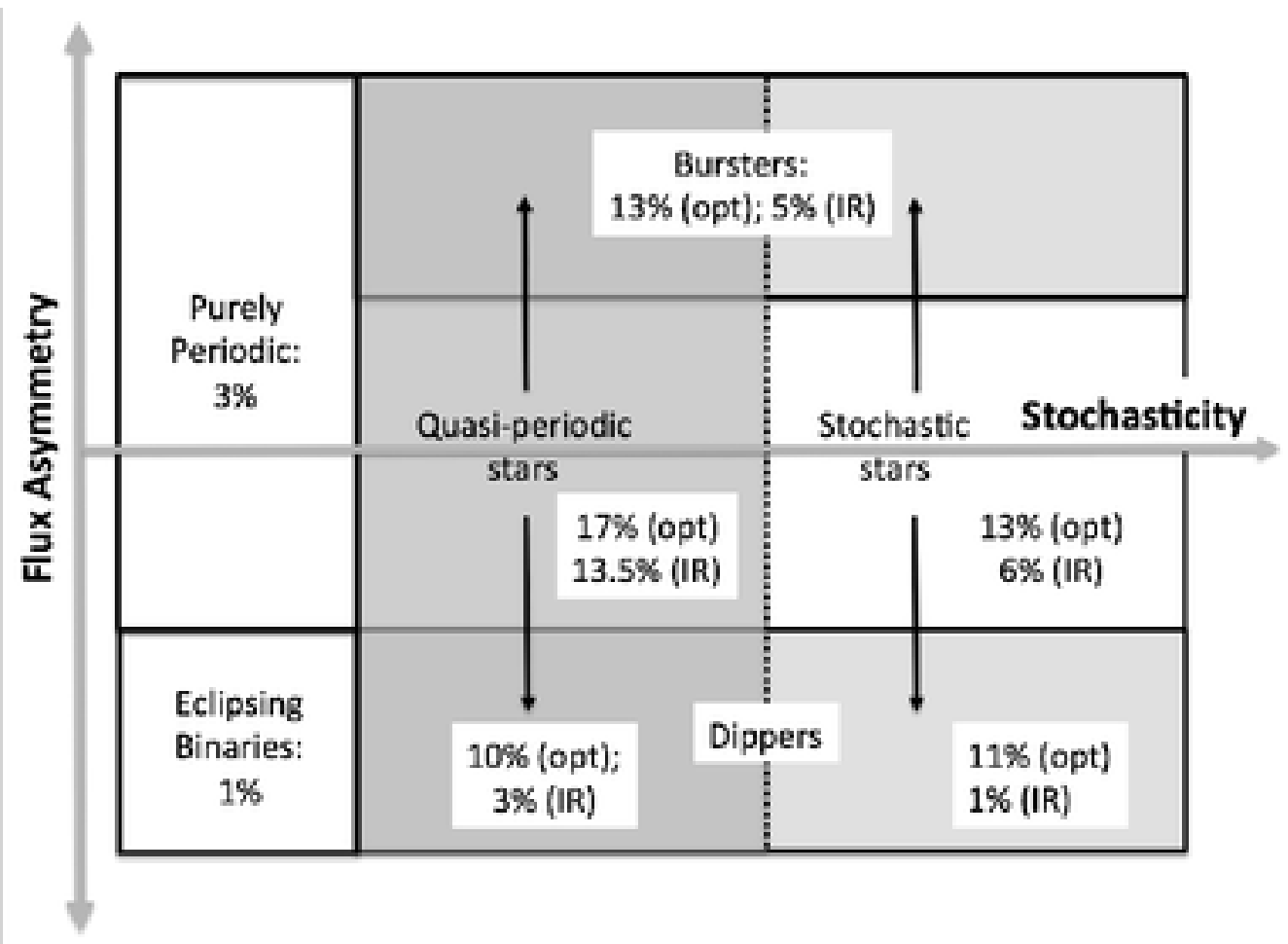}
\end{center}
\vspace{-0.5cm}
\caption{\label{vartypes1} Schematic of the optical and infrared
  variability types detected in our 162-member disk bearing dataset, with optical (``opt'')
  and infrared (``IR'') fractions. Not noted here are the percentages of long timescale and unclassifiable
variables, as well as non-variables; hence the listed values do not add up to 100\%.}
\end{figure}

The fractions of different variability types offers insights into the
diversity and relative importance of different physical mechanisms at
work in the inner disks, magnetospheres, and stellar photospheres of
young stars. We reproduce the schematic of the different variability types in
Figure~\ref{vartypes1}, with their fractions overlaid.

The fractions for different types of optical variability
are also presented in Table~5. A subset of lightcurves did not fall into any of the
morphological classes described above, either because the data were
noisy or the light curve shape was unusual. We denote these as
unclassifiable in Tables 3 and 5.

The non-variable stars are those sources that did not pass the Stetson test (infrared)
or RMS test (optical and single band infrared). Among the
{\em CoRoT} light curves, we are only able to distinguish variability down to RMS values
of 0.02 at $R$=14 and 0.06 at $R$=16. It is likely that the 19\% of optical sources
that are not variable according to our statistical measures simply have lower
level brightness fluctuations.

Of note, the total fraction of dippers (both periodic and aperiodic) is just over 21\%,
a value that significantly exceeds the prediction of \citet{2000A&A...363..984B} (15\%),
but is not as large as the $\sim$30\% found by \citet{2010A&A...519A..88A}. These results
suggest that the frequency of dipper behavior is surprisingly large, given typical inner
disk scale height estimates (e.g., $H/R\sim$0.05).

Assessment of our infrared dataset revealed that long timescale
variability was the most common type, and the other variability groups were less populated. The
associated fractions for each infrared morphology class are 
also included in Table~5. Evidently, infrared variability is more common than optical variability,
at least at the wavelengths and timescales sampled by these
observations of disk bearing sources. 

\LongTables
\addtocounter{table}{+2}
\begin{deluxetable}{ccc}
\tabletypesize{\scriptsize}
\tablecolumns{10}
\tablewidth{0pt}
\tablecaption{\bf CSI 2264 morphology fractions in disk bearing stars}
\tablehead{
\colhead{Morphology class} & \colhead{Optical} & \colhead{Infrared} 
}
\startdata
Bursters & 13$^{+3}_{-2}$\% (21) & 5$^{+2}_{-1}$\% (8)  \\
Periodic dippers & 10.5$^{+3}_{-2}$\% (17)  & 3$^{+2}_{-1}$\% (5)  \\
Aperiodic dippers & 11$^{+3}_{-2}$\% (18)  & 1$^{+2}_{-1}$\% (2)  \\
Quasi-periodic symmetric & 17$\pm$3\% (27)  & 13.5$^{+3}_{-2}$\% (22)  \\
Stochastic & 13$^{+3}_{-2}$\% (21)  & 6$^{+2}_{-1}$\% (9)  \\
Long timescale & 1$^{+2}_{-1}$\% & 30$^{+4}_{-3}$\% (48) \\
Periodic &3$^{+2}_{-1}$\% (5)  & 2.5$^{+2}_{-1}$\% (4)\\
Multiperiodic & 1$^{+2}_{-1}$\% (2)  & 1$\pm$1\% (1)  \\
Eclipsing binary & 1$\pm$1\% (1)  &1$\pm$1\% (1)  \\
Unclassifiable & 11$^{+3}_{-2}$\% (18)  & 29$^{+4}_{-3}$\% (47)  \\
Non-variable & 19$\pm$3\% (30)  & 9$^{+3}_{-2}$\% (15)   
\enddata
\tablecomments{We list the fractions of objects in each variability
  class, along with the number (out of the 162 member disk bearing
  object set) in parentheses. These classifications have been made by
  eye, whereas we provide statistical support for them in Section~6.4}
\end{deluxetable}

\subsection{Assessment of sample bias}

In order  to compare our identified variability
types with models, we must first account for any biases introduced by the
{\em CoRoT} sample selection. Targets for this telescope were selected in advance, whereas 
we observed all stars in a $\sim$1 square degree region with {\em Spitzer}/IRAC.
{\em CoRoT} targets were selected primarily with NGC 2264 membership in mind,
and priority was given to known CTTSs. Since we have already restricted our
light curve morphology sample using membership and infrared excess, there should be no 
associated bias. There was, however, knowledge of variability from {\em CoRoT}'s SRa01 observations in 2008, 
and one might ask whether this contributed to more variables being observed than in a 
randomly selected sample of CTTSs. We argue that the answer is no, for the following reasons.

First, we have examined the input target priority list for the 2011 {\em CoRoT} observations. It is this set from which
the final target list and pixel masks were selected. Of 772 possible high priority
targets, 350 were not previously observed by {\em CoRoT}, and hence should have no variability bias.
Of the 474 targets that were already observed in 2008, {\em none} had variability as the sole signature of youth.
All input stars had prior membership information consistent with our criteria described in the Appendix and therefore
should not be biased by previous knowledge of variability. 

To probe further for systematics in our variability class fractions, we have selected a set of disk bearing NGC 2264 
members that were observed with {\em CoRoT} in 2008 and {\em Spitzer} in 2011.  In this way, we assemble a sample 
similar to the 2011 dataset highlighted in the rest of this paper, but with no prior knowledge of variability (apart 
from some sparser ground-based data, such as that of \citet{2004A&A...417..557L}). We classified the optical morphologies into 
the same categories described in this section and computed their fractions. We find that the fraction of 
non-variables is actually {\em higher} in 2011 than 2008 (19\% vs.\ 11\%), again refuting the idea that variability 
knowledge biased the input catalog, thereby affecting the results reported here. On the contrary, we attribute the 
higher fraction of non-variables in 2011 to a slightly fainter sample. The total number of disk bearing stars 
approximately doubled from the 2008 to the 2011 {\em CoRoT} run, and many of the newly observed sources were fainter 
and thus had lower signal-to-noise than the typical SRa01 target.

Comparing the distribution of variable types in 2008 versus 2011, we find that 
in every one of the categories presented in Sections 5.1--5.6, {\em except} for the aperiodic dippers,
the fraction of stars in each class for 2008 is the same to within the uncertainties presented above in Section 5.8.
The fraction of aperiodic dippers decreased from $\sim$20\% in the 2008 dataset to 11\% in the 2011 dataset.
This change is offset by the increase in the fraction of non-variable sources in 2011, leaving all of the 
other class fractions the same to withint the 1-$\sigma$ errors. We speculate that the slightly fainter (and hence lower mass) 2011 sample 
may bias against the detection of aperiodic dipper behavior. However, it is not clear as to why the {\em periodic}
dipper fraction is then similar in both samples. Because of this discrepancy in the aperiodic dipper fraction,
we suggest that its uncertainty be increased from $\sim$3\% to $\sim$10\% in any theoretical attempt to 
account for the distribution of variability types.

\section{Statistical identification and properties of variability}

By eye, both the optical and infrared
light curve sets display a wide variety of behaviors, including varied
morphologies and timescales. Yet classification and understanding of the variability properties of our
disk-bearing stars benefits from a quantitative approach.
At some level, we are likely to identify all accreting stars as
variable, given sufficient precision.  We wish to determine
the fraction of our light curves that are variable at the $\sim$1\%
level, thereby defining a subsample of objects to which we
can attach a confident morphological classification. We therefore
measure a suite of light curve statistics for further classification, including
amplitudes, standard deviations, symmetry metrics, and timescales.

\subsection{Statistical selection of variables}

\subsubsection{IRAC dataset}
The high precision and cadence as well as large sample size of our {\em Spitzer} and {\em CoRoT}
datasets permits detection of stellar variability down to fairly low
amplitudes in comparison to previous photometric monitoring
studies. Since our focus here is on disk bearing stars, 
we might expect the majority of targets to display non-sinusoidal or
aperiodic variability dominated by accretion and circumstellar effects
rather than the cool spot rotational modulation that explains well the
periodic variability of the WTTSs, as pointed out by \citet{1994AJ....108.1906H}. We have therefore taken several approaches 
to identifying photometrically variable objects in the photometry.

For IRAC data, we take advantage of the fact that nearly simultaneous light curves are
available at both 3.6 and 4.5~$\mu$m for the majority of
objects (151/162 stars in the disk-bearing {\em CoRoT}/IRAC sample). Futhermore, the behavior in these two bands should be
similar, since their emission regions in the disk are at comparable radii. This enables identification
of correlated variability via a Stetson cross-correlation index
\citep{1996PASP..108..851S}, the use of which was previously promoted
for variability detection in young stars by \citet{2001AJ....121.3160C} and \citet{2008ApJS..175..191P}. 
For data in two bands, this index is defined in the following way:
\begin{equation}
Stetson=\Sigma_{i=1}^{N}{\rm sgn}(P_i)\sqrt{|P_i|},
\end{equation}
where $N$ is the number of simultaneous pairs of observations in the
two bands, sgn is the sign function (i.e., -1 for negative input values, 0 at zero,
and 1 for positive input values), and $P_i$ is the product of the normalized residuals of two
observations:
\begin{equation}
P_i=\frac{N}{N-1}\left(\frac{d_{1i}-<d_1>}{\sigma_{1i}}\right)\left(\frac{d_{2i}-<d_2>}{\sigma_{2i}}\right),
\end{equation}
where $d_{1i}$ and $d_{2i}$ are simultaneous photometric points in bands 1 and 2,
respectively, $<d>$ refers to their means, and $\sigma_{1i}$,
$\sigma_{2i}$ their uncertainties, including systematics. 

Since data taken at each wavelength is not exactly simultaneous, we interpolate
one of the light curves onto the time stamps of the other. 
We find that the Stetson index has a roughly constant baseline 
regardless of stellar brightness. Objects with Stetson values well
above this level have a degree of correlation in the two bands that
cannot be accounted for by random noise. We determine a threshold for
variability by examining the Stetson index distribution of likely
field stars that do not meet any of the NGC~2264 membership criteria
outlined in the Appendix, as displayed in Figure \ref{stetdist}. These have a low probability of being
variable and should therefore reflect the intrinsic noise spread in Stetson
index. We fit a Gaussian profile to the distribution, finding a
1--$\sigma$ width of 0.07. We therefore adopt a variability threshold
of 0.21, for 3--$\sigma$ confidence; this is denoted by the dotted line in Figure~\ref{stetdist}.  
The distribution of Stetson indices for likely cluster members (also shown in Figure \ref{stetdist}), 
displays a break at this value, confirming that this cut-off is a reasonable dividing line
between variable and non-variable objects. 

\begin{figure}
\begin{center}
\includegraphics[scale=0.50]{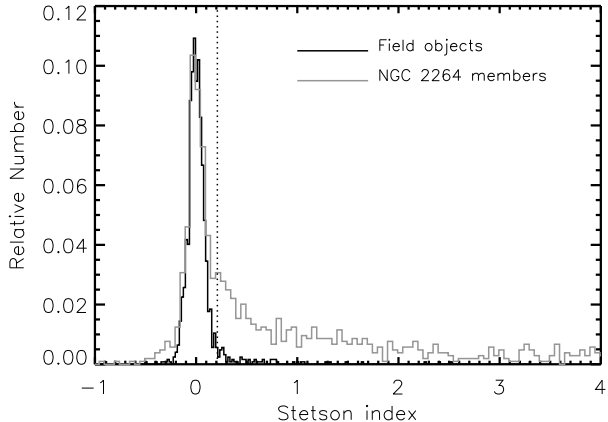}
\end{center}
\vspace{-0.5cm}
\caption{\label{stetdist} Distribution of Stetson indices for objects with mapping
 light curves in both IRAC channels. Light curves of known cluster
 members are clearly more correlated than those of field stars. Our adopted 3-$\sigma$ cut-off for variability,
as determined from the field object distribution, is shown as a dotted line.}
\end{figure}

To select infrared variables among the set of targets with only
single-band IRAC mapping or staring data, we must rely on another variability
criterion. Light curve RMS is a suitable metric, as long as we take
the systematic noise contribution into account. We incorporate the 
noise model determined in Section 2.2 to compare the measured RMS
values against the expected noise level in each mapping light
curve. For staring light curves, we use the 
estimated uncertainties, since these are a good approximation to
the errors. Each of the four staring time series is treated separately. 

\begin{figure}
\begin{center}
\includegraphics[scale=0.51]{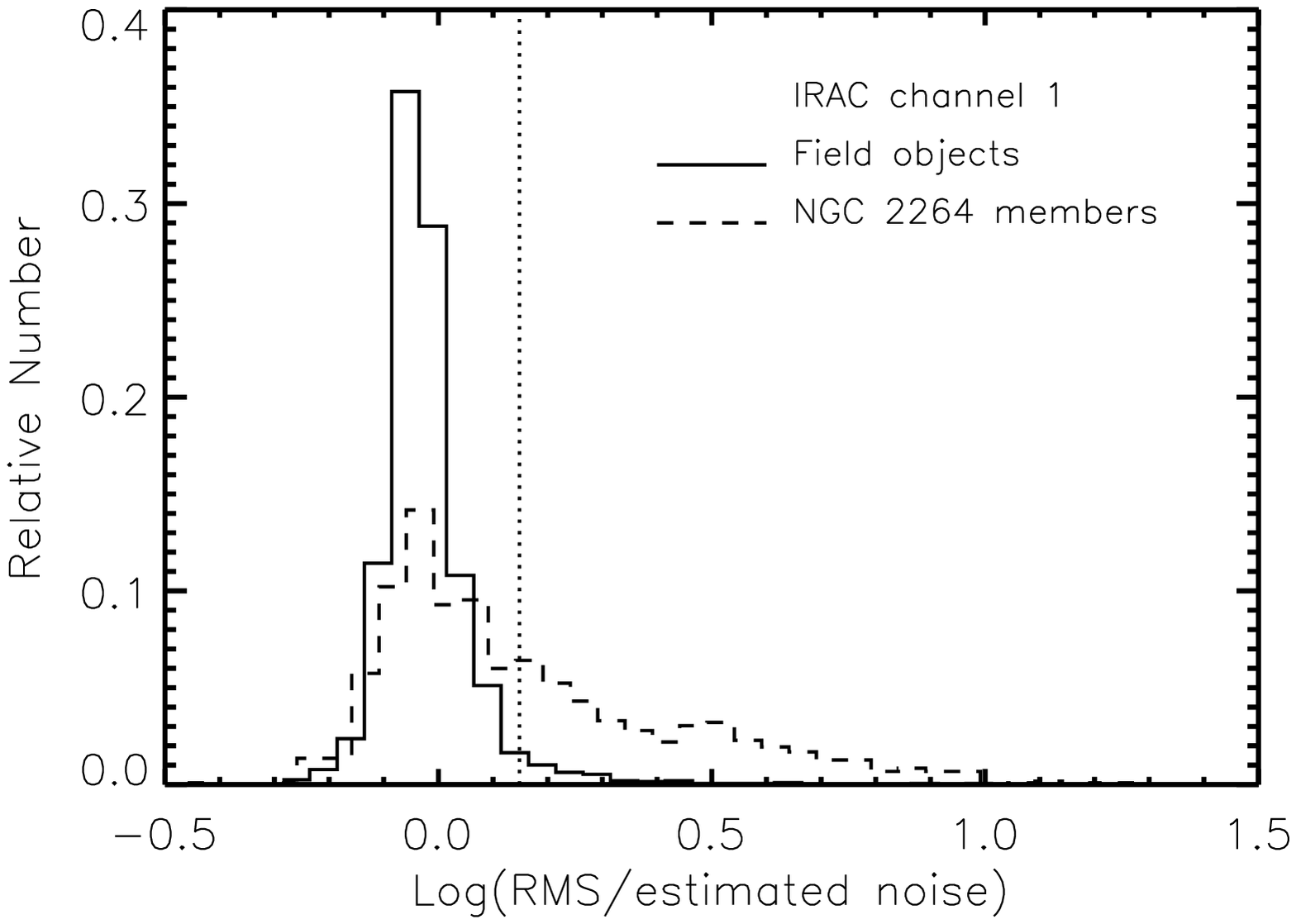}
\includegraphics[scale=0.51]{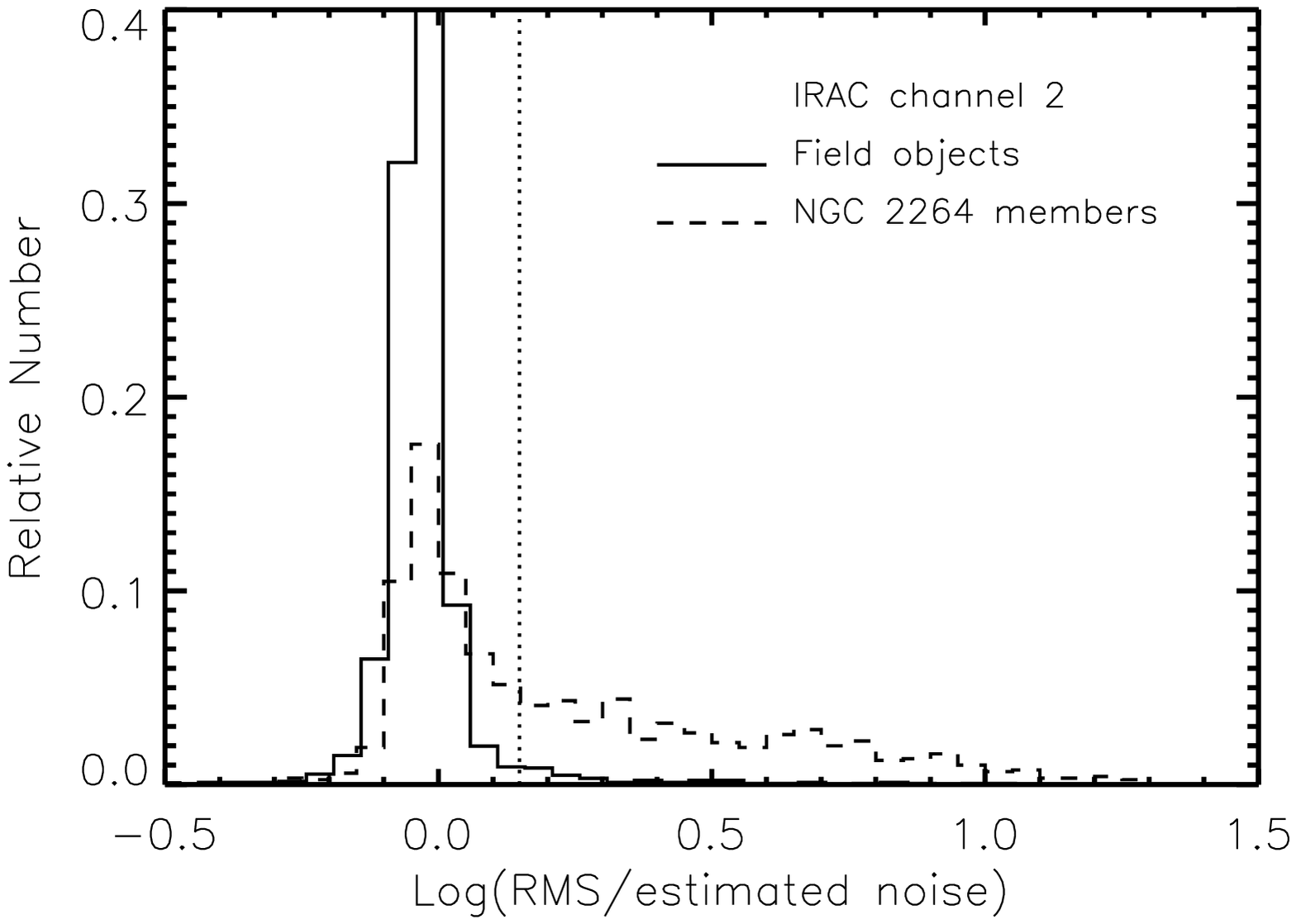}
\end{center}
\vspace{-0.5cm}
\caption{\label{iracrmsdist} Histograms of RMS values from IRAC mapping light curves in channel 1 (top) and
channel 2 (bottom). The dotted vertical line indicates the 3--$\sigma$ variability cut-off determined from the
RMS distribution of field objects.}
\end{figure}

We determine a cut-off value
in the difference between measured and expected $\log$(RMS).
As with the assessment of systematic errors, we have divided our
sample into sets of field stars and confident NGC 2264 members. 
Stars that are candidates but not confident members are not included
in the determination of the cut-off, but they are evaluated for variability
afterwards. We take the
distribution of RMS values for the group of field stars to be representative
of non-variable behavior, and find it to be approximately
Gaussian. The excess of RMS beyond expectations is much
larger for the second group, as shown in Figure \ref{iracrmsdist}. We adopt as our
single-band variability criterion a cut-off of 3--$\sigma$, or 0.15, in the
distribution of $\log$(RMS/RMS$_{\rm expected}$). This is indicated by dotted 
vertical lines in Figure~\ref{iracrmsdist}. For stars with data
in both IRAC bands, we find that the RMS test is not as sensitive as
the Stetson index in detecting variability.

Out of the 162-member disk-bearing dataset considered here, 11 do not
have multi-band {\em Spitzer} data and nine (82\%) of these are variable by the RMS
criterion. In the subset that {\em do} have data in both bands, 137 of
151 (91\%) are variable by the Stetson index. In contrast, only 36\% of stars 
{\em without} disks detected by {\em Spitzer} are variable in the infrared, and
many of these display sinusoidal patterns attributable to rotating starspots.

Objects with amplitudes larger than 0.2 magnitudes were inspected on the raw IRAC images to
determine whether stray light from a bright neighboring star might be
inducing artificial variability via background contamination. This is
often the case when an object with both 3.6 and 4.5~$\mu$m data is detected as variable based on its
RMS, but not based on the Stetson index. A handful of candidate
variables were eliminated in this way.

\subsubsection{{\em CoRoT} dataset}
Selection of variables in the {\em CoRoT} dataset is more difficult than
with {\em Spitzer}/IRAC given the systematic
effects (e.g., jumps, hot pixels) present in some of the light curves. 
Defects remained even after correction of the two most prominent
jumps in the light curves (see Section 3.2). 

For variable selection, we first considered the full set of YSOs
observed by {\em CoRoT} and then narrowed our focus to the 162
disk-bearing stars with simultaneous {\em Spitzer} data.
We determined the median trend in RMS as a function of
magnitude, using 0.5 mag bins and omitting objects that were
flagged or known NGC~2264 members (implying high likelihood of
variability). Despite the unexplained structure at $R>14$ (see Figure~\ref{corotrms}), the median
fit in log space is well modeled by the Poisson noise expectation,
shifted upward by a constant value of 0.35. We therefore adopted this
as the underlying noise distribution. The fit breaks down for $R<12$,
and here we simply adopt the median trend in RMS, which is roughly
linear. 

\begin{figure}
\begin{center}
\includegraphics[scale=0.5]{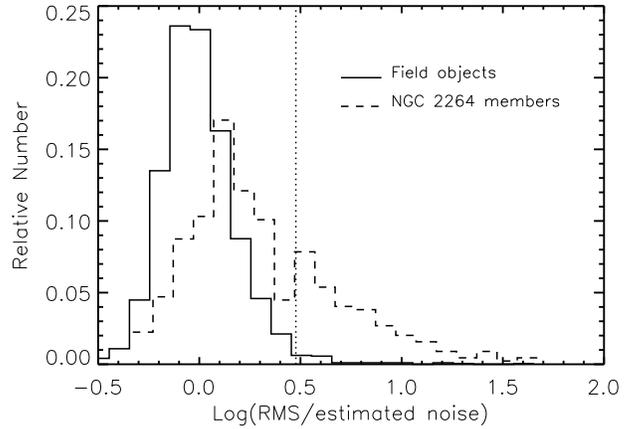}
\end{center}
\caption{\label{corotrmsdist} Distribution of {\em CoRoT} light curve RMS values divided by estimated uncertainty.
Cluster members are clearly more variable than field stars, although
the distribution is determined in part by pre-selection of targets, some of which were observed to be
variable during the 2008 {\em CoRoT} run. The vertical dotted line
marks the 3--$\sigma$ cut-off that we have determined from the
non-member RMS distribution; this was used for variable selection.}
\end{figure}

We require that variables have RMS values at least three times the
median noise level for their magnitude. In Figure \ref{corotrmsdist}, we display the
distributions of $\log$(RMS/noise) for likely cluster members and field
stars. The cluster members are clearly more variable, and the chosen
cut-off at RMS/noise=3.0 selects variables at the $\sim$3--$\sigma$ confidence level, according
to a fit of the non-member distribution. Additional low-amplitude periodic objects that did not
meet this RMS selection criterion were identified via periodogram and
autocorrelation function analysis, which we discuss below.

\begin{figure*}
\begin{center}
\includegraphics[scale=0.7]{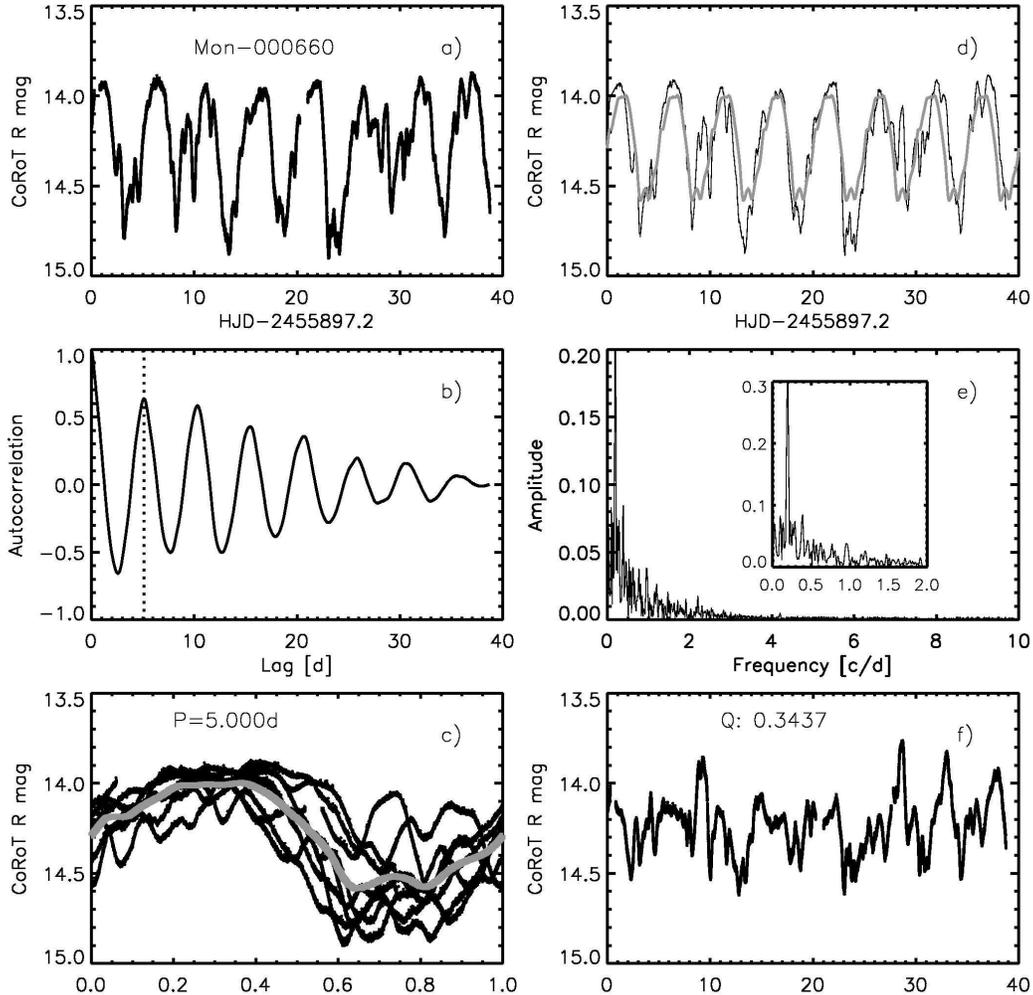}
\end{center}
\caption{\label{acfexample}Example of the process we have used to
  identify periodic and quasi-periodic variables. In panel a) we show
  the raw light curve for Mon-000660, and in panel b) its
  autocorrelation function with the highest peak marked. Zooming in on
the corresponding frequency in the periodogram in panel e), we refine
the frequency by noting the maximum here. We then phase the light
curve to the corresponding period in panel c), producing the boxcar
smoothed trend seen in grey. We overplot that trend on the raw light
curve in panel d), and subtract it from the raw light curve in panel
f). This remaining flux is then used to compute the $Q$ value and determine
that substantial residuals make this light curve quasi-periodic.}
\end{figure*}

We find the chosen threshold to be adequate in that all of the selected objects are
clearly variable by eye. However, there is a collection of faint
($R>16$) objects in the {\em CoRoT} sample that are also clearly variable
by eye, but do not meet the selection criterion since their RMS values are $<$0.05~mag.
The brightness fluctuations in these cases consist of short duration upward
or downward events that depart significantly from the median value but
are too transient to contribute significantly to the overall variance.
The coherence of the light curves during the events causes them to stand
out in comparison to the artificial fluctuations caused by {\em CoRoT}
systematics. In a number of cases, this variability is supported by strong
correlation with infrared behavior in the corresponding IRAC light curve.
We identified a handful of additional such variables by eye, as
indicated in Table 4. 

Disregarding flagged light curves, we find 218 definitively variable stars
in the entire {\em CoRoT} dataset, circled in Figure~\ref{corotrms}. Although the 
weak-lined T Tauri stars are not the subject of this paper, we find that $\sim$45\%
of them are variable by the RMS criterion alone (this fraction would increase were
periodicity detection to be employed as well). Considering hereafter only the ones
in the 162 object disk-bearing sample with both IRAC and {\em CoRoT} data, we are left with 83 variable by the RMS
criteria. A further 30 faint stars mainly with $R>16$ were added to this set based on
variability identified by eye. We identified 19
additional variables based on periodicity or quasi-periodicity alone,
as explained below. The variability status of all objects in the disk-bearing
dataset is indicated in Table~4.

\subsection{Identification of periodic and quasi-periodic variables}

\begin{figure*}
\begin{center}
\includegraphics[scale=0.8]{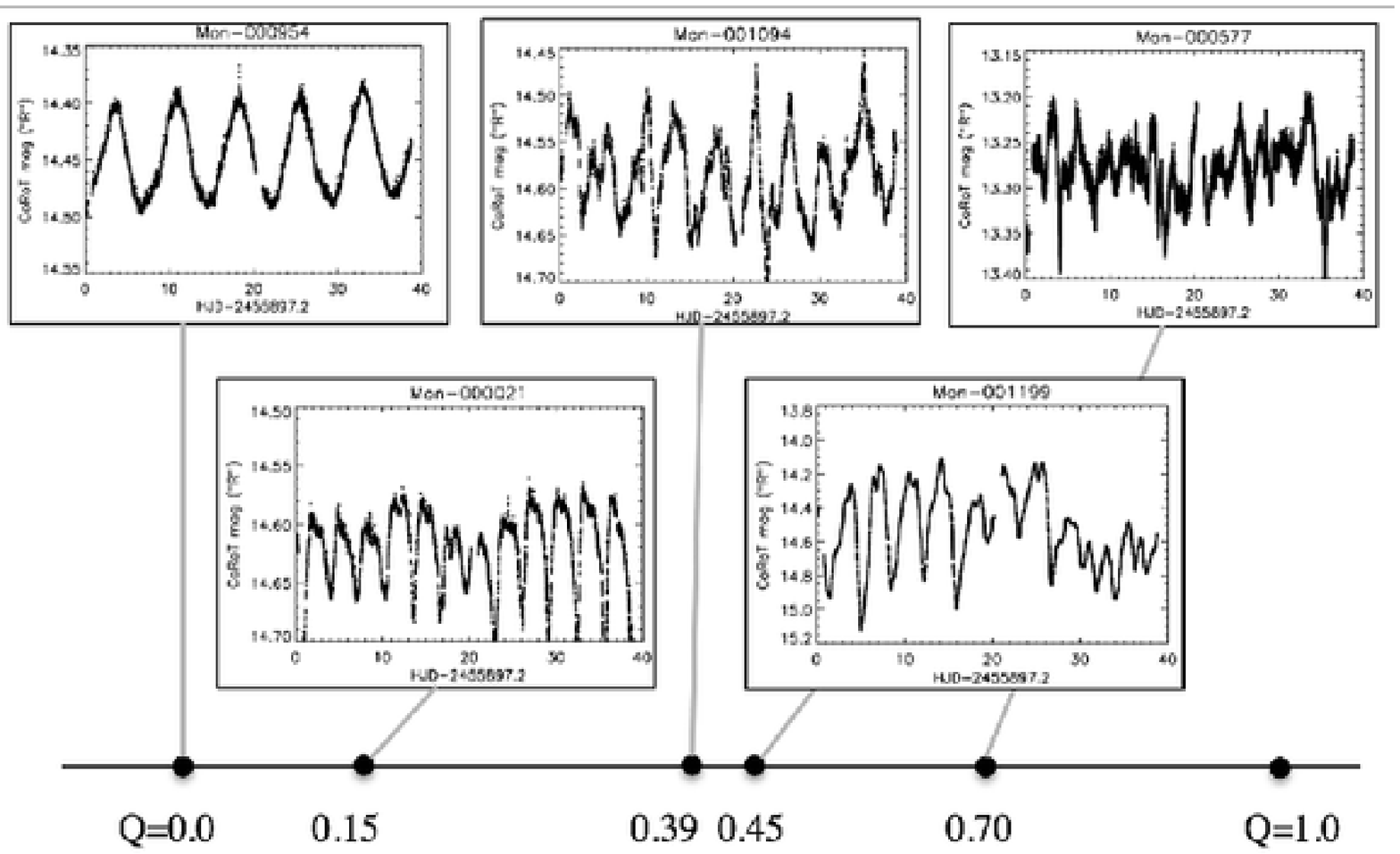}
\end{center}
\vspace{-0.5cm}
\caption{\label{Qfig}{\em CoRoT} light curves with representative values of the
  $Q$ parameter, ranging from periodic ($Q$=0--0.15) to quasi-periodic
  ($Q$=0.15--0.5), to aperiodic $Q>0.5$.}
\end{figure*}

Periodic photometric behavior is common among YSOs, although it is
mostly associated with spotted WTTSs. Our disk-dominated sample does
include a number of stars with quasi-periodic light curves, 
few of which are sinusoidal. We have developed a criterion below to
differentiate between ``periodic'' and ``quasi-periodic'' stars. By
``quasi--periodic,'' we refer to light curves that have a stable period
but whose shape and/or amplitude changes from one cycle to the next.
We use ``periodic,'' on the other hand, to denote stable repeating patterns
with shapes that evolve minimally over the 40 days of
observation. 

To select these objects from the IRAC and
{\em CoRoT} datasets and identify legitimate variables that did not
meet the RMS or Stetson criteria, we have developed a period search
technique. \citet{2013MNRAS.432.1203M} showed that the
autocorrelation function (ACF) is a particularly good tool for
selecting the correct period from a non-sinusoidal light curve, by
considering the larger of the first two local ACF peaks. The
commonly used periodogram, on the other hand, tends to display many
peaks corresponding to harmonics and aliases that may be confused with
the true signal. We have therefore carried out a preliminary period
search by interpolating all light curve magnitude values onto evenly spaced time grids
with interval $\Delta\tau$ and $N$ points, and computing the ACF 
based on the following equation:

\begin{equation}
{\rm ACF}(\tau)=\frac{\sum_{i=0}^{N-\tau/(\Delta\tau)-1}(d_i-<d>)(d_{i+\tau/(\Delta\tau)}-<d>)}{\sum_{i=0}^{N-1}(d_i-<d>)^2}.
\end{equation}

Here $d_i$ are the light curve datapoints, $<d>$ is their mean,
$\tau=n\Delta\tau$ is the time lag, and N is the total number of
points in the interpolated light curve.
We let the time lag run from zero to the maximum baseline of the time
series; peaks in the ACF indicate lag values for which the light curve
is self correlated. While interpolation may alter the light curves slightly, the sampling
is dense enough compared to variability timescales that we expect any
resulting inconsistencies to be small. We typically oversample the light curve by a factor
of 1.5, corresponding to $\Delta\tau\sim 6$ minutes for {\em CoRoT} data and $\Delta\tau\sim 1.5$ 
hours for IRAC data. For each ACF, we note all local maxima occuring at  
time lags greater than zero and less than half the total time baseline
(i.e., $\sim$15 days for IRAC light curves and $\sim$20 days for {\em
 CoRoT} light curves). We required the amplitude of any such peaks to be greater than 0.05 over the
surrounding local minima. We then select the first or second local maximum,
depending on which is higher. An example of this process and the succeeding steps
are illustrated for Mon-000660 in Figure~\ref{acfexample}.

To check whether the selected ACF peak
corresponds to a significant periodicity in the light curve, we then
compute a Fourier transform periodogram and search for peaks within
15\% of the frequency expected from the period of the ACF peak. Upon
identifying the periodogram peak, we phase fold the light curve
around the selected period. Based on the coherence of phased light curves, we find that periods extracted from the
periodogram are more accurate than those adopted from the
ACF; the latter is more sensitive to long-term trends in the light
curve. Initial use of the ACF is nevertheless vital to determining which of multiple peaks is the correct period to phase around,
as shown by \citet{2013MNRAS.432.1203M}.

Once the light curve is folded, we generate a smoothed phase-folded light curve 
smoothing over a boxcar with width 25\% of
the period. This approach is similar to that of
\citet{2008ApJS..175..191P}, except that we only phase the light curve
to a single period, as opposed to a continuum of periods. We overlay the smoothed phase curve on the original light
curve, repeating it once per period. Comparison of the phase trend curve
with the raw light curve provides an impression of how well a periodic
model explains the behavior. We subtract the two curves to produce a
residual as a function of time. For strictly periodic light curves,
the remaining points should consist of noise, and indeed the residuals
are consistent with the uncertainties shown in Figures~2 and \ref{corotrms}.  However, most of
our light curves are better described as quasi-periodic: the amplitude
of the residuals is significantly reduced compared to the raw data,
but there remain strong trends not attributable to systematic
errors. This is particularly the case for objects that display
repeating flux dips of varying amplitude. 

We have adopted a metric to assess the degree of periodicity
in the light curves, by comparing the RMS value before and after subtraction of the smooth
phased curve. Since we are only considering timescales up to half the
data length, we first remove any long-term trends on timescales over
15 days (IRAC) or 20 days ({\em CoRoT}) by subtracting a boxcar smoothed version with a
window of 10 days. We compute a periodicity metric, ``Q,'' by assessing how close the
light curve points are to the systematic noise floor before and after the
phased trend is subtracted from the light curve:
\begin{equation}
Q=\frac{({\rm RMS}_{\rm resid}^2-\sigma^2)}{({\rm RMS}_{\rm raw}^2-\sigma^2)},
\end{equation}
where RMS$_{\rm raw}$ and RMS$_{\rm resid}$ are the RMS values of the
raw light curve and the phase subtracted light curve, respectively,
whereas $\sigma$ is the estimated uncertainty including the systematics
(e.g., Section 3.3). Testing on sinusoidal light curves
typical of WTTSs, we find $Q$ to be a few percent or even
negative (i.e., when the uncertainty is an overestimate). However, for light curves 
that appear to contain multiple sources of variability, the value is larger, $\sim$0.15--0.60. 
Light curves with no detectable periodicity have $Q$ values of
$\sim$0.6--1. Examples of different $Q$ values are shown in
Figure~\ref{Qfig}. We will test the dependence of $Q$ on parameters such as
period and time sampling in future work.

As an independent check on the periodicities, we note whether the corresponding peak in
the Fourier transform periodogram exceeds the local noise level by at
least 4.0. This criterion was put forth by \citet{1993A&A...271..482B} as
indicating a 0.1\% false alarm probability;  see \citet{2010ApJS..191..389C} for further discussion. All of the periodic sources and many of
the quasi-periodic sources have significant periodogram detections. 
A more extensive analysis of the periodic sources in the 2011 {\em CoRoT} sample will be
presented in a forthcoming paper in the context of an updated
rotation rate analysis for NGC~2264.

Once a significant periodic or quasi-periodic behavior has been
identified and the phased trend subtracted out of the light curve, we
recompute the periodogram to assess whether further periodicities are
present. We once again use the criterion of periodogram
local signal-to-noise greater than 4.0. We mask out all frequencies
associated with harmonics of the first detected signal, as well as the
aliases. The frequencies of aliases are determined by overlaying on
the primary peak the window function associated with the periodogram. If one or more significant signals remain
outside of the masked values, we then repeat the period search as
outlined above. 

\begin{figure*}
\begin{center}
\includegraphics[scale=0.8]{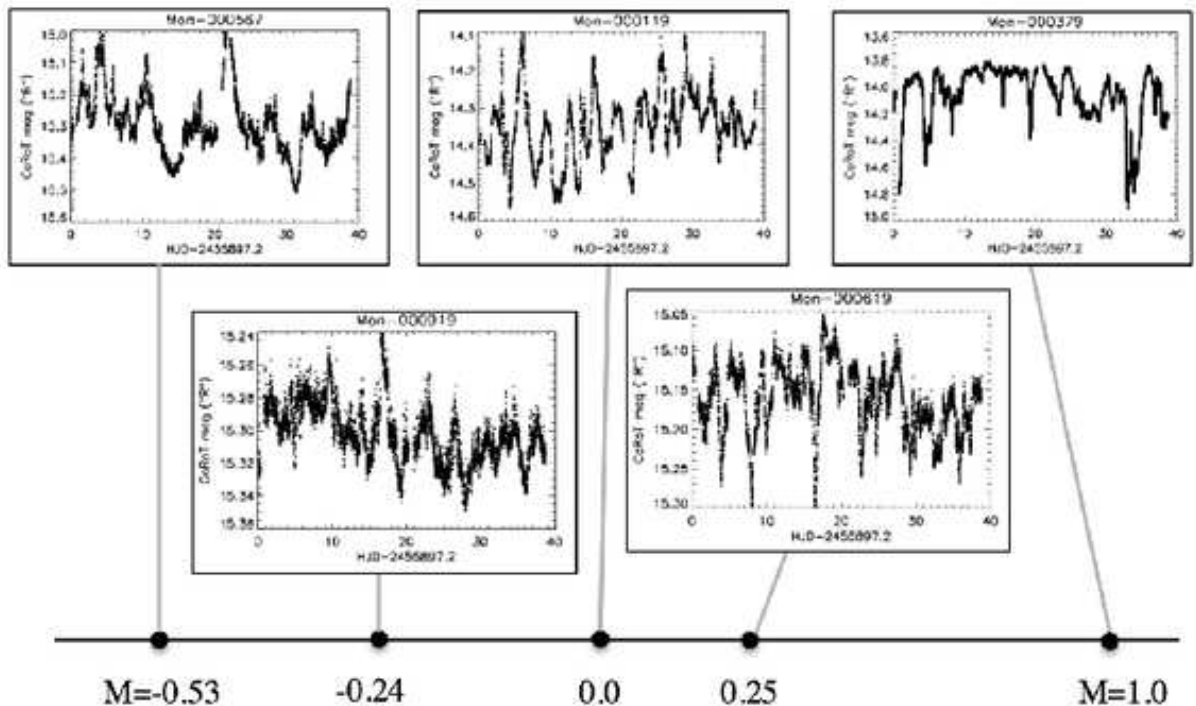}
\end{center}
\vspace{-0.5cm}
\caption{\label{Mfig}{\em CoRoT} light curves with representative values of the
  $M$ parameter, ranging from bursting ($M<-0.25$) to symmetric
  ($M$=-0.25--0.25), to dipping $M>0.25$.}
\end{figure*}

We report the results of our period search in Table 4, noting in the morphology 
column which objects are significantly periodic, quasi-periodic, and
multi-periodic. We identify two optical light curves with multiple periodicities
(Mon-000434 and Mon-000164), as well as one infrared light curve with two periods, only 
one of which is observed in the optical (Mon-001181). These could indicate binary
systems in which both stars show spot modulation.

Among the stars that are periodic or quasi-periodic, the distributions of periods in the optical and in the infrared both peak
near five days, but we find a significant dearth of infrared periods
beyond nine days compared to a steady decline in the number
of optical periods out to fifteen days. Of note, some of the infrared
periods may originate at the stellar surface, if stellar emission
dominates the disk flux at these wavelengths (as is the case for
very weak disks).

\subsection{Light curve flux asymmetry}

In addition to the cases of quasi-periodicity, we observe that many of the light curves
are asymmetric with respect to a reflection along the magnitude axis.
Some stars have prominent downward flux dips, while others have abrupt increases (see Section~5). We believe 
this behavior is connected with the physical mechanisms causing variability. To quantify the degree of
flux asymmetry, we have developed a metric, $M$. To determine its value,
we first select the 10\% highest and 10\% lowest magnitude values in each light curve, after removal
of 5--$\sigma$ outliers. This process is carried out by first smoothing the light curve on 2 hour
timescales ({\em CoRoT}) or 6 hours timescales (for the more sparsely sampled IRAC data), and subtracting
the smoothed trend from the raw data. Outliers are then measured on the residual light curve.
After their removal, we compute the mean of the remaining points and compare with the median of 
the {\em entire} outlier filtered light curve. We define the asymmetry metric via:
\begin{equation}
M=(<d_{\rm 10\%}>-d_{\rm med})/\sigma_{d},
\end{equation}
where $<d_{\rm 10\%}>$ is the mean of all data at the top and bottom decile of light
curve, $d_{\rm med}$ is the median of the entire light curve, and $\sigma_d$
is its overall RMS.

In some cases, there is a clear asymmetry in the light curve, but it is superimposed on a longer
timescale trend. Our asymmetry metric is only sensitive to asymmetries on timescales less than
about half the light curve duration, or 15--20 days. We therefore remove trends on longer
timescales by subracting out a smoothed version of the light curve before computing $M$. We find
that a smoothing window of a few days is sufficient; in cases where short-term light curve features 
were oversubtracted, we retain the $M$ value computed on the raw
data. These cases were identified visually comparing the raw light
curve with the long-term trend overplotted. All light curves were also
visually inspected to determine which value was most appropriate. 

We encounter a range of $M$ values in our disk bearing dataset, from approximately -1 (prominent upward
flux peaks) to just over 1 (flux dips). Examples are shown in Figure~\ref{Mfig}.
Examining the light curves by eye, we observe the most obvious 
asymmetric behavior to occur for values of $M$ greater than 0.25, or less than -0.25. 

\subsection{Division of light curve morphologies by $Q$ and $M$}

\begin{figure*}
\begin{center}
\epsscale{1.17}
\includegraphics[scale=0.7]{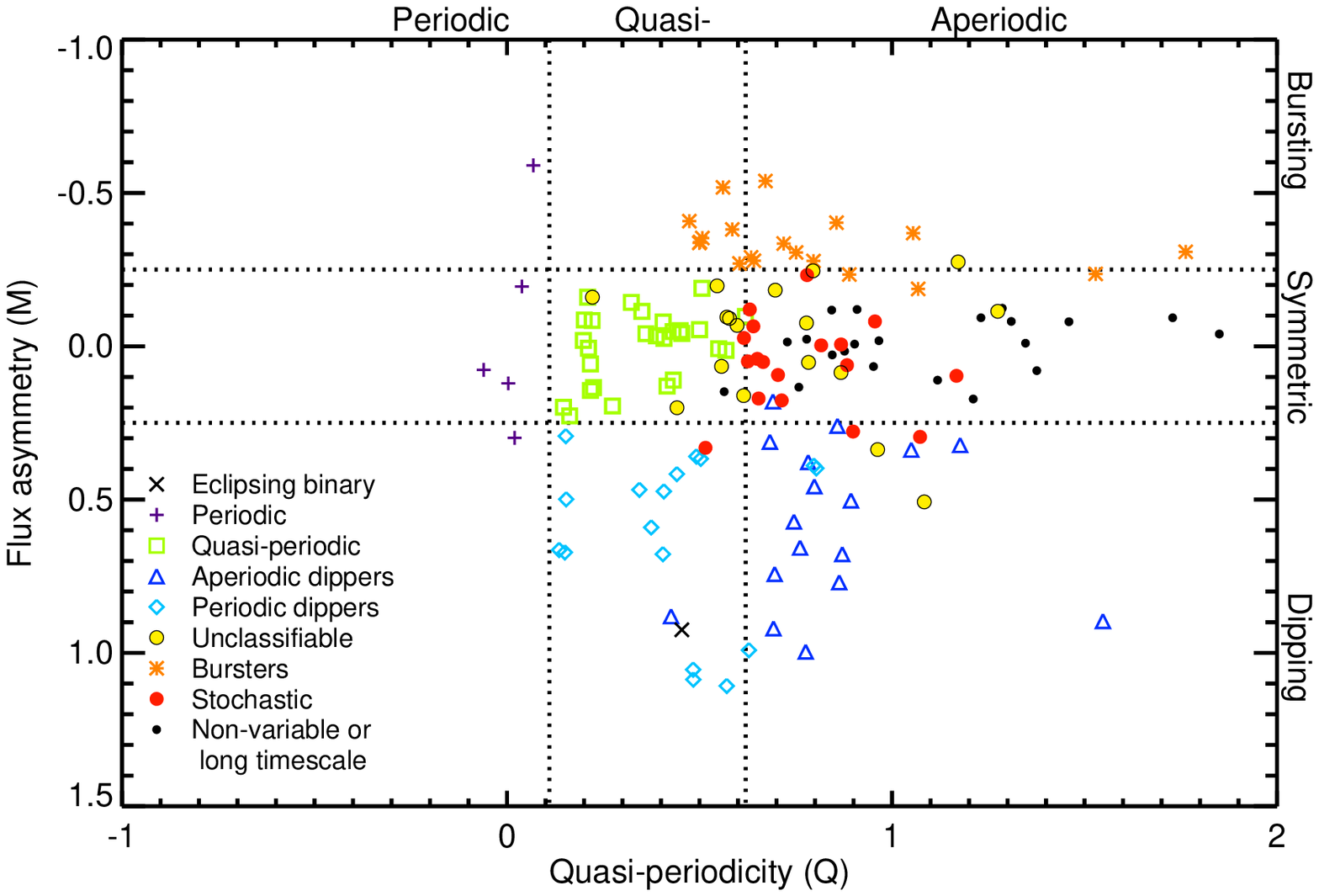}
\includegraphics[scale=0.7]{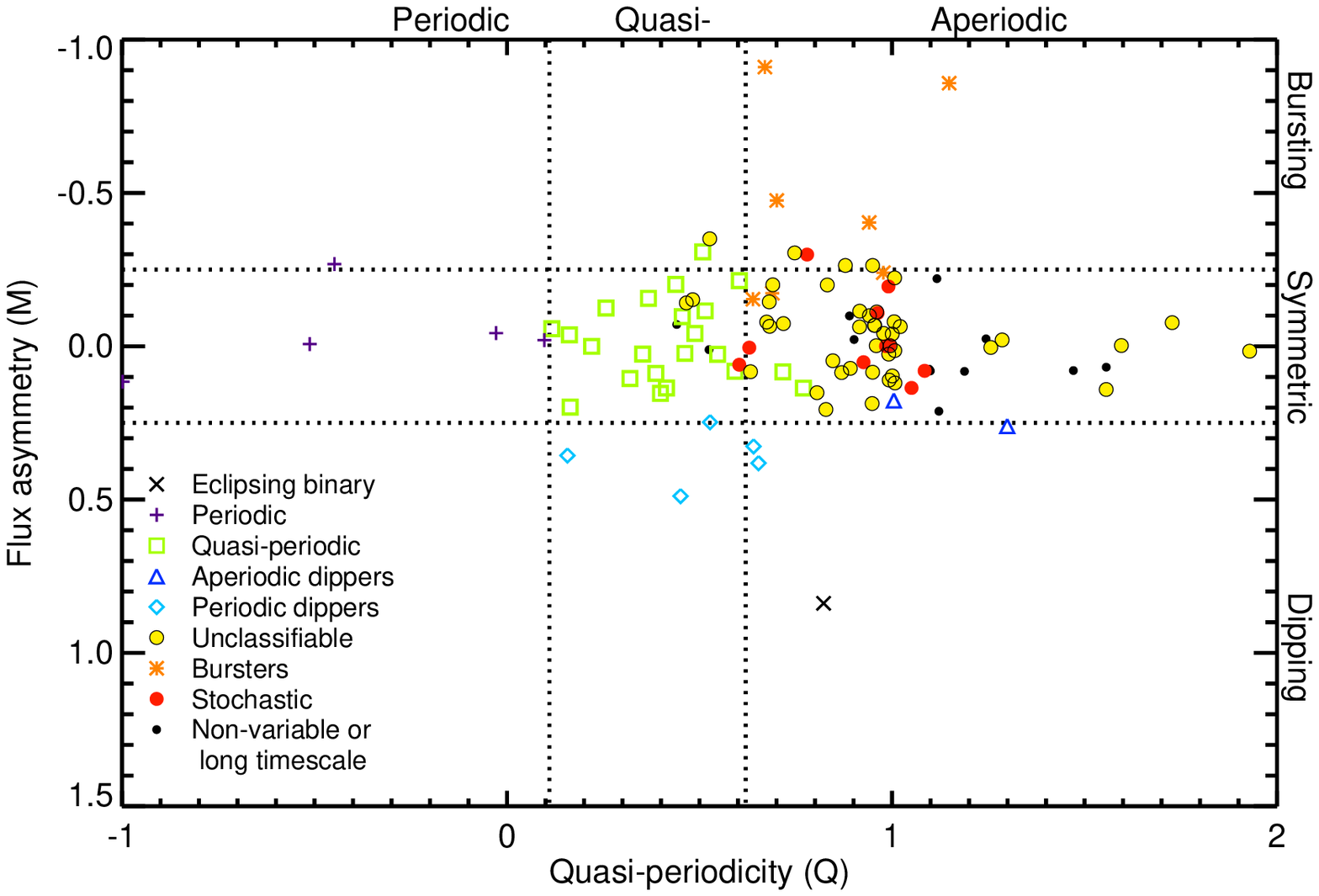}
\end{center}
\vspace{-0.5cm}
\caption{\label{qmplots}{\em Top}: Light curve morphology classes, as divided by
  the quasi-periodicity ($Q$) and flux
  asymmetry ($M$) parameters for optical light curves from {\em CoRoT}
  in our disk-bearing sample. Color coding indicates the variability
  classification chosen by eye, before statistical assessment. The
  eclipsing binary is not strictly periodic because its light curve
  contains aperiodic fluctuations out of eclipse. {\em
    Bottom}: Same as the top, but for infrared light curves acquired from
  {\em Spitzer}/IRAC. The pile-up of points at $Q\sim 1$ occurs
  because the subtraction of an incorrect phase curve tends to leave
  the RMS unchanged from its raw value.}
\end{figure*}

The $M$ metric, in combination with timescale (Section 6.5) and the periodicity measure, $Q$, enables us to 
quantitatively retrieve the morphology classes that were first established by eye, and that presumably 
represent different variability mechanisms. We plot $Q$ against $M$ in Figure~\ref{qmplots}, along with 
suggested boundaries to divide the different variability types. Examining the classifications made by eye 
(color coded points), we find that the selected boundaries in $Q$ (0.11, 0.61) and $M$ ($\pm$0.25) are quite 
successful in separating the variability sample into classes. They appear slightly less useful in the infrared 
since many more sources in this band vary on long time scales and relatively fewer exhibit dipping or bursting 
behavior. The lower infrared cadence may play a role here as well. Nevertheless, the fact that two parameters 
can divide our sample so well into the predetermined groups is promising for future variability classification 
efforts based on sparser data.

The $Q$-$M$ diagram reveals several facets of the variability in our sources.
First, while we have selected boundaries between different classes, the
statistics show that there is a continuum of light curve behavior along both the
$Q$ and the $M$ axis. This suggests that sources on the boundaries of variability
classes may be characterized by multiple physical mechanisms. 
Second, the $Q$-$M$ diagram displays several areas devoid of points. There are relatively
few variables that are both bursting and periodic. This may point to the stochastic
nature of accretion. Likewise, there are relatively few dipper objects with highly
periodic variability. The highest amplitudes of these light curves correlate with
unstable variability patterns, such that they are quasi-periodic but not periodic.

\subsection{Aperiodic timescales}
Timescale is an important quantifiable aspect of our light curves. Visual inspection
of the data reveals that some objects oscillate quite rapidly (i.e., multiple
zero crossings in a week), while others show only long-term trends. We wish to 
define a timescale measure for the purposes of examining correlations with physical parameters.
While it is easy to attach a value to the periodic and quasi-periodic light
curves, a timescale for aperiodic stars is not so obvious. Tools such
as the autocorrelation function and periodogram have traditionally
been used to assess whether there is a characteristic timescale for
aperiodic variability. However, it is not immediately clear what
amplitude level one should choose to extract a timescale from the ACF,
or how to determine whether there is such a special time at
all. Furthermore, the time sampling and baseline may strongly
influence the appearance of the ACF. 

\begin{figure}
\begin{center}
\includegraphics[scale=0.43]{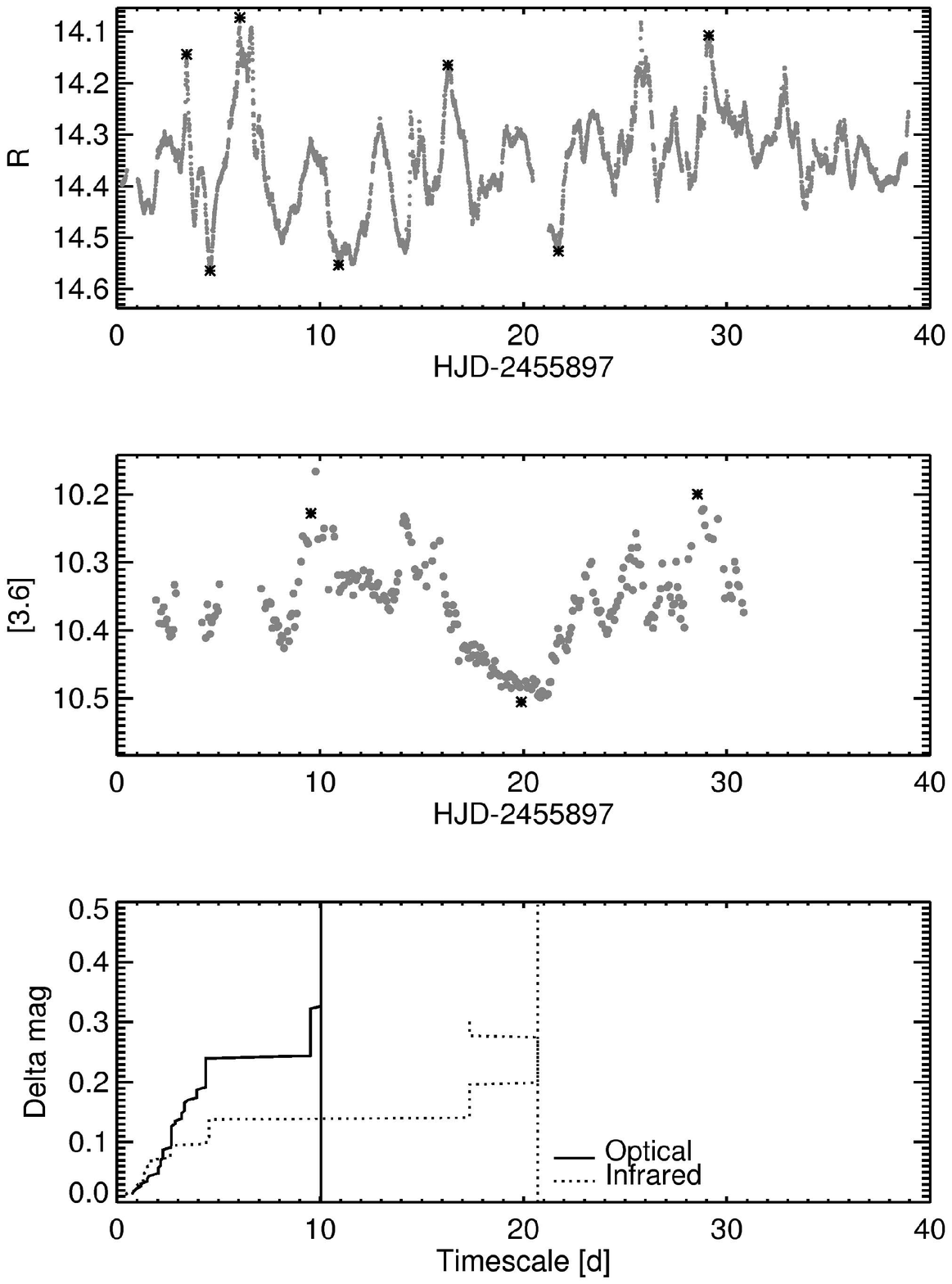}
\end{center}
\vspace{-0.5cm}
\caption{\label{timescaleex} An example of how we compute optical and infrared
timescales for the aperiodic variable Mon-001054. The raw {\em CoRoT} (top) and IRAC 3.5~$\mu$m (bottom)
light curves are shown in grey, whereas peaks selected as being separated by at least 80\% of the maximum
minus minimum of the light curve are marked with black asterisks. The bottom plot shows
the resulting amplitude-timescale trend, with the 80\% value marked with vertical lines. See the text for
further details.}
\end{figure}

\begin{figure*}
\begin{center}
\includegraphics[scale=0.7]{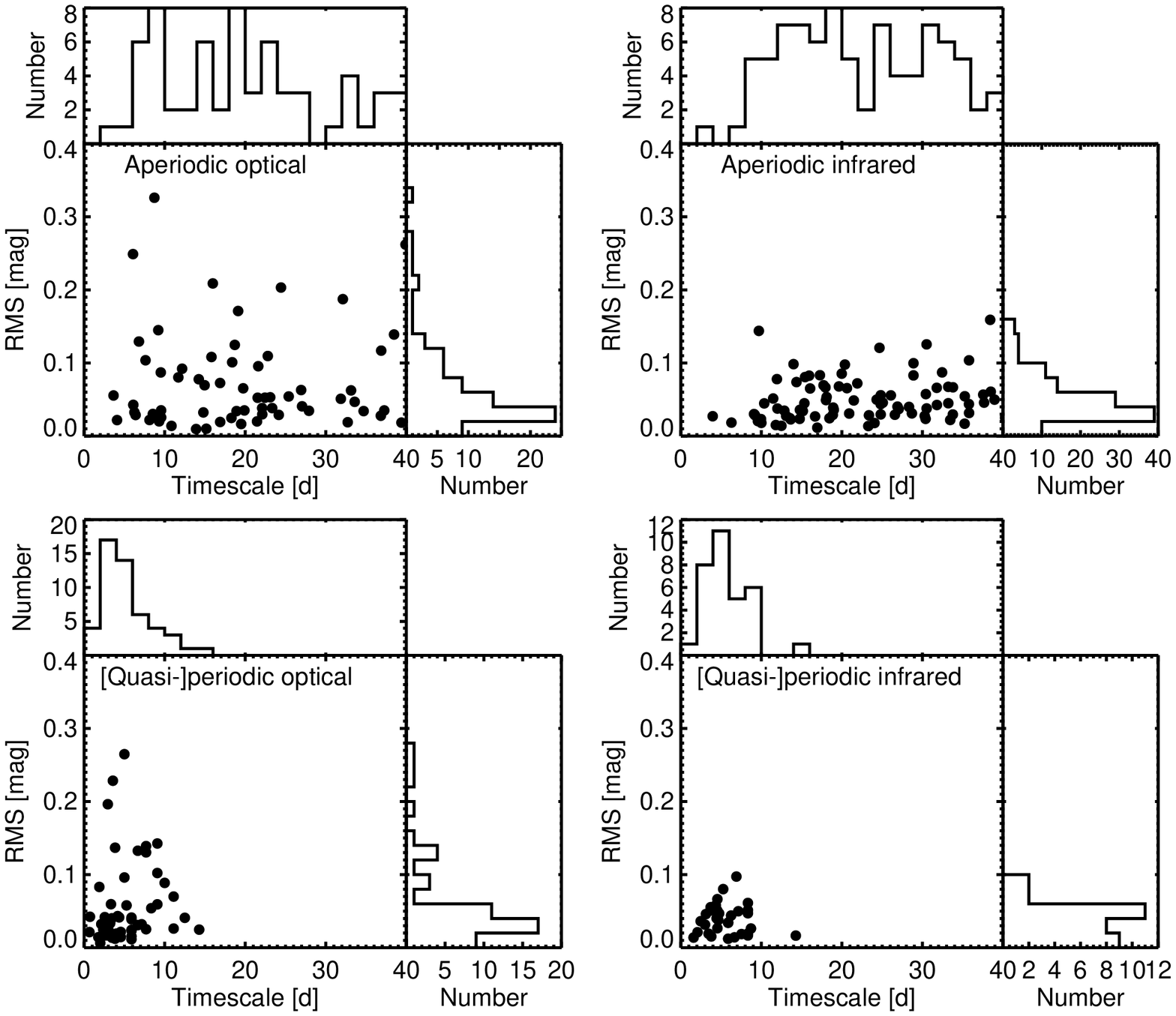}
\end{center}
\caption{\label{rmstime}Timescale versus RMS, separated into [quasi-]periodic and aperiodic behavior,
in both bands. We also display the distributions of each set of points.}
\end{figure*}

We have adopted a different strategy, which is to identify the median timescale separating the
largest consecutive peaks in a light curve. This process consists of
two steps: 1) for each light curve, we calculated how the inferred
timescale varies as a function of the amplitude threshold applied in
selecting peaks to calculate timescales between, and 2) we collapse
this set of timescales into a single value representing the highest
amplitude variations within the light curve.  For the first step, we
calculate timescales for amplitudes as small as the light curve's
noise level, and as large as the light curve's full range (i.e., the
absolute maximum minus the absolute minimum, after filtering for outliers).
For each of these values, we run through
the light curve, marking maxima and minima that differ
from surrounding peaks by more than that amplitude (see the top and middle panels of Figure~\ref{timescaleex}). 
We start by selecting the absolute maximum, and count all peaks preceding it. We then repeat the process to 
identify peaks succeeding it. From this ``PeakFind'' algorithm, we estimate a characteristic timescale for each amplitude by
computing the median time difference of all consecutive pairs of peaks, and then multiplying by
two. The normalization by 2.0 ensures that the derived timescale will match
the period for periodic objects.  This process results in a trend of timescale versus amplitude, which we invert to amplitude as a
function of timescale, as shown in the bottom panel of Figure~\ref{timescaleex}.
We note that this function is not necessarily monotonic, because of the smaller number statistics in the case of large amplitudes
and small numbers of peaks. 

For many of our light curves, the PeakFind timescale reaches a maximum value that is much shorter than the duration of the 
light curve. This implies that the full variability amplitude is accounted for by a relatively short timescale phenomenon. 
To extract a single timescale from the trend-- the second step in the process-- we note the maximum peak-to-peak amplitude 
of the light curve and adopt the timescale corresponding to 70\% of this (i.e., the verical lines in 
Figure~\ref{timescaleex}). If this timescale is comparable to the duration of the time series, or $\sim$30--40 days, then it is 
a lower limit on the true maximum variability timescale. We emphasize that this ``70\% amplitude'' timescale is not 
necessarily a {\em characteristic} timescale, since variability may be produced on a continuous spectrum of timescales, and 
we are not sensitive to those longer than 30--40 days. However, for fairly rapid variability we consider it an approximate 
upper limit to the timescales on which variability is generated.  Simulations using damped random walks show that, for 
aperiodic signals with characteristic time scales of 0.1-5 days and amplitudes of ~0.2 mag or more, the peak-finding time 
scale is well correlated with the true time scale on average but shows scatter comparable to the true time scale for any 
individual source \citep{Fin14}. Therefore, the time scales are best interpreted in an ensemble sense. Using them as 
guidance, we are able to distinguish between aperiodic light curves that oscillate on day to week timescales and those that 
wander up and down over the course of a month or more.  We find that
the distribution of aperiodic stochastic timescales from the PeakFind method centers
around 5--10 days, indicating that short timescales are dominant in the optical, at
least among the spectral types encompassed by our dataset.

With two types of timescales (i.e., quasi-periodic and aperiodic) in hand for the variables in our
sample, we can compare them with the RMS values in the optical and infrared, as shown
in Figure~\ref{rmstime}. Separating objects into their variability
type (aperiodic or quasi-periodic), we do not find any obvious correlations, apart from
a subtle rise in variability amplitude from one to five day timescales. Aperiodic infrared variables show higher
amplitudes than the periodic varieties, with many of the former achieving RMS values between 0.1 and 0.2 magnitudes. 
Both optical and infrared variables achieve their highest quasi-periodic amplitudes on timescales
between five and ten days. The amplitudes of aperiodic light curves have a much wider distribution of amplitude
versus timescale.

\section{Correlation of optical and infrared variability}

We expect correlation between the optical and infrared variability in cases
where the dominant variability mechanisms take place on or near the
stellar surface. We predict much less correlation when
the mid-infrared flux is dominated by disk emission.
Ultimately, the degree of optical/IR correlation is a function of
many factors, including viewing angle, number of distinct variability
mechanisms and their amplitudes, disk flux, and disk geometry (e.g., inner
wall radius, flaring, gaps). While we do not have enough information
to break these degeneracies, we can examine the multiwavelength
correlation properties in the context of the different light curve
morphologies that we have identified. In Section 8, we explore further
connections with available stellar and disk properties.

\begin{figure*}
\begin{center}
\epsscale{1.17}
\plottwo{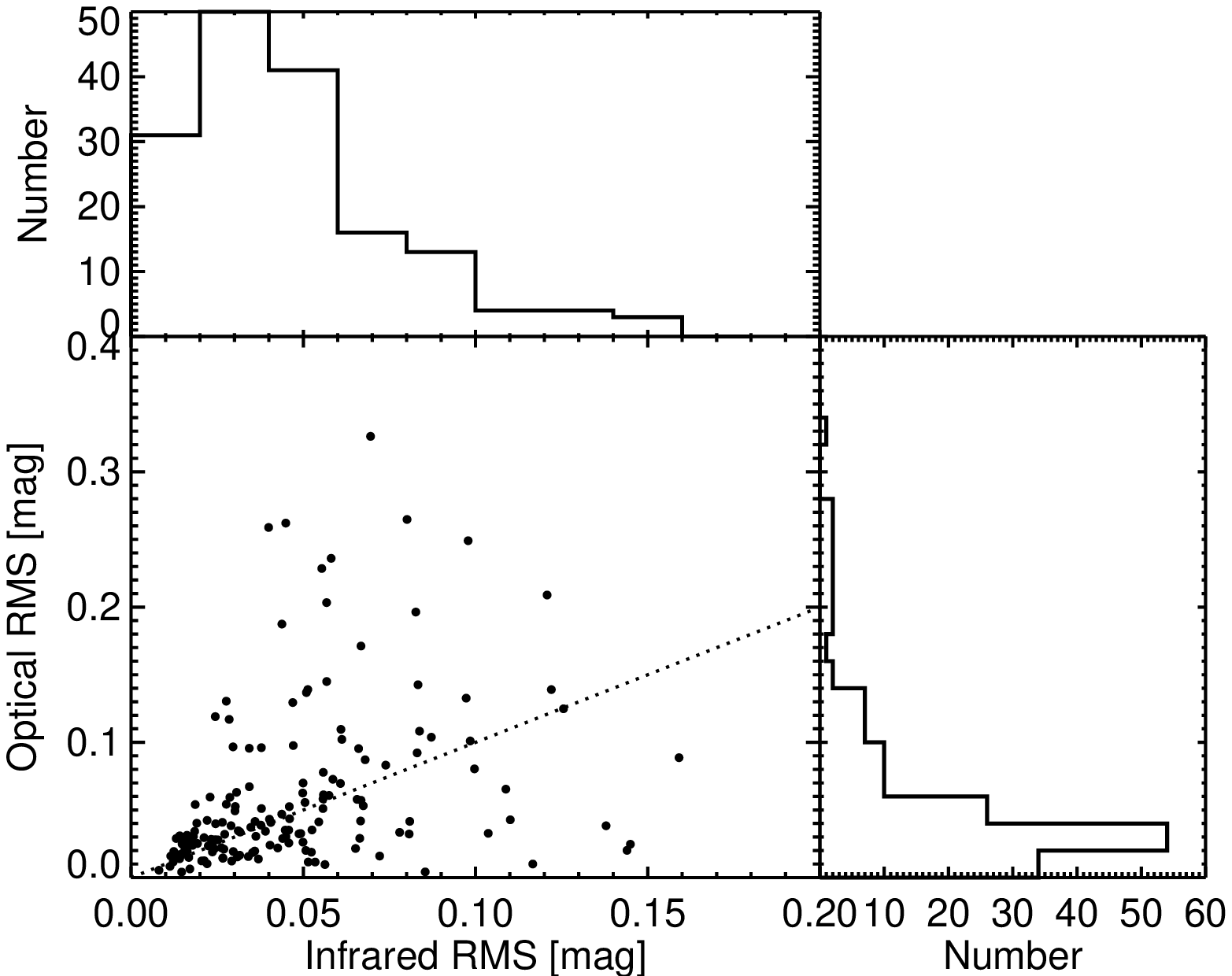}{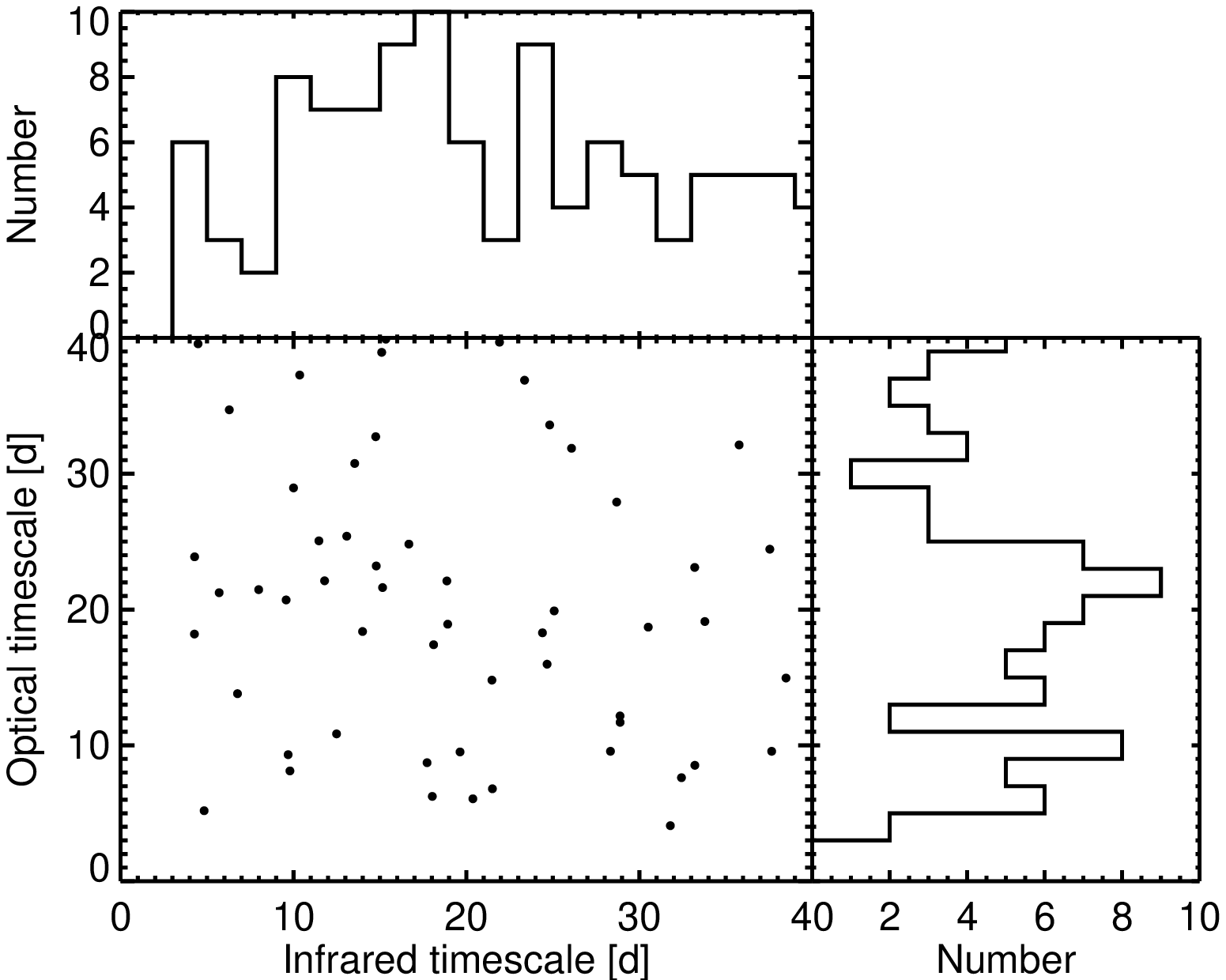}
\end{center}
\caption{\label{optir}{\em Left}: Correspondence between optical and
infrared RMS values for all objects in our dataset. {\em Right}: Optical and
infrared aperiodic timescales in each band. These are quantized, because the
aperiodic timescale involves division of the total time baseline (the same for all
sources) by the number of peaks larger than a particular amplitude. Here the points
at timescales of $\sim$40 days are lower limits. Timescales are not shown for objects that have a
quasi-periodic optical or infrared light curve, since we find that the detected periods,
if present in both bands, are always similar.}
\end{figure*}

Of particular interest is how the statistics compare in the two
different bands, optical and infrared. To portray the diversity of
light curve behavior, we have plotted in Figure~\ref{optir} the RMS values of
each light curve. There are no distinct clusters in this diagram, and none of the
morphology groups described in Section 5 occupies any particular region, apart from periodic
and unclassifiable variables having preferentially low RMS values. The
latter are likely noise dominated, whereas the former could be the
result of cool starspots, for which RMS values of order a few percent
are consistent with previous results in the literature.

We also present in Figure~\ref{optir} a comparison of the inferred
optical and infrared timescales for aperiodic variables. The values in
both bands cover most of the parameter space from one to 40 or more
days. Short-timescale behavior in the optical usually corresponds to
relatively short timescale behavior in the infrared, but short-timescale behavior in the
infrared does not necessarily mean a short timescale in the optical.

\subsection{Optical/infrared correlation via the Stetson index}

To measure in detail the degree of correlation between optical and infrared time series, we
first interpolate each {\em CoRoT} light curve onto the same
timestamps as the IRAC 3.6 and 4.5~$\mu$m mapping data. Since the {\em CoRoT}
observations were obtained at such high cadence, the effect of
interpolation on the variability is negligible. The result is pairs of
optical and infrared light curves with typically $\sim$300 points. 

We have performed a cross correlation of the {\em CoRoT}/IRAC sets using the 
Stetson index \citep{1996PASP..108..851S}, as well as subtraction to generate color trends. We find that many sets of light
curves are well correlated for part of the time series but less
correlated or varying with a different color slope in other parts. 
This makes it difficult to quantify correlation via a single
parameter. 

\begin{figure}
\begin{center}
\includegraphics[scale=0.51]{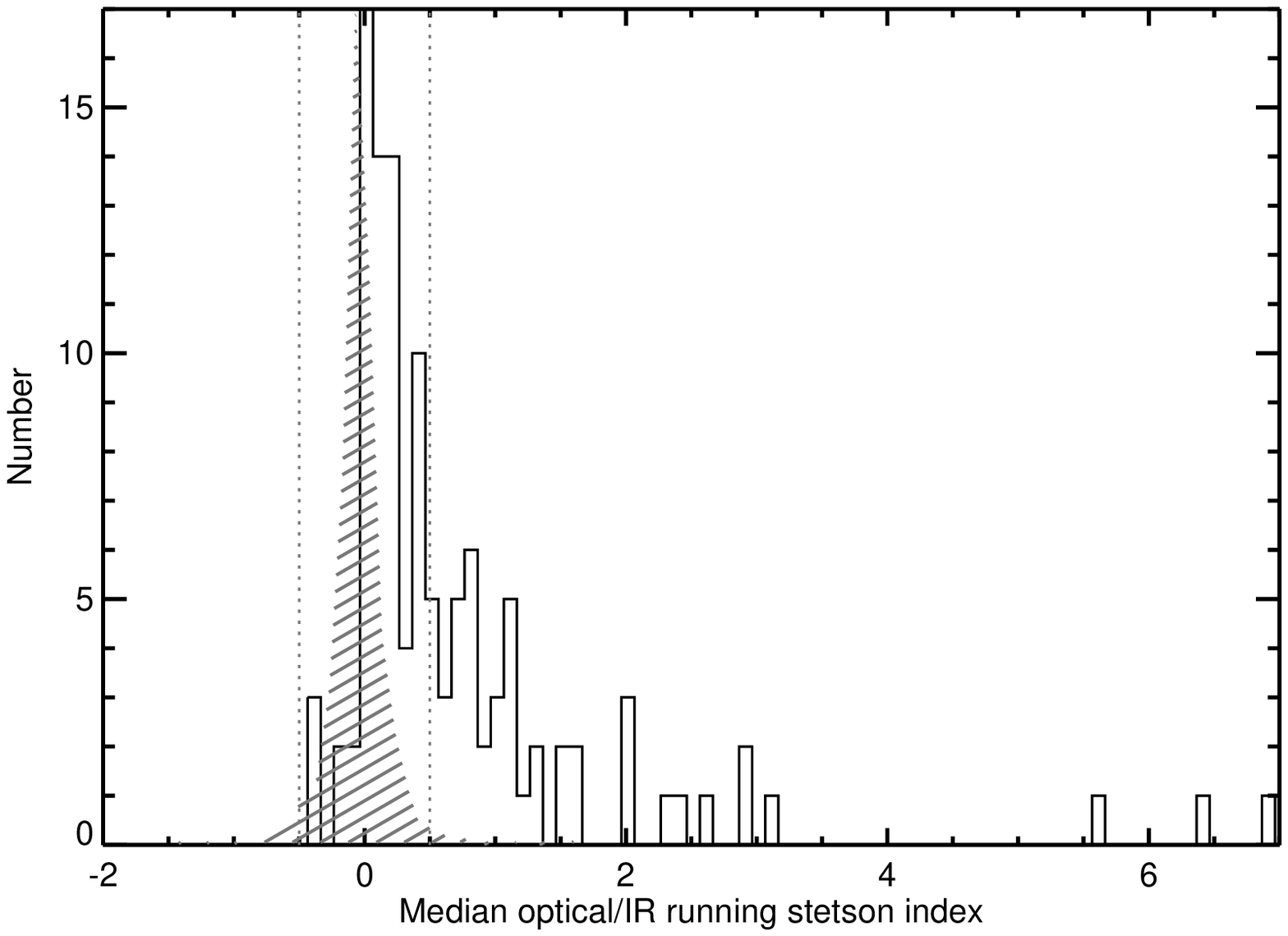}
\end{center}
\vspace{-0.5cm}
\caption{\label{runstet} Distributions of median running Stetson
  index for pairs of {\em CoRoT} and IRAC light
  curves in our disk-bearing sample. Here the values from the 3.6 and 4.5~$\mu$m bands are
averaged. The hashed region shows the distribution for randomly
selected light curve pairs, and dotted lines marked the 95\%
confidence interval for random Stetson indices. Values greater than
0.6 indicate well correlated optical and infrared light curves.}
\end{figure}

As an alternative, we have computed a time-dependent Stetson index by comparing 20-point 
sections (1.7 days) of the IRAC and interpolated {\em CoRoT} light curves. The 20 point step size was
motivated by the typical duration of short-timescale light curve fluctuations, which is a few days.
Well correlated light 
curves should display a positive Stetson index with little dependence on time.  Light curves 
that are correlated at some times but not others will instead display a fluctuating Stetson 
index, reaching large values at some times and dropping to zero or below at others. To 
differentiate this behavior from chance correlation between sections of two light curves, we 
have performed simulations using time series from different stars. We match optical and 
infrared light curves from {\em different} stars, thereby guaranteeing that there should be no 
true correlation. Only light curves with at least 250 points are
allowed, since the more sparsely sampled HDR mode light curves will
have a broader Stetson index distribution.
One thousand simulations of randomly matched light curve pairs revealed that 
95\% of uncorrelated light curves have a median running Stetson index between -0.5 and 0.5 when 
sampled at every 20 points. Repeating the simulations on HDR light
curves for every five points, the value increased to 1.0.
We also ran the simulations for a sampling step size of 
50 points, but found that the running Stetson index showed more systematics, and its median was less 
consistent between the two IRAC channels, despite flux behavior being nearly the same in these 
bands. The distribution of median running Stetson index for {\em CoRoT}
light curves paired with IRAC light curves is shown in Figure~\ref{runstet}.

We report the median running Stetson index for all objects in Table 4.
Stars with large negative running Stetson indices exhibit
anticorrelated behavior in the optical and infrared.

The fractions of correlated versus uncorrelated and
anticorrelated light curves may indicate the percentage of sources in
which variability is disk dominated. To determine this number, we
assemble the distribution of median runnning Stetson indices in
Figure~\ref{runstet}, along with the distribution expected for uncorrelated
light curves, as derived in our simulations. Although we have drawn a
cut-off of 95\% confidence, it is clear that the set of Stetson indices is
skewed with respect to the distribution defined by the simulations. 
We conclude that there is low-level optical/infrared correlation in
many of the objects; this is not surprising since stellar variability
will generally have a small infrared contribution due to emission from 
the long-wavelength tail of the star's spectral energy distribution. 
In general, however, the larger infrared contribution from the dust means
that variability originating in the disk will dominate the light curve.

\begin{figure}
\begin{center}
\includegraphics[scale=0.87]{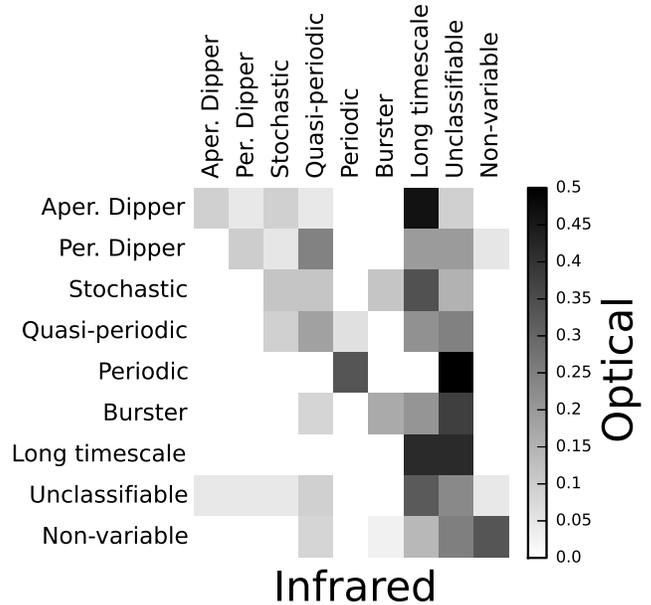}
\end{center}
\vspace{-0.5cm}
\caption{\label{corrmatrix} The correlation matrix between assigned optical and infrared variability types.
For each optical morphology class, this diagram shows the fractions of objects that occupy each infrared
class.  It should be read along rows. For example, just over 10\% of optical variables classified as aperiodic dippers are 
also aperiodic dippers in the infrared; the reverse (i.e., 10\% of aperiodic infrared dippers being aperiodic dippers 
in the optical) is not necessarily true. Continuing along the same row, we also see that just over 50\% of aperiodic 
optical dippers are long timescale variables in the infrared. The remaining 34\% of optical variables are a mixture of
periodic dipper ($\sim$6\%), stochastic ($\sim$11\%), quasi-periodic ($\sim$6\%), and unclassifiable (11\%) in the infrared.}
\end{figure}

Overall, 38\% of our sample shows evidence of correlation at the 2-$\sigma$ level, and 58\% shows correlation at the 
1-$\sigma$ level. There is generally a correlation between the shapes in the optical 
and in the infrared, but in many cases it is quite weak (i.e., it consists of transient dips or bursts matching up). To further 
illuminate the relationship between optical and infrared behavior, we have plotted a correlation matrix comparing the 
morphologies assigned in each band. Figure~\ref{corrmatrix} confirms a low-level optical/infrared correlation at best 
in most sources.

\subsection{Well correlated optical and infrared variability}

We have identified light curves that are correlated in the optical
and infrared by selecting those with a large median running Stetson
index.  Some of these light curves display partial correlation in that
the infrared flux trends mimic those in the optical over part of the
time series, but at other points there is no resemblance between the
two bands. 

There are two scenarios for correlation of behavior in the optical and
infrared. First, if there is only one variability processes affecting
the star, then the light curves should exhibit a relatively simply
wavelength dependence. Second, if the disk flux dominates in
the infrared bands but variability reflects mainly reprocessed stellar
light, then we would also expect to see correlated behavior, either in
or out of phase. We test these scenarios in Section~8 by comparing the
Stetson index to measures of disk to star flux.

\begin{figure*}
\begin{center}
\includegraphics[scale=0.7]{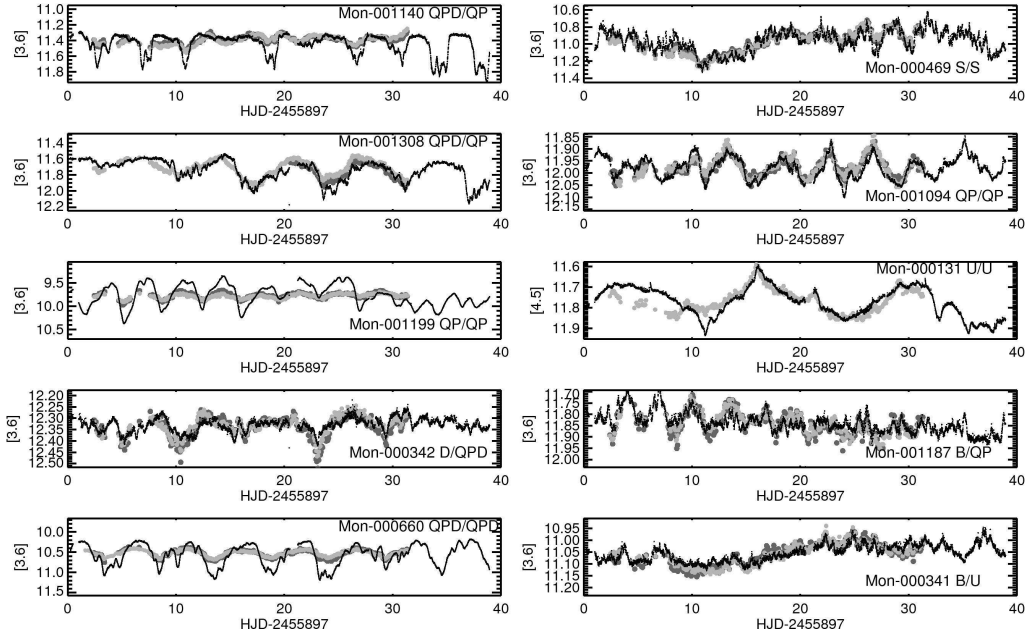}
\end{center}
\caption{\label{corr} Light curves with correlated optical
  and infrared behavior. Small black points are {\em CoRoT} data,
  light grey points are 3.6~$\mu$m data, and dark grey
  points are 4.5~$\mu$m data (sometimes hidden behind the 3.6~$\mu$m
  points). Labels show the Mon id along with the optical and infrared
  morphologies, respectively; morphology abbreviations are the same as
in Table 4.}
\end{figure*}

We have identified three broad categories of morphological behavior
among the well correlated light curves: optical dippers, bursters and
stochastic stars with similar infrared amplitude, and optical
stochastic stars with lower amplitude infrared light curves. A
selection of these is presented in Figure~\ref{corr}, while optical
and infrared light curves for the entire
162-member disk bearing dataset are provided in the Appendix, Figure~\ref{iroptall}. 
Some of these have behavior that makes sense in the context of the extinction dominated variability model. Here,
fluctuations in the light curve are caused by changing amounts of dust blocking the
star, and the optical/infrared correlation simply reflects the
wavelength dependence of extinction, diluted by any flux from the
inner disk. Confronting this model are the two objects Mon-000183
and Mon-000566 which display {\em deeper} dips in the infrared than in
the optical. This behavior is not expected from any standard extinction law
and may require unique geometry, such as occultations of the disk by itself
(i.e., the front blocks flux from the back wall). 

The remaining set of well correlated optical and infrared
light curves displays very similar morphology and amplitude in each
band. Half of these are members of the optical bursting class, and the rest are stochastic
or unknown morphology. Strong optical/infrared correlation for the
these would be expected if flux changes are reflective of increases
in accretion, the radiation from which is absorbed and reemitted by
the inner disk. 

\subsection{Uncorrelated optical and infrared variability}

Some 62\% of the variable light curves in our sample display little 
correlation between the optical and infrared bands.
In these cases, there are likely two or more variability mechanisms at
work, connected separately with the star and the disk. Alternatively, a single
variability mechanism could be at work if it only affects one wavelength region. Since the infrared
flux is in most cases not dominated by stellar or accretion emission, disk-driven mechanisms are necessary
to explain the often large amplitude ($\sim$0.1 mag) infrared modulation on week or longer
timescales. One possible origin is magnetic turbulence, proposed by
\citet{2013AAS...22120505T} to modulate the inner disk scale
height and thereby alter the observed mid-infrared emission.

We have assembled a set of the most prominent uncorrelated optical and
infrared behavior in Figure~\ref{uncorr}. The wide range of
morphologies in both bands is evident.
Particularly interesting examples of uncorrelated behavior occur in
Mon-000185, Mon-000273, Mon-000357, Mon-000876, and Mon-000928, for which there is almost no optical
variablity, but high-amplitude (0.1-0.3 mag) infrared light curve
excursions on 5--10 day timescales. We have checked the individual images of these
objects for erroneous cross-matching of {\em CoRoT} and {\em Spitzer} sources, but in all
cases there is no nearby star that could be a better match. This type of behavior
occurs in a few percent of our disk bearing stars.

The selection of objects with uncorrelated behavior is particularly
helpful for investigating the distribution of infrared variability
timescales, since we do not have to worry about much contamination
from optical processes. We have measured the aperiodic infrared
timescales of stars with median running Stetson indices less than 0.6.
As illustrated in Figure~\ref{IRtimescale}, the distribution displays a clump around 5--15 days, 
although there are many additional objects with infrared timescales at 20 days and longer. The individual examples 
of rapid infrared changes as well the overall distribution lend support to
the idea that disk structural changes may be occurring close to or even
faster than the local dynamical timescale in some cases. The latter is between a few days
and a few weeks, depending on stellar mass and dust properties.

\begin{figure*}
\begin{center}
\includegraphics[scale=0.6]{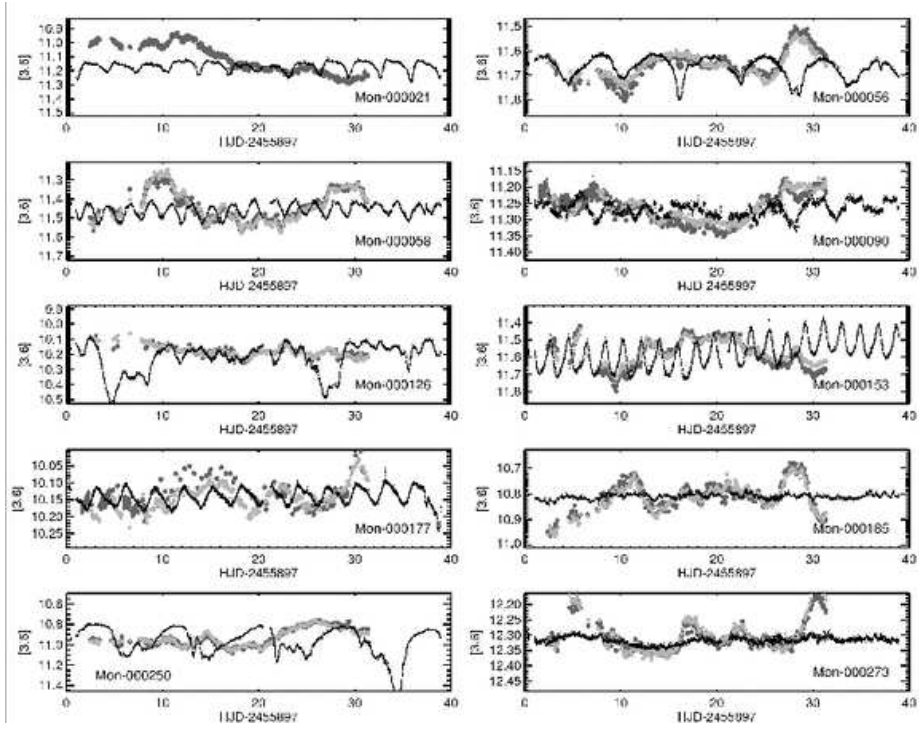}
\end{center}
\caption{\label{uncorr} Light curves with uncorrelated optical
  and infrared behavior. Small black points are {\em CoRoT} data,
  light grey points are 3.6~$\mu$m data, and dark grey
  points are 4.5~$\mu$m data (sometimes hidden behind the 3.6~$\mu$m points).}
\end{figure*}

\addtocounter{figure}{-1}
\begin{figure*}
\begin{center}
\includegraphics[scale=0.6]{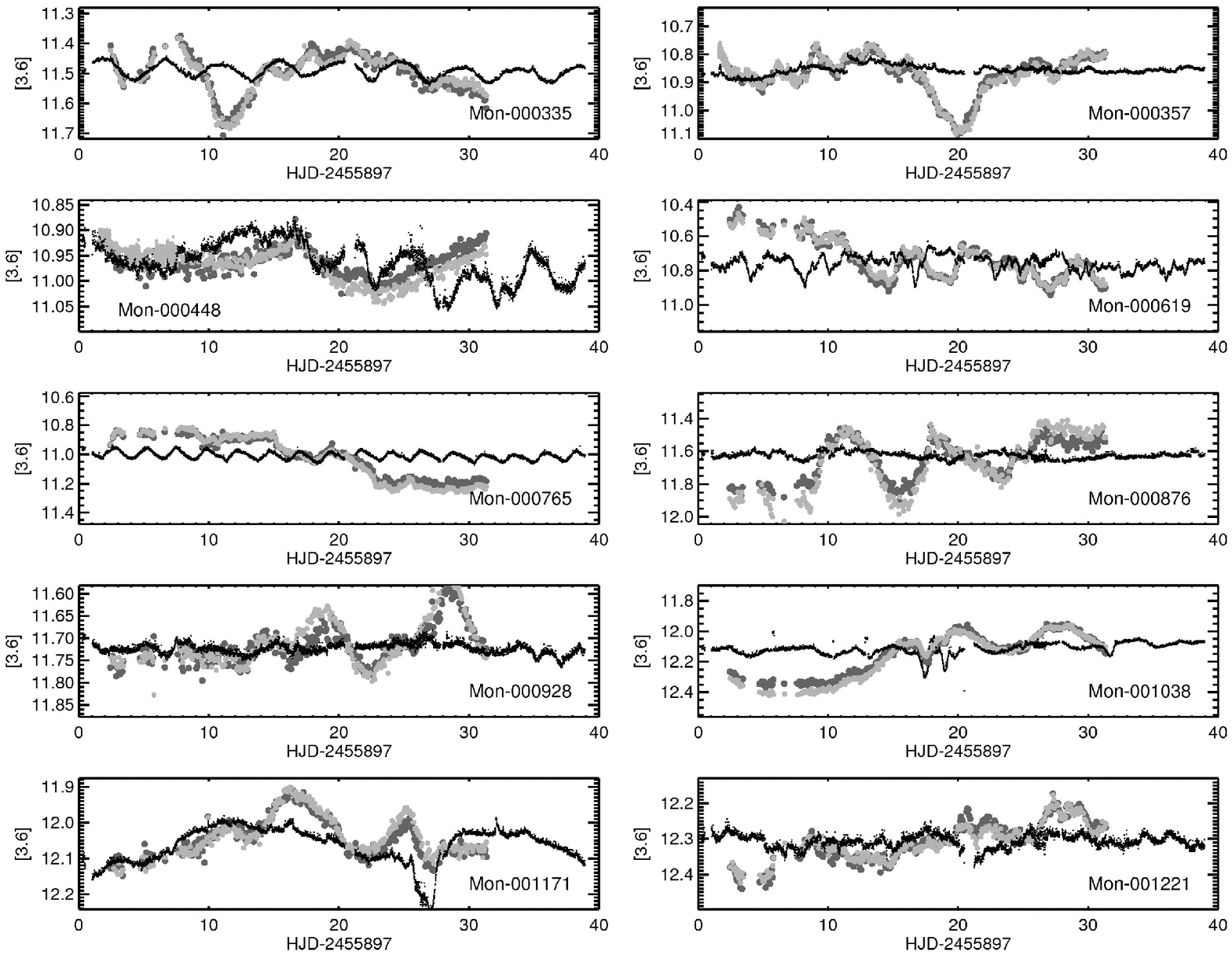}
\end{center}
\caption{--continued.}
\end{figure*}

\begin{figure}
\begin{center}
\includegraphics[scale=0.5]{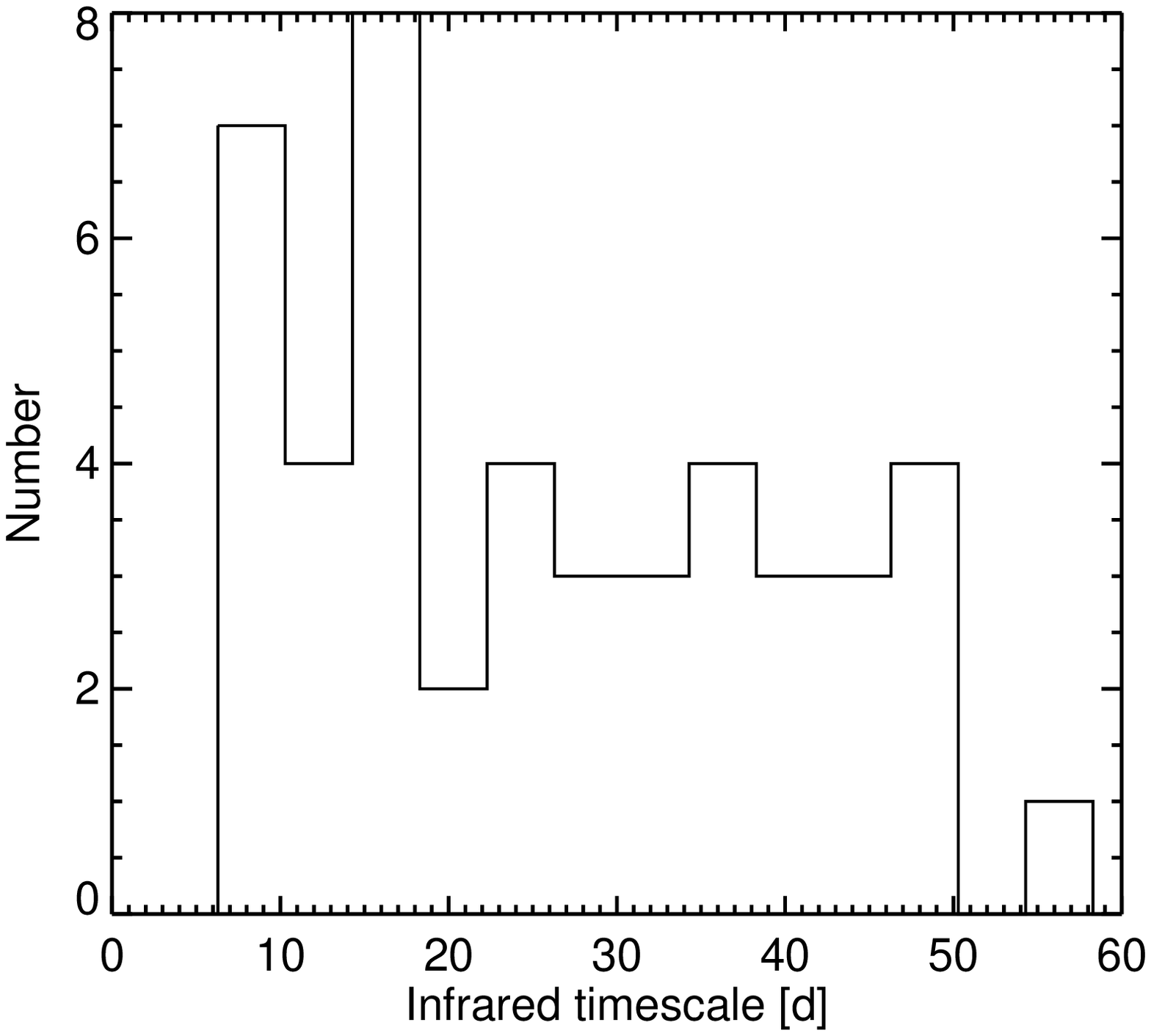}
\end{center}
\vspace{-0.5cm}
\caption{\label{IRtimescale}The distribution of timescales measured
  with the PeakFind algorithm for infrared light curves that are not
  well correlated with their optical counterparts.}
\end{figure}

\subsection{Inverse correlation: optical and infrared phase shifts}

A prediction of YSO variability models for edge-on systems \citep[e.g.,][]{2013AAS...22125610K} is
that inner disk regions may receive non-uniform illumination from the
central star, thereby reradiating infrared flux in time-variable
manner. An example is an accretion hot spot on the stellar surface,
radiation from which interacts only with the region of the inner disk
on that hemisphere of the star. In this case, we expect to observe the
infrared emission 180 degrees out of phase with the hot spot
signature, since we do not view the heated region of the disk wall until it
has rotated behind the star. 

\begin{figure}
\begin{center}
\includegraphics[scale=0.44]{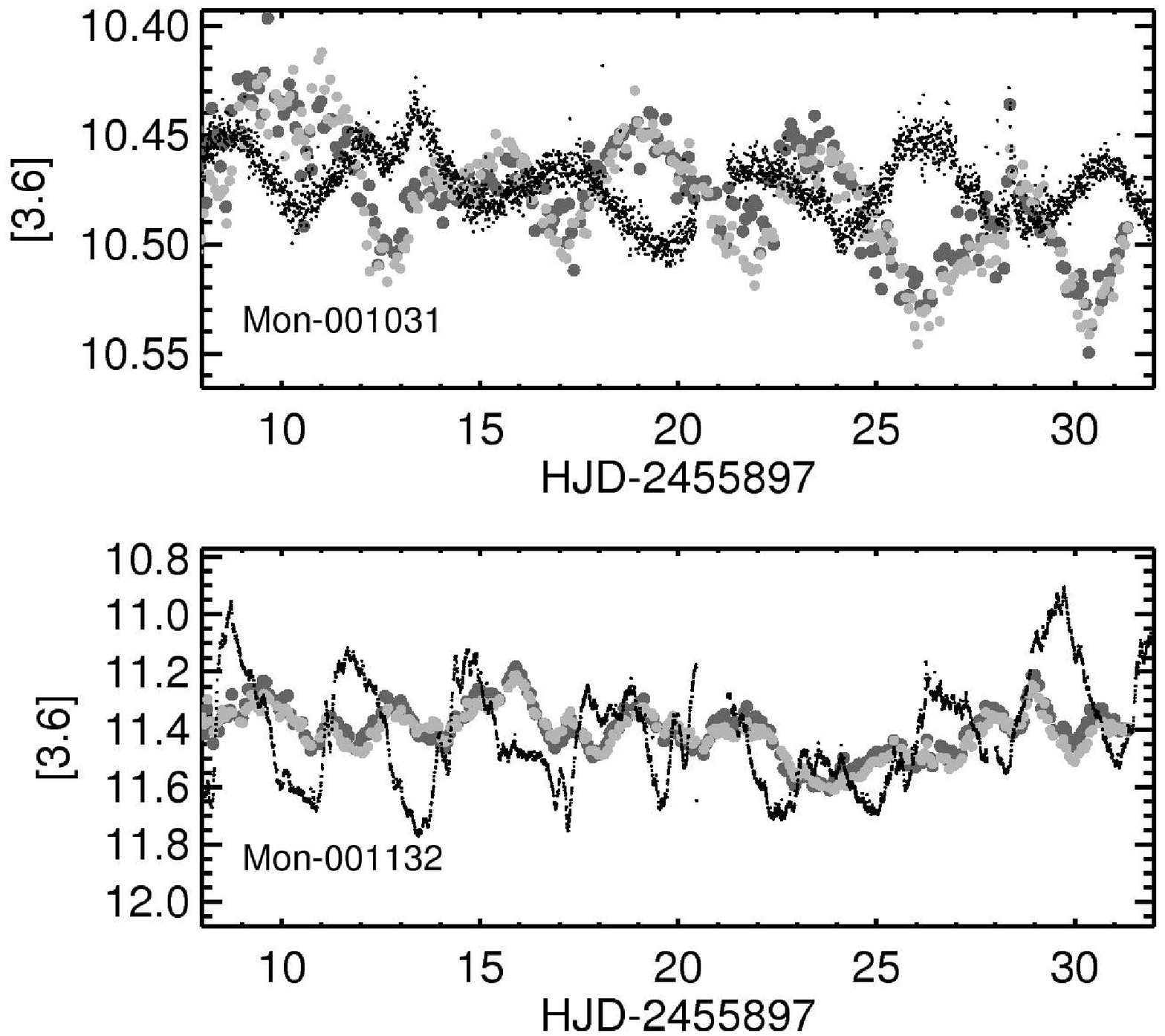}
\end{center}
\caption{\label{anticorr} Light curves with potential phased shifts between
  their optical and infrared behavior. Small black points are {\em CoRoT} data,
  light grey points are 3.6~$\mu$m data, and dark grey
  points are 4.5~$\mu$m data (sometimes hidden behind the 3.6~$\mu$m points).}
\end{figure}

Surprisingly, we encounter very few such examples of optical/infrared
phase shifts in our time series. The only object with a clear inverse
relationship between the two bands is Mon-001031. Mon-001132 displays
a more subtle phase lag, but this shift does not appear
to be constant across the entire observation time. We present the two
light curves in Figure~\ref{anticorr}.

\section{Variablity in the context of stellar and disk properties}

For most of the {\em CoRoT}/{\em Spitzer} dataset, we have available spectral types,
classification of the infrared excess, and optical through infrared photometry enabling
determination of position on the H-R diagram. In many cases, we also have measurements of the
equivalent width of prominent emission lines, such as H$\alpha$. 
We have correlated the variability properties measured in Section~6 with a number of the above-mentioned
parameters and searched for combinations that may offer physical insight. 

\subsection{Relationship of variability to stellar parameters}

\begin{figure*}
\begin{center}
\includegraphics[scale=0.75]{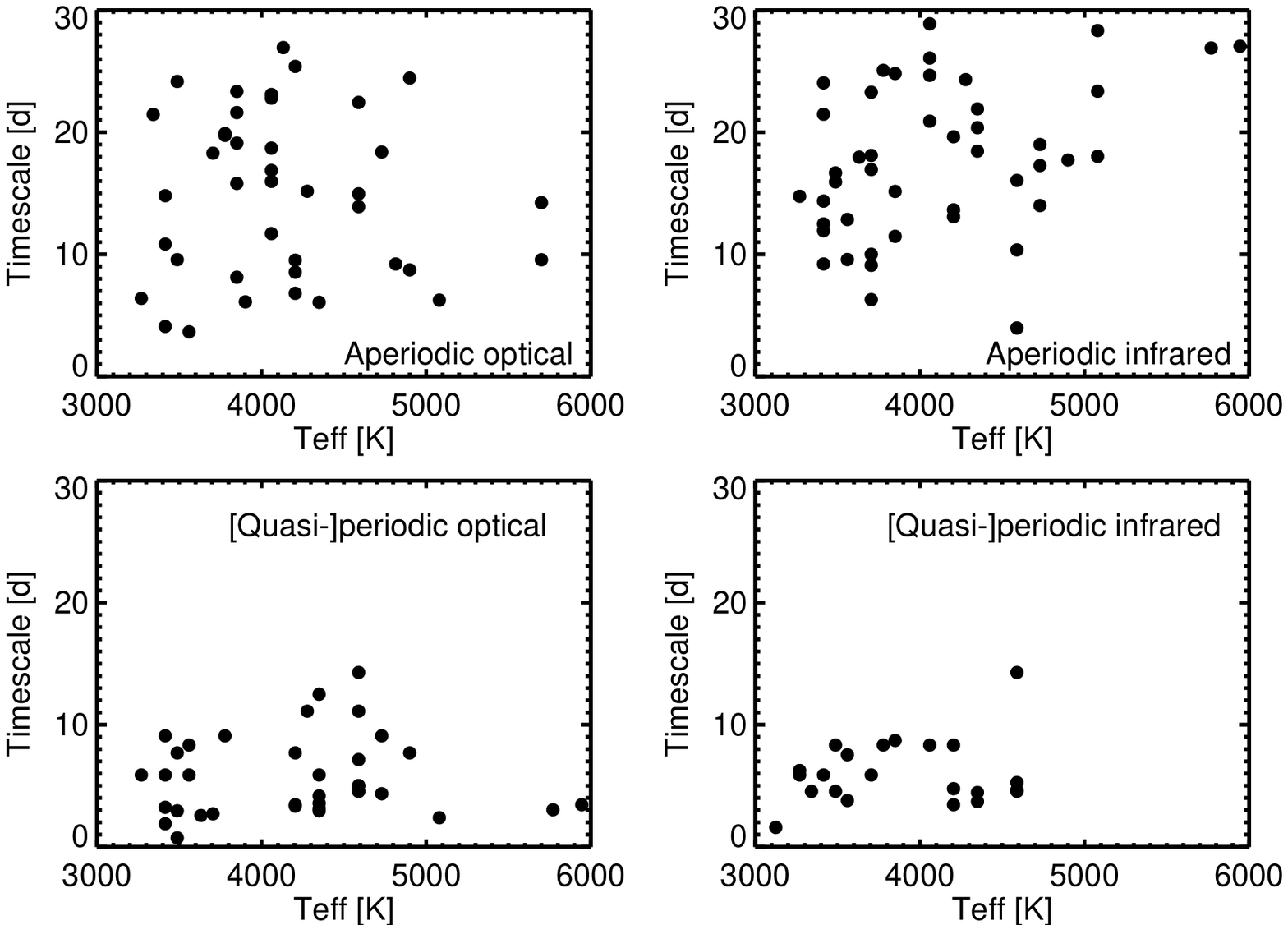}
\end{center}
\caption{\label{timescalemass}We plot variability timescale against effective temperature (where spectral types are available), as
a proxy for mass. We do not see correlations between these two parameters, although the lower envelope of aperiodic infrared
timescales is roughly consistent with the orbital period at the disk sublimation radius.}
\end{figure*}

The main stellar parameter of interest is the effective temperature, which serves as a
proxy for mass. We have estimated this for the 112 stars in our sample that have available
spectral types in Table~3. We then plotted the variability properties involving timescale
and amplitude against temperature. One might expect infrared variability timescale to scale with 
temperature since it could reflect dynamics near the sublimation radius, which is in turn dependent on stellar
mass. For completeness, we have separated the timescale measurements into quasi-periodic and aperiodic
sets, as well as divided them into the two wavelength bands before comparing with temperatures.
The correlation diagrams resulting from this exercise are shown in Figures~\ref{timescalemass}.
Timescale, whether periodic or aperiodic, does not show any prominent trends as a function of temperature. 
However, there is a lower envelope to the distribution of aperiodic infrared timescales versus temperature;
this could be consistent with the orbital period at the sublimation setting the minimum timescale for 
disk variability.
In addition, the implied lack of a global dependence of the variability timescale on mass does not necessarily mean
that there is no relationship between these two parameters. We suspect that there are subclasses of variability
that exhibit different behavior as a function of stellar mass. For example, the dipper objects show larger infrared 
amplitudes in comparison to the optical for preferentially later spectral types. 
Finally, there may be currently unobservable factors, such as disk inclination, that
influence the variability properties. 

\begin{figure*}
\begin{center}
\includegraphics[scale=0.75]{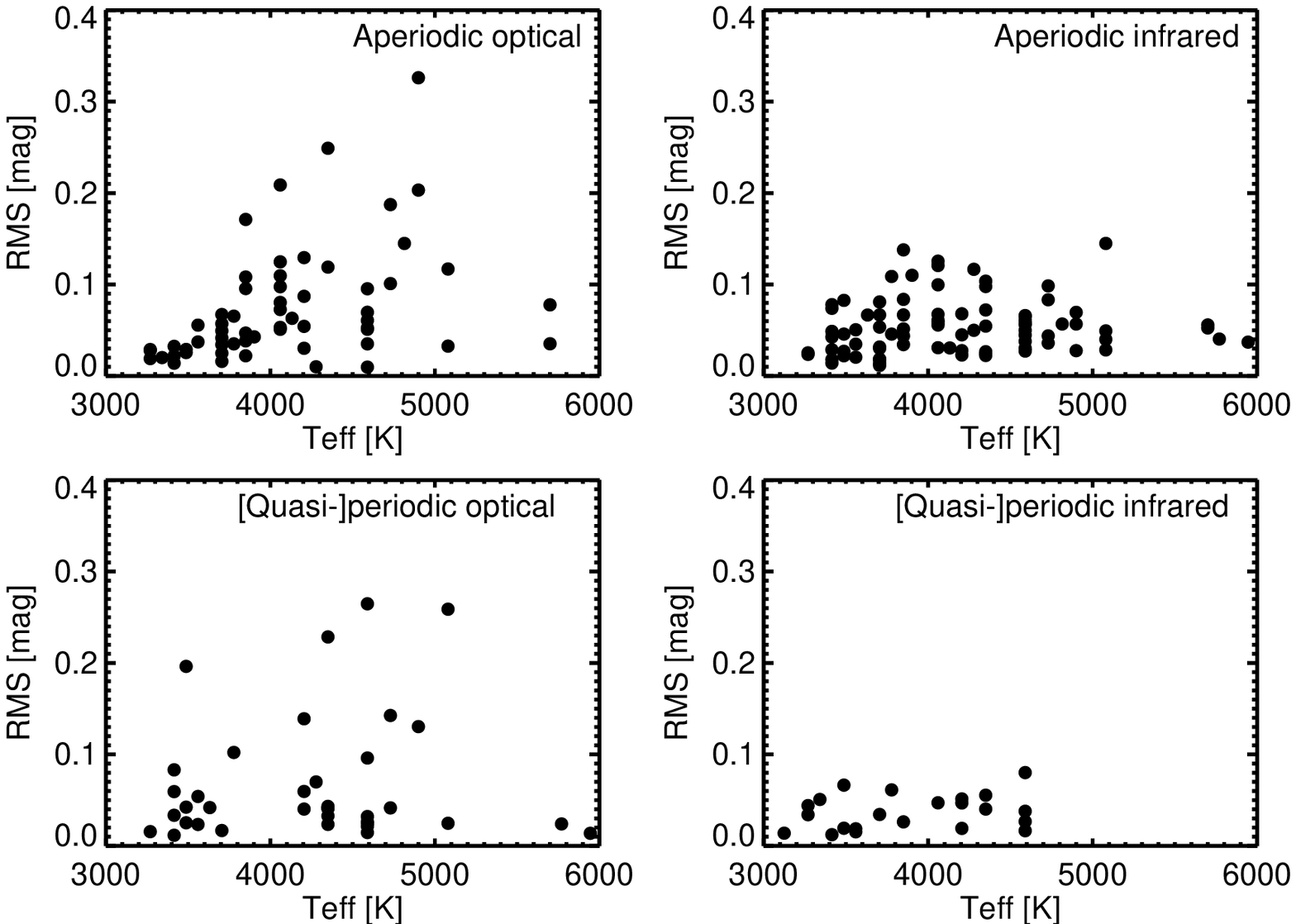}
\end{center}
\caption{\label{rmsmass}RMS versus stellar effective
    temperature, for [quasi-]periodic and aperiodic variables in both
    bands. The amplitudes of optical variability appear to grow with
    mass, while no such trend is seen in the infrared.}
\end{figure*}

We have also carried out a similar comparison of variability amplitude versus temperature, as shown
in Figure~\ref{rmsmass}. Here, we find the surprising result that aperiodic variability amplitudes are significantly 
lower around the cool stars in our sample. This trend is the opposite of what one would expect from a detection bias, in 
which it is harder to detect low amplitude variability around fainter stars. The effect is stronger in the optical, but
also seems to appear in the infrared. This is in contrast to the periodic variability, which shows no appreciable
mass dependence. The rms/temperature trend in aperiodic variables may reflect a correlation of magnetic
field strength and/or configuration with mass. However, this may be contradicted by recent results from Zeeman 
doppler imaging which suggest that the dipole field component {\em weakens} with increasing mass \citep[e.g.,][]{2012ApJ...755...97G}.

\begin{figure}
\begin{center}
\includegraphics[scale=0.5]{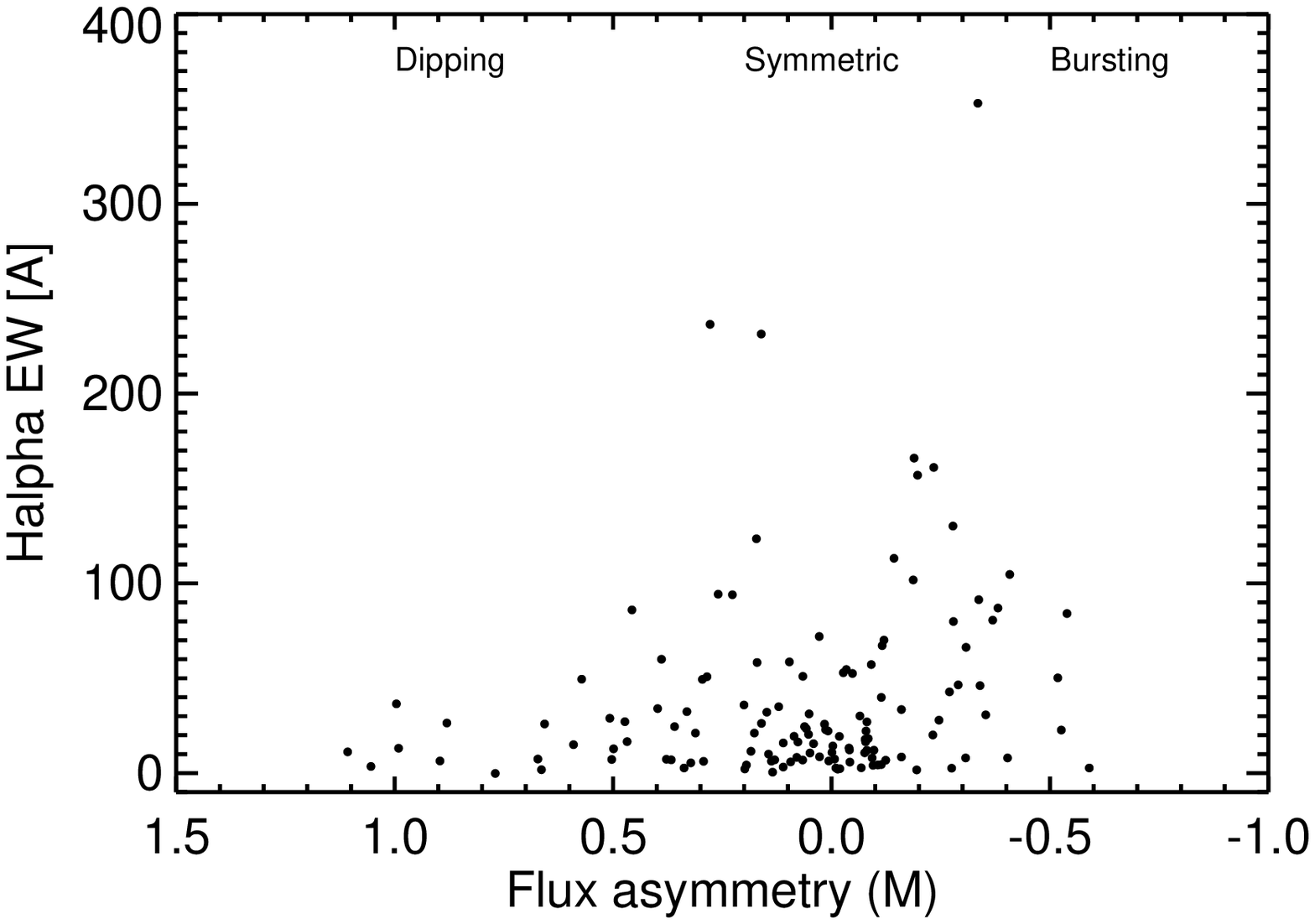}
\end{center}
\caption{\label{MHalpha}Flux asymmetry ($M$) against the equivalent width of
H$\alpha$ emission. The more negative $M$ values correspond to
bursting behavior in the light curves.}
\end{figure}

Perhaps the most telling correlation we have noted is the connection
between light curve flux asymmetry ($M$) and H$\alpha$ equivalent
width. We plot these two parameters in Figure~\ref{MHalpha}. While
there is not a one-to-one correspondence between them, it is clear
that the more negative $M$ values correspond to stronger H$\alpha$
emission. This finding supports the idea that the burster class of
light curves represents the most strongly and unstably accreting stars
in our dataset (see also \citet{Stauff13} for confirmation of this
idea). 

\subsection{Infrared excess}

\begin{figure}
\begin{center}
\includegraphics[scale=0.5]{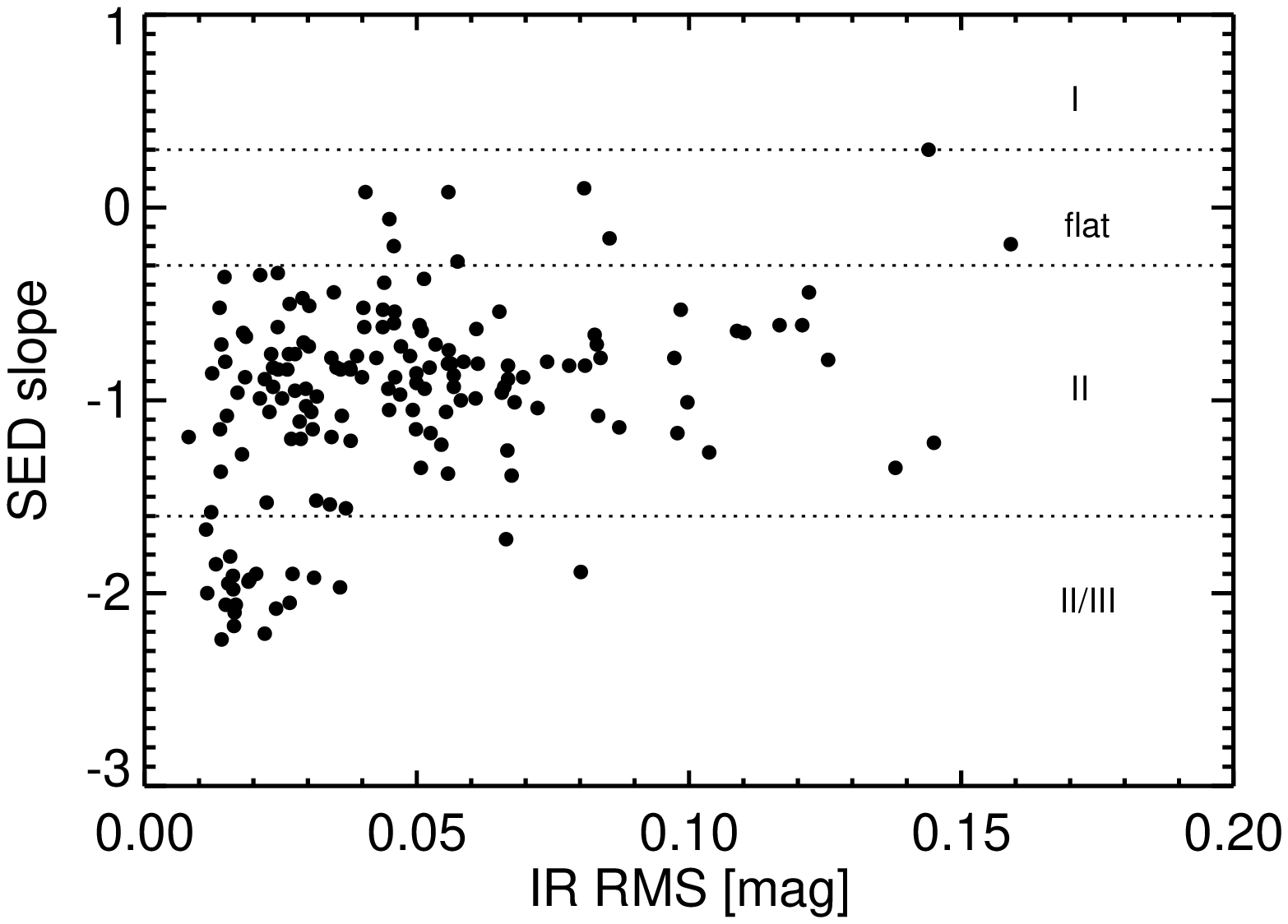}
\end{center}
\vspace{-0.5cm}
\caption{\label{iralpha} RMS value of infrared light curves versus their SED slope, $\alpha$. 
More positive $\alpha$ values are correlated with larger infrared RMS, suggesting a relationship
between disk variability and evolutionary state. Dotted lines mark the boundaries between various
SED classes, which are labeled; most of our sample is class II.}
\end{figure}

In addition to searching for global correlations between
variability and stellar properties, we can also ask whether
variability is well connected to {\em disk} properties. 
We first compared the RMS light curve value for infrared variables
with the slope of the SED, $\alpha$ (as calculated in Section 2.1). 
The result, shown in Figure~\ref{iralpha},
does not display a strong dependence on morphology type, but there
does appear to be a subtle lower envelope, with few high-amplitude
variables at low $\alpha$ values. This suggests that disks with earlier
classes can achieve higher levels of variability, a result that has also been borne
out in other clusters \citep[e.g.][]{2009ApJ...702.1507M}. It also makes sense
given that the class II/III objects in our sample have less dust to generate variability.

\begin{figure}
\begin{center}
\includegraphics[scale=0.5]{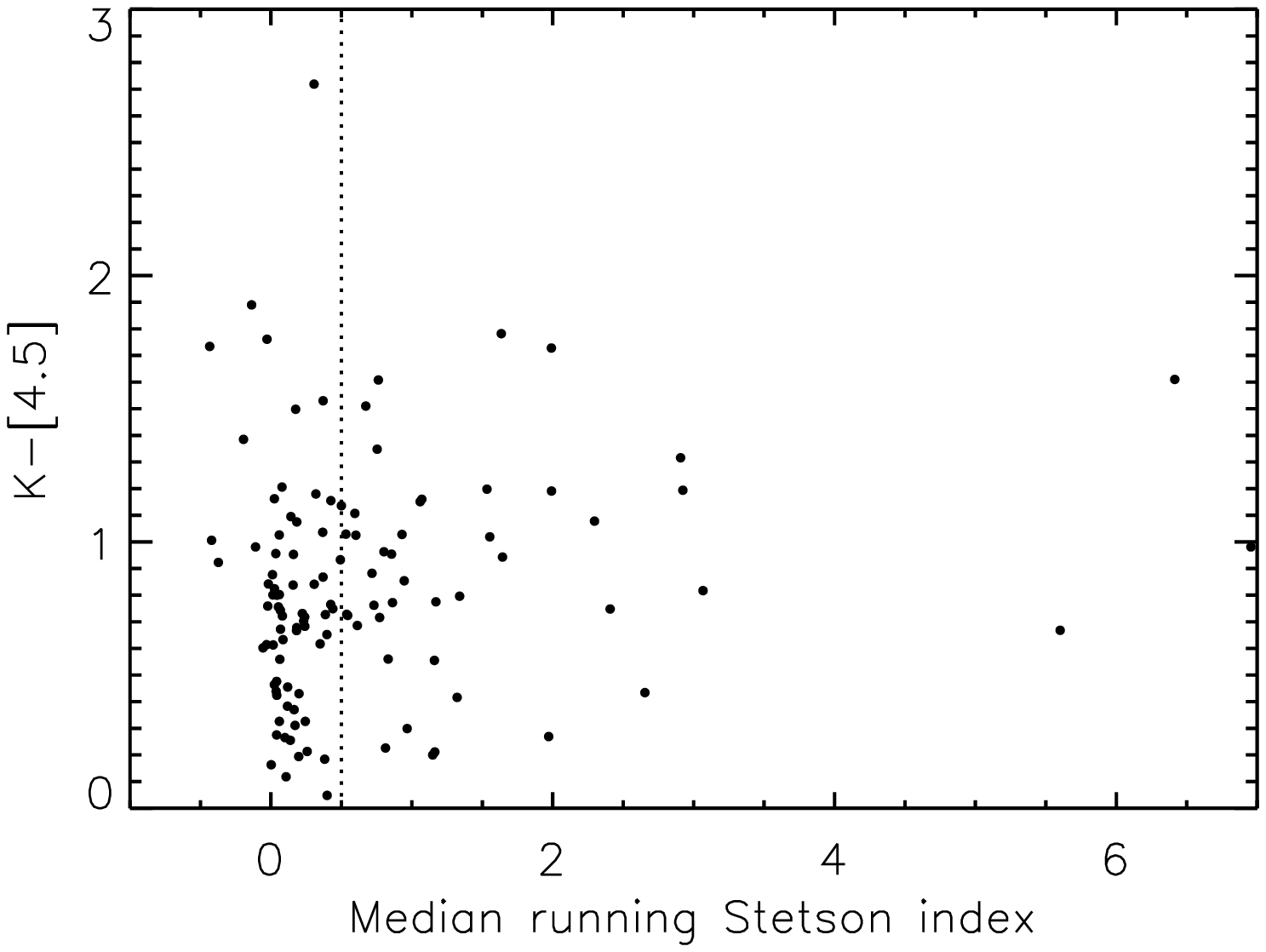}
\end{center}
\vspace{-0.5cm}
\caption{\label{stetsonds}The degree of optical/infrared correlation
    (i.e., Stetson index) does not
  have a one-to-one relationship with the strength of disk emission,
  as parameterized here by the $K$-[4.5] color.}
\end{figure}

We also suspect that the ratio of disk excess to stellar flux plays a
significant role in light curve morphology.  As a proxy for this ratio,
we have computed the $K$-[4.5] color of all sources in our sample.
We compare this with the median running Stetson index in
Figure~\ref{stetsonds}. Surprisingly, there is no strong dependence of 
optical/infrared correlation on disk strength either. However, the
light curves of a number of individual sources do make
sense in the context of their SEDs. For example, several stars (e.g.,
Mon-001094) display well correlated behavior at similar amplitudes in the two bands.
We suspect in these cases that there is a single variability process at work,
and it is associated with the stellar surface. The disk, on the other hand,
is weak and has little emission compared to the stellar flux in all bands.
We see this is the case from the small $K$-[4.5] values of these objects.
Estimates of their disk to stellar flux ratios in the infrared also support this
idea, since the values are less than 10$^{-3}$.

There are other objects in our diagram for which the high degree of optical/infrared correlation is not
expected to relate strongly to the disk flux strength. These are the dippers, which are suspected
to be caused by dust extinction events. This phenomenon is mostly dependent on disk
inclination, and can occur even for transitional disks, as is the case for
prototype Mon-000660 (also known as V354~Mon).
Therefore, as with the lack of variability and temperature trends, the fact that 
there is no global relationship with disk flux is not particularly surprising.

\section{Summary}
We have analyzed a sample of 162 disk bearing members of NGC~2264 with
high cadence time series photometry in the optical from {\em CoRoT}
and in the mid-infrared from {\em Spitzer}. 
Overall, we find that 81\% of our disk bearing sample is variable in
the optical, and 91\% is variable in the infrared. While the stars without
disks will be the subject of a future paper, we have found that the
infrared variability level among these is much lower, at 36\%.
These statistics confirm that accretion disk processes are strongly associated with flux
changes.  The timescales comprising the aperiodic variability span a range from a few days
out to the duration of the dataset.

We identify seven light curve morphology classes in each band, including quasi-periodic 
variability, dippers, bursters, stochastic behavior, strictly periodic, and long timescale 
behavior. Among the optical variables, quasi-periodic phenomena are the most common, whereas 
long timescale and unclassifiable behavior are the most common in the infrared. Periodic 
and aperiodic dipper behavior is surprisingly common in the optical, at over 21\% of the
variability sample. Comparing with the \citet{2000A&A...363..984B} prediction of a 15\% 
occurrence rate for occultations by circumstellar material, we infer that either the disk
scale height has to be higher than previously assumed, or the disk obscuration model may
be invalid in part. A further puzzle came in the form of two stars (Mon-000183 and Mon-000566)
displaying dips that are {\em deeper} in the infrared (nearly 10\%) than in the optical
($\sim$2--3\%). A very non-standard extinction law, or geometric peculiarities will have to
be invoked to explain these two enigmatic objects.

Pure periodicities are very rare in the disk bearing sample. We argue that each of these classes 
represents a distinct physical variability mechanism on the star or in the disk. While 
follow-up studies including spectroscopy are needed to confirm physical mechanisms, the data 
are consistent with variable circumstellar obscuration, unsteady accretion, rotating starspots, 
and rapid structural changes in the disk-- and likely multiple of these processes happening 
simultaneously. Some of the variability may be due directly to the effects of these phenomena
on physical properties, while some may be due to radiative transfer effects on the spectral
energy distribution.

The clear asymmetries in the dipper and burster type light curves
motivated us to develop a statistic measure of asymmetry, ``M.''  
In combination with the quasi-periodicity, ``Q,'' we can uniquely
identify the variability type of each star in each band without
relying on subjective evaluation. Although we
set out to understand the wavelength dependence of variability in our
sample, we have found that in fact optical and infrared variability
behavior in young disk-bearing stars is not well correlated in over 50\% of cases. 
Most of our stars have different optical and infrared classes, with
dipping and bursting behavior becoming less common at longer
wavelengths and long timescale variability becoming more common.

We highlight the set of high amplitude infrared variables for which
there is little to no corresponding variability in the optical. We
have measured timescales of order $\sim$5--10 days for these variables,
providing evidence that they represent structural rearrangements in
the disk. Among the entire sample, we find that high amplitude ($>$0.1 magnitudes)
infrared variability appears in objects with SED class of type II and earlier,
whereas the majority of type II/III transitional disk systems have RMS infrared amplitudes 
less than 0.05 magnitudes. Other correlations
identified between variability properties and star/disk parameters
include increases in aperiodic amplitude with stellar effective
temperature, as well as increased flux asymmetry with H$\alpha$ equivalength
width. These correlations hint at a connection between variability
properties and magnetospheric structure. 

Overall, we have developed a process for automatically classifying
variability that can be applied to other datasets and clusters as
well. We have focused exclusively on time series photometry from {\em
 CoRoT} and {\em Spitzer} here, but in future papers we will explore
variability at the full range of available wavelengths.


\appendix

\subsection{NGC 2264 membership criteria}
NGC~2264 has been the subject of many young cluster studies, from
photometric and H$\alpha$ censuses
\citep[e.g.][]{2002AJ....123.1528R,2004A&A...417..557L} to X-ray 
\citep[e.g.,][]{2004AJ....127.2659R,2006A&A...455..903F,2013arXiv1309.4483F} and radial velocity surveys
\citep[e.g.,][]{2006ApJ...648.1090F}. To cull a reliable membership list from candidates reported in the literature,
we required that objects meet at least two of six criteria: 1) photometric data consistent with the $V-I$ or
$R-I$ cluster locus defined by \citet{2006A&A...455..903F} (see their
Section 3.2), 2) strong photometric H$\alpha$ emission, according to
the criteria of \citet{2008AJ....135..441S}, {\em or} spectroscopic H$\alpha$
equivalent width (EW) larger than 10\AA, 3) X-ray detection at a flux
greater than 10$^{-4}$~$L_{\rm bol}$ \citep{2004AJ....128..787R,2004AJ....127.2659R,2006A&A...455..903F}, 4)
radial velocity consistent with NGC 2264 membership, as classified by \citet[][]{2006ApJ...648.1090F} 5)
mid-infrared excess indicating a disk (i.e., class I, II, or flat
SED, according to selection methods described in Section 4.2), and 6) spatial location coinciding with an $A_V>7$ region of
NGC~2264, if an object displays an infrared excess or X-ray emission but is not detected
in the optical. 

We identified highly embedded objects for criterion 6) by plotting
X-ray and mid-infrared source locations on the extinction map produced
by \citet{2012A&A...540A..83T}. These obscured stars were added to our overall
membership list for further study, but not included in this paper
since we are focusing specifically on targets with {\em CoRoT} detections.

The above requirements eliminate most field dwarf and extragalactic
contaminants from our membership sample. We have avoided selecting
members based on variability detection so as to not bias our
statistical analysis of the light curve types.

\subsection{IRAC staring data}

While not used extensively in this paper, the high-cadence staring data from
{\em Spitzer}/IRAC offers an important window into short-timescale
infrared variability in YSOs. In preparing it for analysis, we performed
a series of procedures to remove systematics. The steps described below
may be of general interest to other {\em Spitzer}/IRAC users.

\subsubsection{Staring data quality and correction of systematics}

The pixel-phase effect, or variation of measured flux as a function of sub-pixel position, is
a known issue affecting the IRAC detector (see the IRAC instrument
handbook, v2.0.3). Intrapixel sensitivity variations introduce two systematic effects to
the staring (i.e., continuous, non-dithered) light curves: first, the flux may vary up to 6\% as the pointing
shifts gradually from one side of a pixel to the other; second, the flux
exhibits oscillations at the percent level on $<$1~hour timescales.
The first effect reflects the pointing stability of {\em Spitzer},
while the second was determined by the engineering team
to be associated with the cycling of a battery heater onboard the
satellite, causing periodic flexure between the star tracker and the
cold focal plane. The result is that the pointing center
on the detector oscillates within individual pixels, convolving their
sensitivity variations with the intrinsic flux. In October 2010, the
{\em Spitzer} engineering team was able to reduce both the amplitude and
period of the heater-associated temperature fluctuations, the
latter from $\sim$60 minutes to $\sim$40 minutes. However, the effect remains
prominent at the 0.5\% level in both channels, comparable to the white
noise contribution for bright stars in the sample.

Our goal of assessing infrared variability of young stars on short
timescales and at low amplitudes neccesitates the removal of pixel
phase variations to the extent possible from the data. A pixel phase correction
exists for Warm {\em Spitzer} observations, but tests on our light curves
revealed that it works for stars on some pixels, but not others. Several techniques have been
developed to provide a more consistent correction, many of which involve fitting a
polynomial to the flux as a function of detector $X$ and $Y$ position \citep[e.g.,][]{2011A&A...528A.111B}.
Unfortunately this approach does not work well for many of our targets, as
the erratic variability that we seek to study cannot be modeled analytically and thus prevents a robust fit.
In \citet{2011ApJ...741....9C}, we adopted a
different technique, involving a Gaussian model of the intra-pixel
sensitivity variation, for which free parameters were fit by minimizing
the height of the pixel-phase oscillation peak in the
periodogram. Maps of the pixel sensitivity were determined
independently for each star, and the inferred flux variation as a
function of $X$ and $Y$ position was then subtracted from the
light curves. 

While this method provided satisfactory corrections without
compromising intrinsic stellar variability, it was
computationally intensive. To apply a similar technique to our $\sim 500$ staring
targets, we require an algorithm with fewer free parameters. Fortunately during our 2011 observations of NGC~2264,
object centroids were stable to $\sim$0.08 pixel (i.e., 0.1\arcsec) in
the $X$ detector position over the course of each 16--26 hour staring mode
observation. Flux variations are much better correlated with the $Y$
centroids, which vary by 0.3--0.4 pixel (0.3--0.5\arcsec). Plots of the normalized flux versus $Y$ for
stars lacking obvious variability by eye revealed nearly linear
trends, as shown in Figure~\ref{fluxposition}.
We therefore chose a basic pixel phase correction that fits a
single slope characterizing the pixel sensitivity as a function of $Y$
position. To determine this slope, $A$, we express the sensitivity, $s$,
as a linear function of  $Y$ position on the detector:
\begin{equation}
s(y)=A(y-y_{\rm med})+1.
\end{equation}
Here we have normalized the sensitivity to a median value of 1.0 in the
area of the pixel where the data fall.
Applying this sensitivity function, the true flux data, $d_{\rm true}(y)$,
will be observed as
\begin{equation}
d_{\rm obs}(y)=d_{\rm true}(A(y-y_{\rm med})+1)
\end{equation}

Note that we have specified the observed and true data to have the same median.
To correct the observed data, we then invert this equation,
exploiting the fact that $y-y_{\rm med}<<1$:
\begin{equation}
d_{\rm true}(y)=d_{\rm obs}/(A(y-y_{\rm med})+1)\sim d_{\rm obs}(1-A(y-y_{\rm med}))
\end{equation}

We wish to determine the value of $A$ so that the corrected
light curve is devoid of pixel-related systematics. To do this, we take advantage of the pixel phase
effect, which causes $y(t)$, the detector position, to oscillate
rapidly as a function of time. The period of oscillation is
approximately 40 minutes. To determine this value more accurately, we
compute a Fourier transform periodogram for each chunk of staring
data, $y(t)$, first subtracting out its median value $y_{\rm med}$ to remove low frequency systematics: 
\begin{equation}
FT(f,y(t_k))=\frac{2}{N}[(\sum_{k=0}^{N}\sin(-2\pi ft_k)(y(t_k)-y_{\rm med}))^2+(\sum_{k=0}^{N}\cos(-2\pi ft_k)(y(t_k)-y_{\rm med}))^2]^{1/2},
\end{equation}
where $f$ is frequency in the periodogram, $k$ is the index of points
in the time series, and $N$ is the total number of points.
We numerically locate the highest peak in this periodogram and note
its frequency, $f_0$. We find a value of $\sim$37.5 cycles per day; this can be
seen with the solid curve in the bottom panels of
Figure~\ref{pixelperiodogram}. The corresponding pixel
phase period is $1/f_0\sim $38.4 minutes. 

We will now use the oscillatory behavior of $y(t)$ to determine the
best value of the slope $A$ in the pixel sensitivity function
$s(y)$. We do this by minimizing the value of the Fourier transform
periodogram of the data itself, $d(y(t_k))$, at $f_0$. The value of $A$ resulting
from this process can then be used to remove the effects of pixel
sensitivity variation from the data without compromising variability. We begin by writing out the
Fourier transform periodogram for the true data, again subtracting
out the median value, $d_{\rm med}$, to eliminate periodogram systematics:
\begin{equation}
\begin{split}
FT(f,d(y(t_k))=\frac{2}{N}[(\sum_{k=0}^{N}\sin(-2\pi ft_k)(d_{\rm true}(t_k) -d_{\rm med})^2+(\sum_{k=0}^{N}\cos(-2\pi ft_k)(d_{\rm true}(t_k) -d_{\rm med}))^2]^{1/2}\\
=\frac{2}{N}[(\sum_{k=0}^{N}\sin(-2\pi ft_k)(d_{\rm obs}(t_k)\cdot (1-A(y_k-y_{\rm med})) -d_{\rm med})))^2+\\
(\sum_{k=0}^{N}\cos(-2\pi ft_k)(d_{\rm obs}(t_k) \cdot(1-A(y_k-y_{\rm med})) -d_{\rm med}))^2]^{1/2}
\end{split}
\end{equation}
We next minimize the periodogram with respect to the slope, $A$, at the peak
frequency $f_0$:
\begin{equation}
\frac{d{\rm FT}(f_0,d(y(t_k))}{dA}=0.
\end{equation}
It can be shown that this equation takes the form
\begin{equation}
0=AC_1+C_2+AC_3+C_4,
\end{equation}
where $C_1$, $C_2$, $C_3$, and $C_4$, are constants, as follows:
\begin{equation}
C_1=-\left[\sum_{k=0}^{N}\sin(-2\pi f_0t_k)\frac{d_{\rm
      obs}(t_k)}{d_{\rm med}}\cdot (y_k-y_{\rm med}) \right]^2,
\end{equation}
\begin{equation}
C_2=\left[\sum_{k=0}^N \sin(-2\pi f_0t_k)\left(\frac{d_{\rm
      obs}(t_k)}{d_{\rm med}}-1\right) \right] \left[\sum_{k=0}^{N}\sin(-2\pi f_0t_k)\frac{d_{\rm
      obs}(t_k)}{d_{\rm med}}\cdot (y_k-y_{\rm med})\right]
\end{equation}
\begin{equation}
C_3=-\left[\sum_{k=0}^{N}\cos(-2\pi f_0t_k)\frac{d_{\rm
      obs}(t_k)}{d_{\rm med}}\cdot (y_k-y_{\rm med}) \right]^2
\end{equation}
\begin{equation}
C_4=\left[\sum_{k=0}^N \cos(-2\pi f_0t_k)\left(\frac{d_{\rm
      obs}(t_k)}{d_{\rm med}}-1\right) \right] \left[\sum_{k=0}^{N}\cos(-2\pi f_0t_k)\frac{d_{\rm
      obs}(t_k)}{d_{\rm med}}\cdot (y_k-y_{\rm med})\right]
\end{equation}

Operating on the four sections of by-BCD staring light curves, we
determine analytically the slope $A$ that minimizes the pixel phase
oscillation signal by computing
\begin{equation}
A=\frac{-(C_2+C_4)}{C_1+C3}
\end{equation}

Finally, we divide out the pixel response from the light curve
sections by calculating $d_{\rm true}$ as in Equation 3. 
Because the $X$ and $Y$ positions are well correlated in time, elimination
of the variation associated with shifts in $Y$ only should effectively
remove flux trends in $X$ as well (see Figure~\ref{fluxposition}).

Application of our algorithm resulted in a substantial reduction in
the level of systematic variability in our light curves,
as indicated by both visual inspection and measurement of the height
of the periodogram peak at the 40-minute pixel-phase oscillation
period. An example comparing our modified correction with the standard
correction and a raw light curve is shown in Figure~\ref{pixelperiodogram}.
For approximately 1\% of light curves, the pixel sensitivity
distribution was not well modeled by a linear fit and our correction
introduced oscillatory features that were not present in the original light
curve. In these cases, we retained the raw light curve or the version
corrected with the standard Warm {\em Spitzer} pixel phase
prescription, depending on which one displayed lower levels of
systematics. 

\begin{figure}
\begin{center}
\includegraphics[scale=0.51]{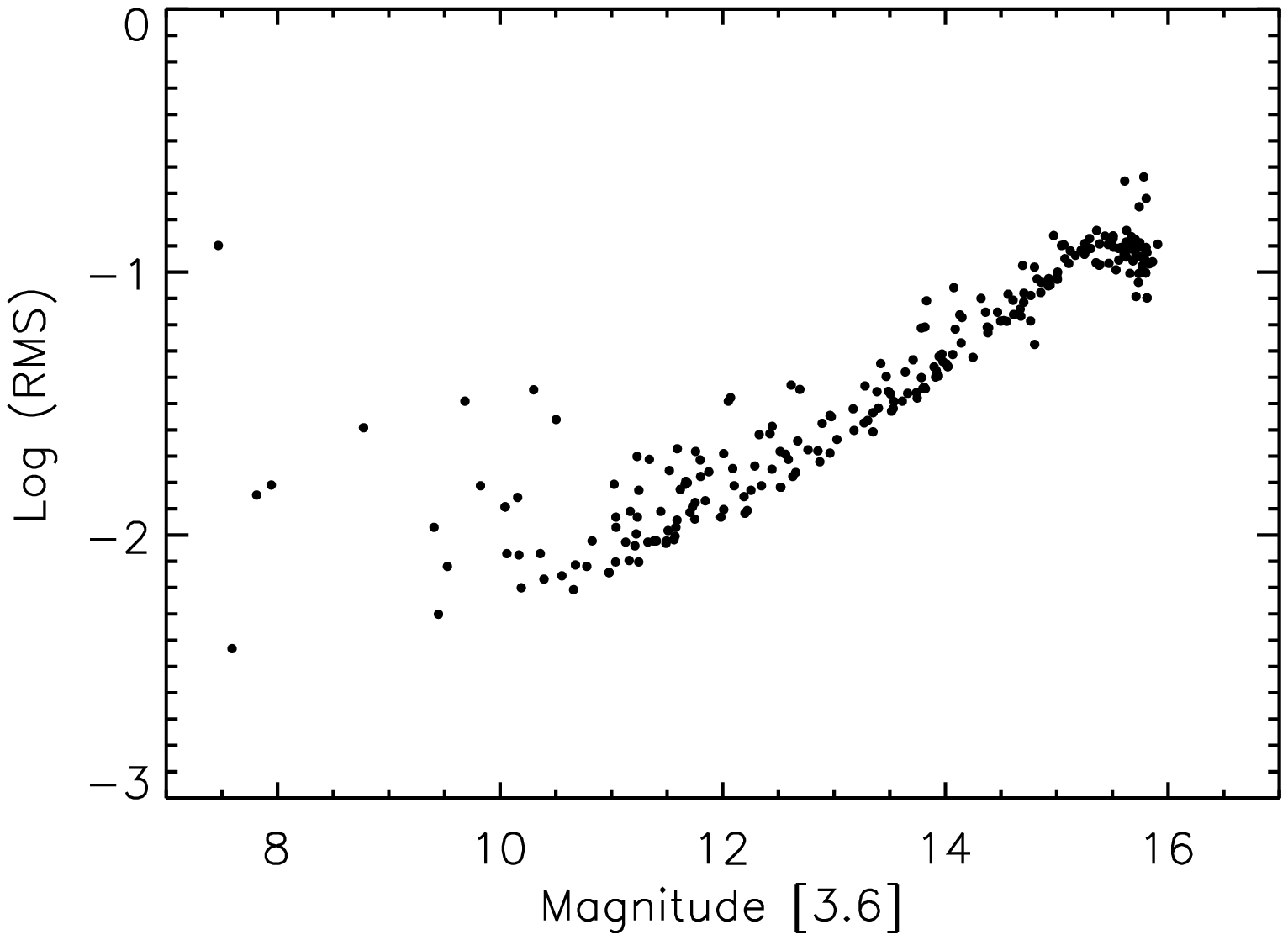}
\includegraphics[scale=0.51]{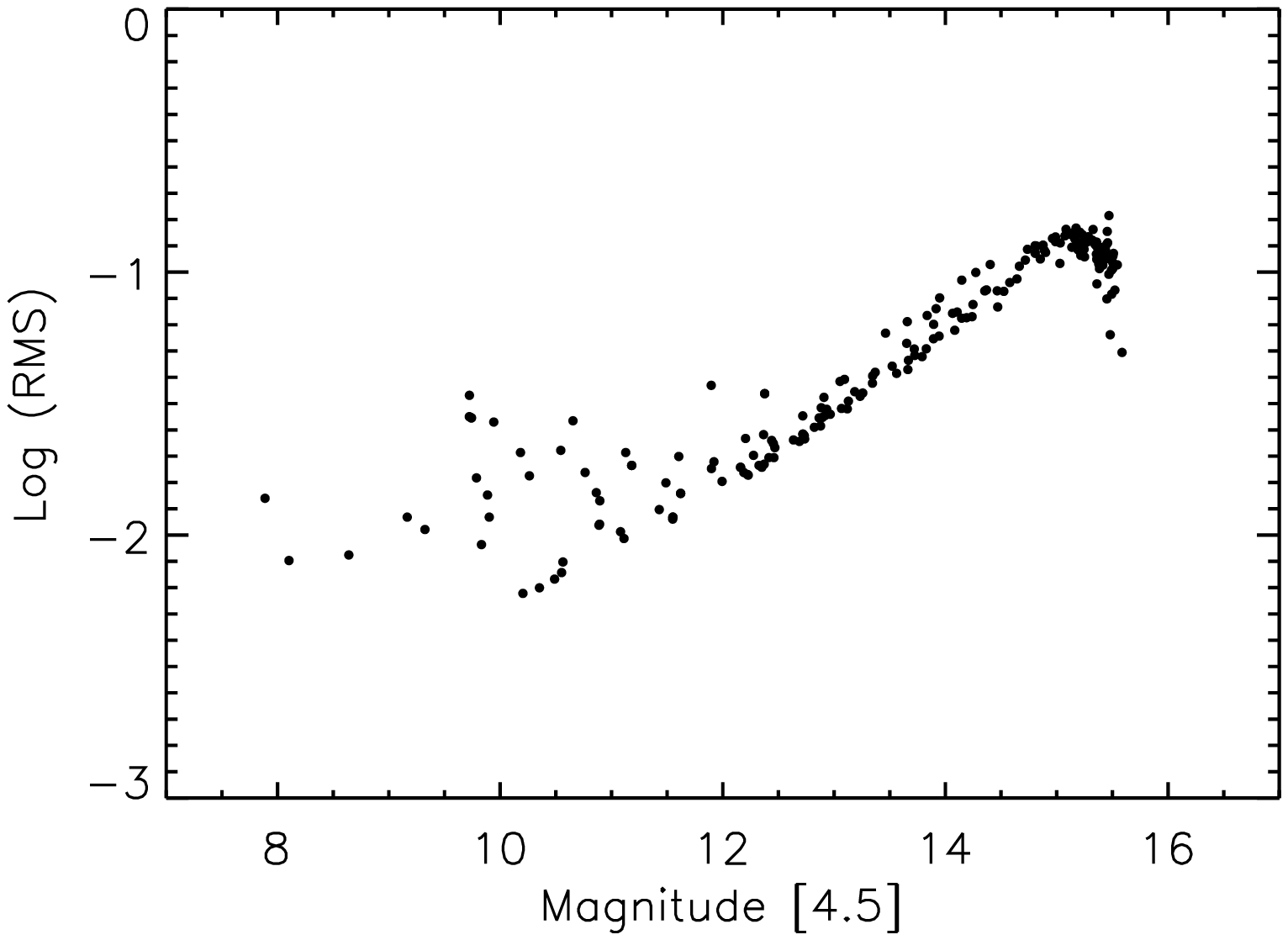}
\end{center}
\vspace{-0.5cm}
\caption{\label{IRACstd} Standard deviations for the first section of IRAC staring light curves in channel 1 (top) and
channel 2 (bottom), as a function of magnitude. The lower envelope of values closely matches the uncertainties predicted
by Poisson noise and sky background, wherease the higher values are due to stellar variability.}
\end{figure}

The RMS values of the corrected 3.6~$\mu$m staring light curves ranged from 0.003 magnitudes (at a magnitude
of 7.5) to 0.1 (at a magnitude of 15.5). The 4.5~$\mu$m light curves had
similar but slightly lower precision (by a factor of $\sim$1.5). RMS
values as a function of magnitude for one of the staring
photometry sections are presented in Figure~\ref{IRACstd}; these are consistent with the predicted
uncertainties based on Poisson noise and sky background.

\subsubsection{Merging the mapping and staring data}

Each section of staring photometry was surrounded by lower cadence
mapping photometry. While both data types contain the same systematics,
those taken in mapping mode are much less correlated in time, since the dithering and other pointing
changes sample the varying pixel sensitivity in a random manner. We
therefore cannot remove the mapping systematics but have accounted for
them (e.g., Section 3.3) for statistical purposes. 
Much of the staring photometry, on the other hand, displays
significant and correctable zero point offsets from the surrounding mapping
points. This effect occurs when the sensitivity of a single pixel
containing the object centroid during a staring observation differs
from the mean sensitivity across the many different pixels occupied
during mapping observations. Since discontinuities in the light curves
create challenges for variability analysis, we have introduced a further set of
corrections to ensure smooth transitions between staring and mapping
photometry. Following pixel phase mitigation, we selected the set of
$\sim$280 by-BCD staring points lying within 1.2 hours of the beginning or end of each
staring data section. Likewise, we select the set of $\sim$6 mapping
points (from by-AOR photometry) lying within 9.6 hours of the beginning or end of each mapping section. For each set of photometry, we
computed a linear fit to the magnitudes as a function of time, to
account for short-timescale variability. Offsets
between adjacent staring and mapping light curve sections were then
determined by subtracting the fit values at the point midway between
them. Each staring section had an offset at both ends, from which
we determined the mean. This final offset was then applied to the entire
staring section, thereby knitting it to the surrounding mapping data. 

The typical offset between staring and mapping data was
$\pm$0.01--0.05 magnitudes. It is unclear as to why it is often
larger than the random scatter in the mapping data. Since the offset fits are undoubtedly affected
by observational noise, we use the combined light curves with caution
and provide statistics separately for mapping data and individual
staring data sections. To produce staring light curves with
lower uncertainties, we further binned sets of 10 points for a final cadence of 2.5 minutes.
In all, there were twenty-six stars in our disk-bearing sample with both IRAC staring data and
{\em CoRoT} observations.

\begin{figure*}
\begin{center}
\includegraphics[scale=0.8]{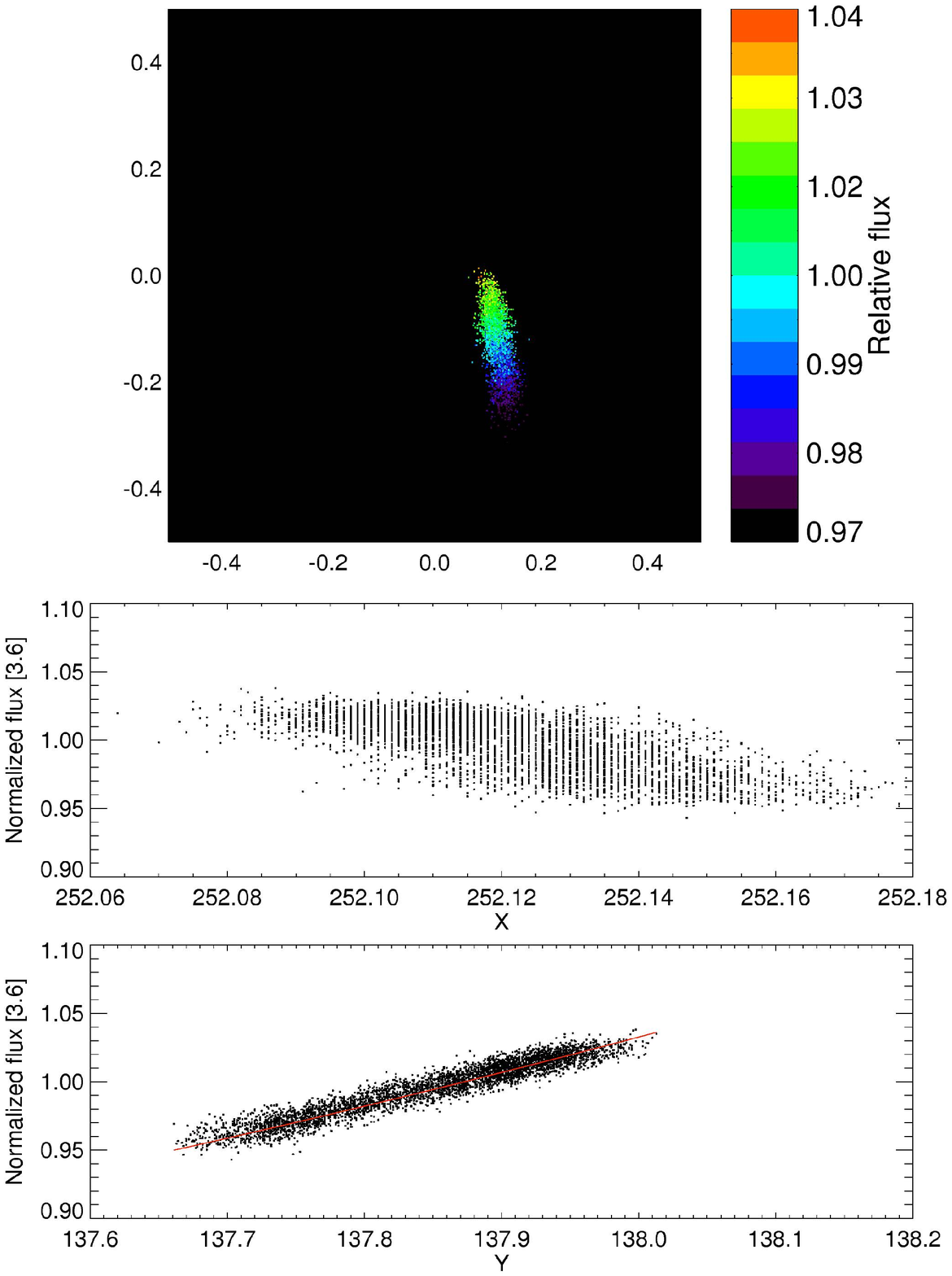}
\end{center}
\caption{\label{fluxposition} Normalized flux values as a function of
pixel position for a non-variable star in our sample (top). The
coordinate {0,0} is the pixel center. The pointing varied primarily in the $Y$ direction,
as shown in the flux vs. $Y$ and flux vs. $X$ plots at middle and
bottom. The red line indicates the slope in flux versus $Y$ that minimizes the pixel
phase peak in the periodogram shown in Figure~\ref{pixelperiodogram}; this is {\em not} a direct linear fit to the data.}
\end{figure*}

\begin{figure*}
\begin{center}
\includegraphics[scale=0.8]{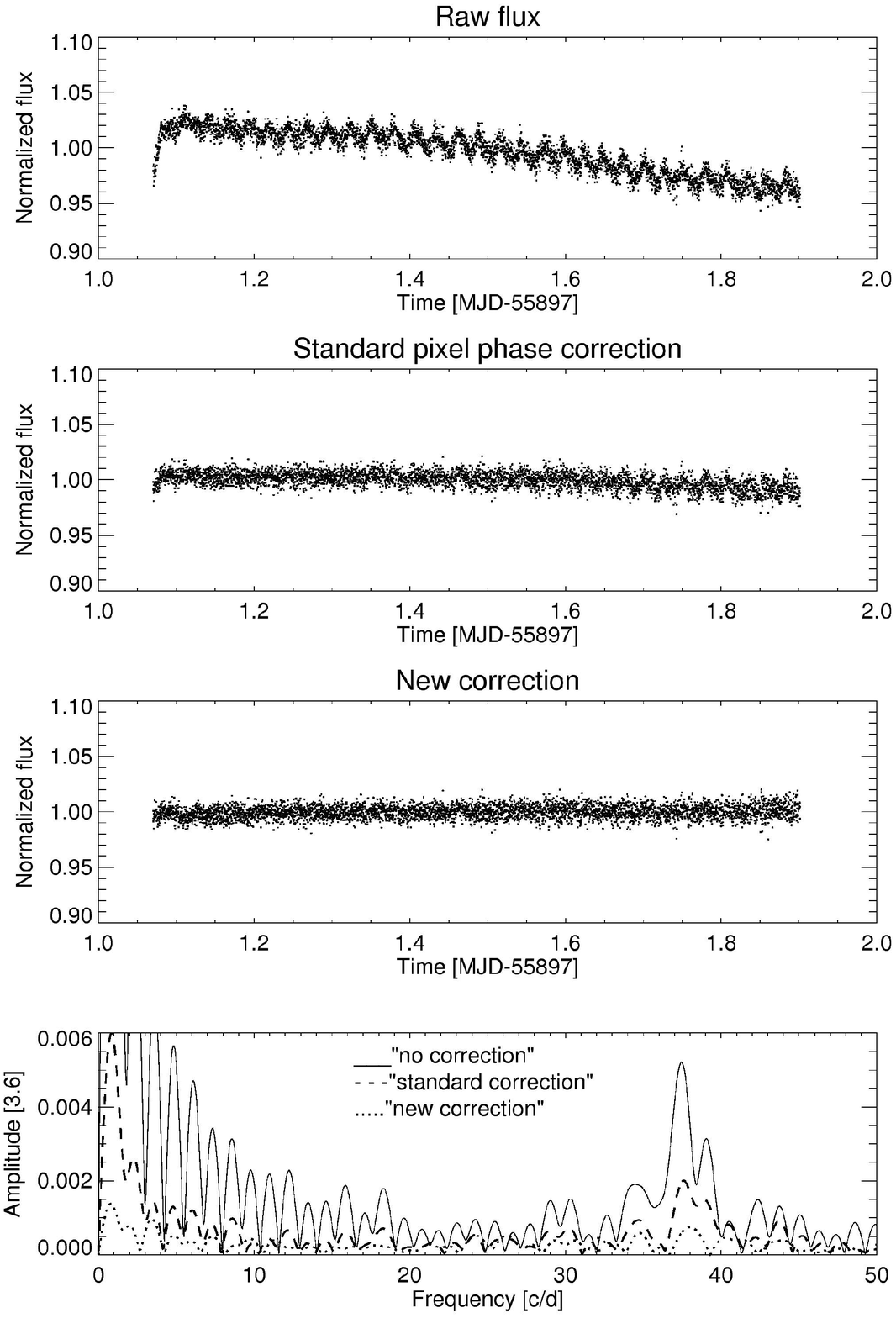}
\end{center}
\caption{\label{pixelperiodogram} {\em Top}: Uncorrected light curve, showing
the pixel phase oscillation, as well as longer timescale
systematics. {\em Middle}: Light curve processed with the standard Warm
{\em Spitzer} pixel phase correction; this algorithm does not fully remove
the systematics. {\em Bottom}: Light curve corrected by the procedure
described in this paper. Below these light curves, we display the
periodograms of the corrected light curves (dashed, dotted curves), as compared with the
periodogram of the raw light curve (solid curve).}
\end{figure*}

\subsection{Complete combined optical/infrared dataset}
We assemble in Figure~\ref{iroptall} the entire 162-member disk bearing
dataset observed by {\em CoRoT} and {\em Spitzer}. Small black dots
are optical data, light grey points are 3.6~$\mu$m data, and dark grey
points are 4.5~$\mu$m data.

\begin{figure*}
\begin{center}
\includegraphics[scale=1.1]{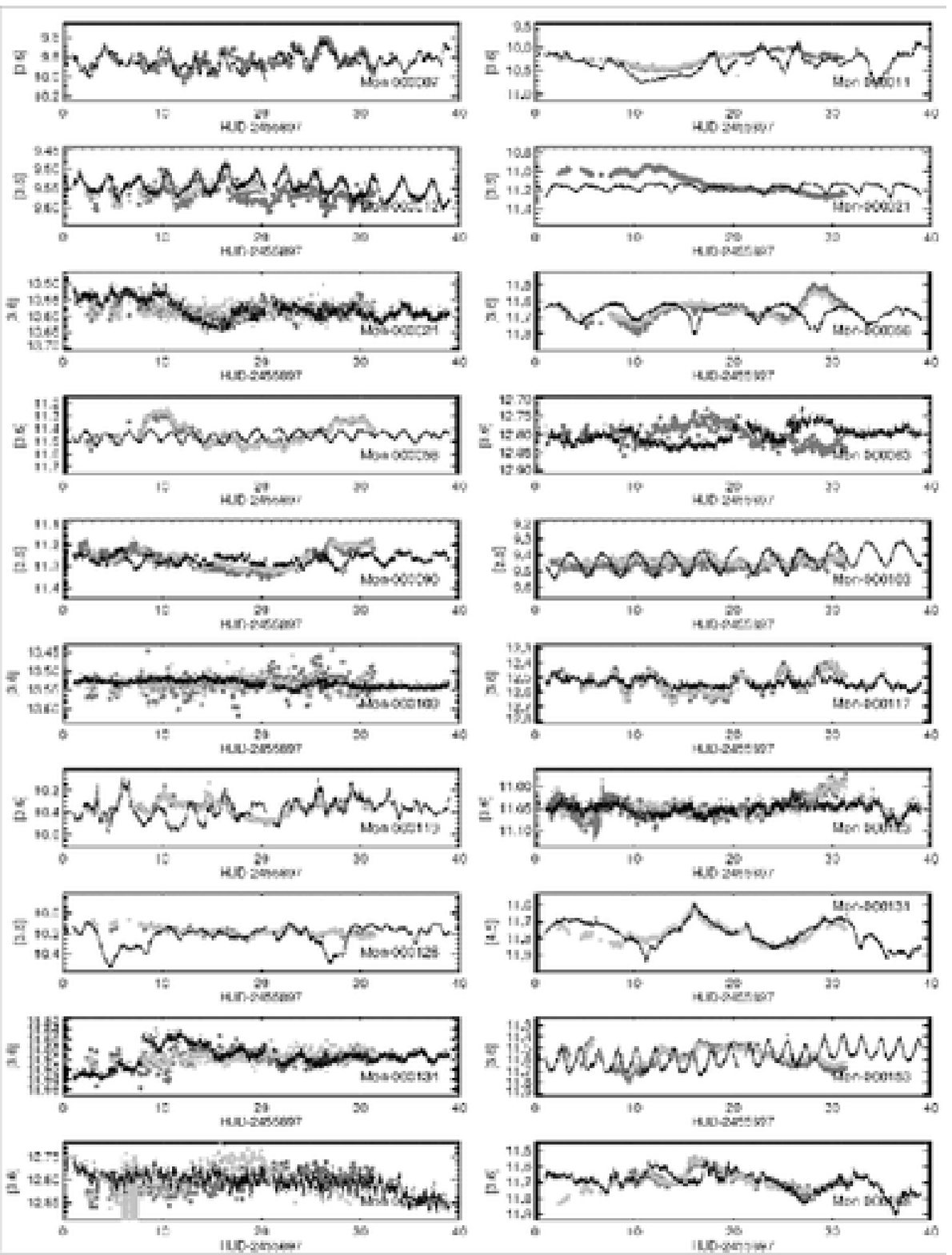}
\end{center}
\caption{\label{iroptall} Light curves for all stars in the 162-member
  disk bearing sample of this paper. Small black points are {\em CoRoT} data,
  light grey points are IRAC 3.6~$\mu$m data, and dark grey
  points are IRAC 4.5~$\mu$m data (sometimes hidden behind the
  3.6~$\mu$m points). More information is available in Tables 3 and 4;
objects are presented in the same order here. The entire figure is
available in the Online Journal.}
\end{figure*}

\addtocounter{figure}{-1}
\begin{figure*}
\begin{center}
\includegraphics[scale=1.1]{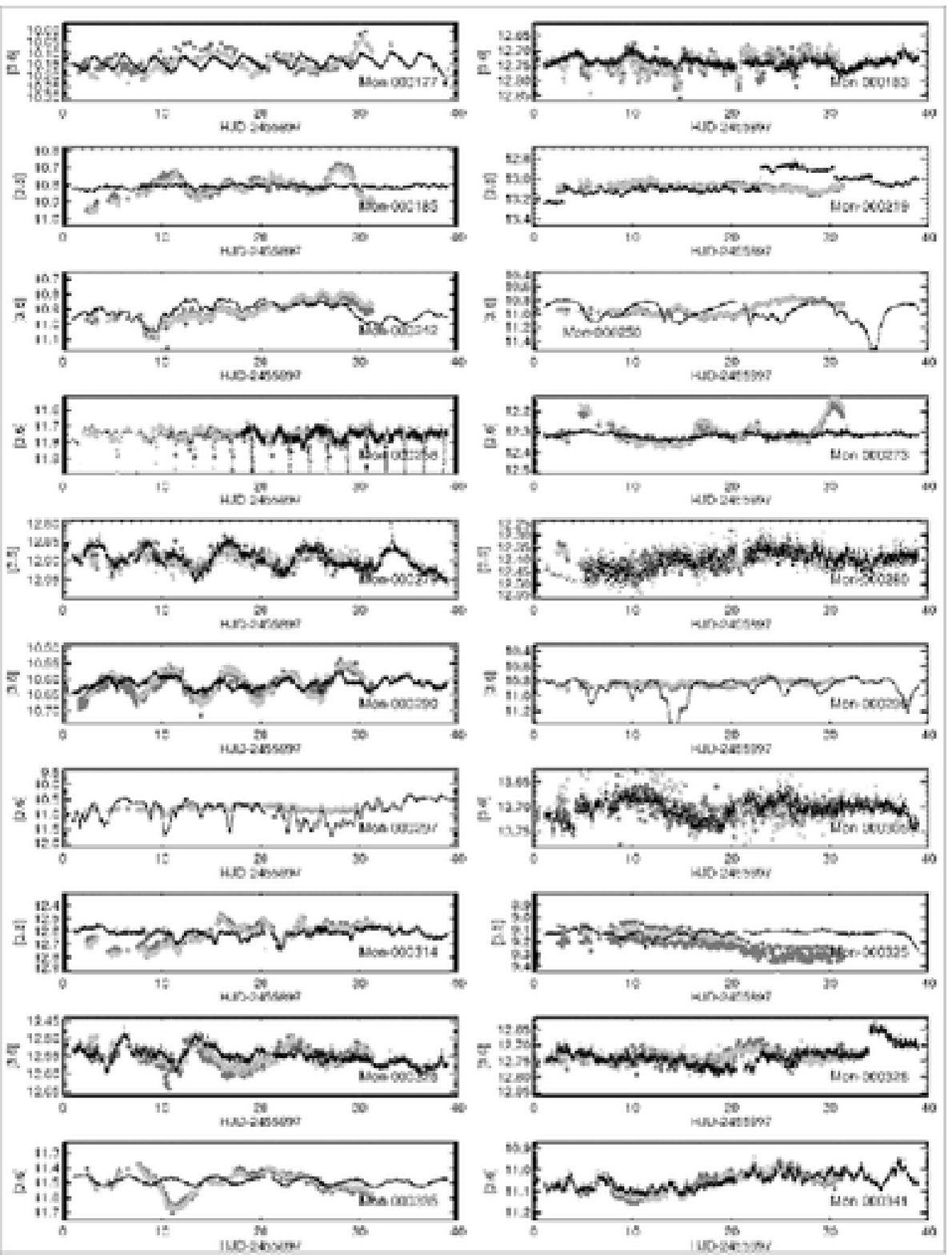}
\end{center}
\caption{--continued.}
\end{figure*}

\addtocounter{figure}{-1}
\begin{figure*}
\begin{center}
\includegraphics[scale=1.5]{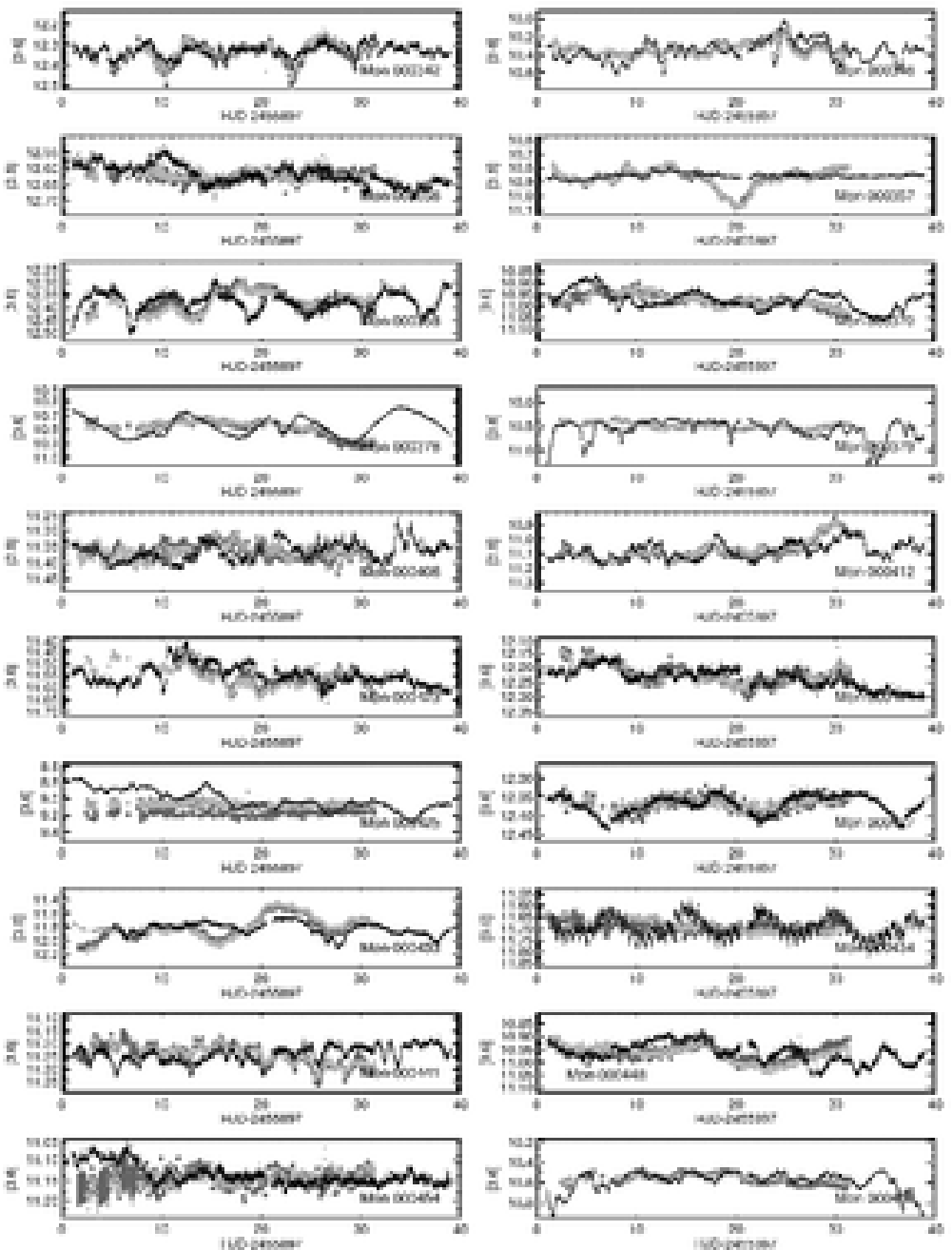}
\end{center}
\caption{--continued.}
\end{figure*}

\addtocounter{figure}{-1}
\begin{figure*}
\begin{center}
\includegraphics[scale=1.5]{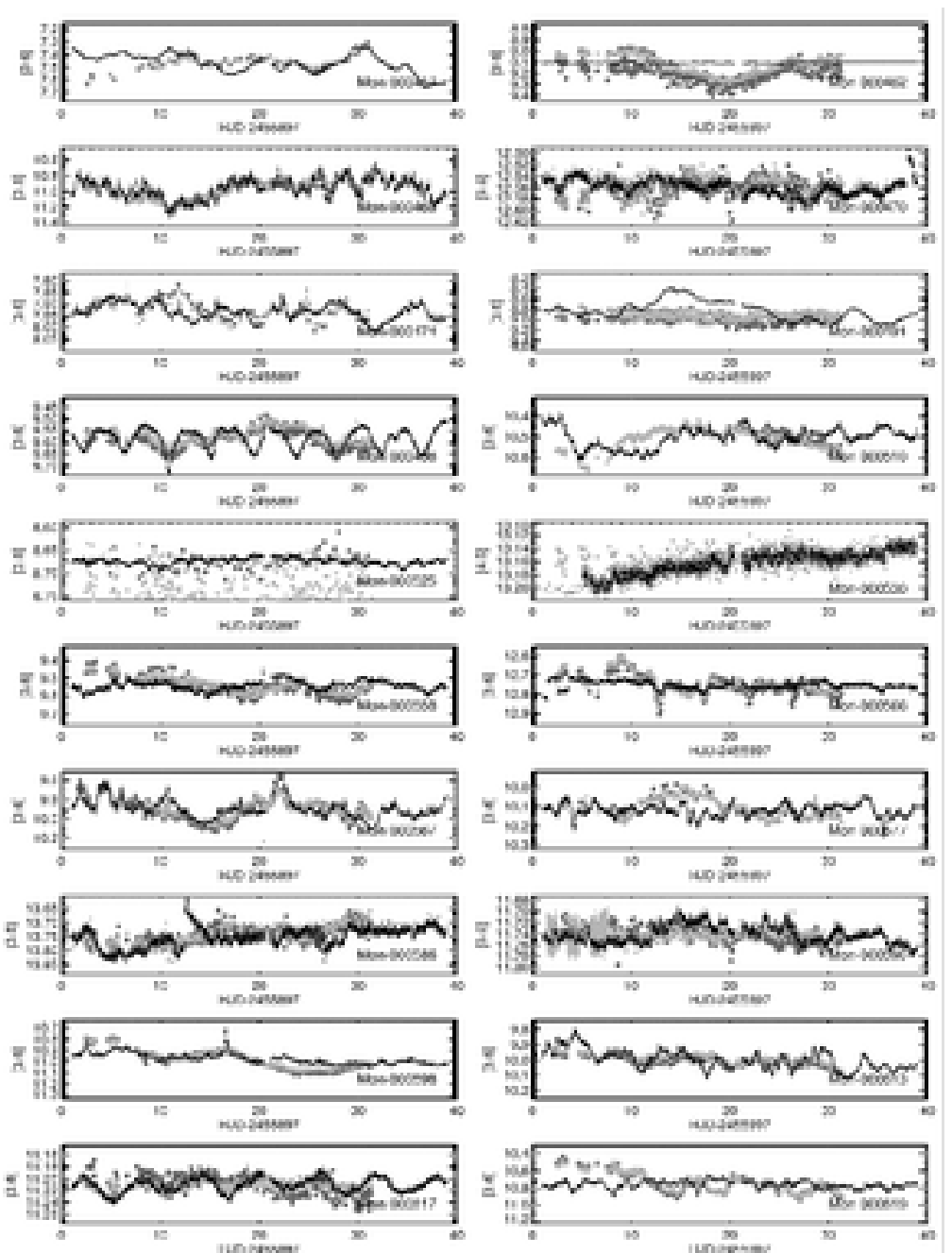}
\end{center}
\caption{--continued.}
\end{figure*}

\addtocounter{figure}{-1}
\begin{figure*}
\begin{center}
\includegraphics[scale=1.5]{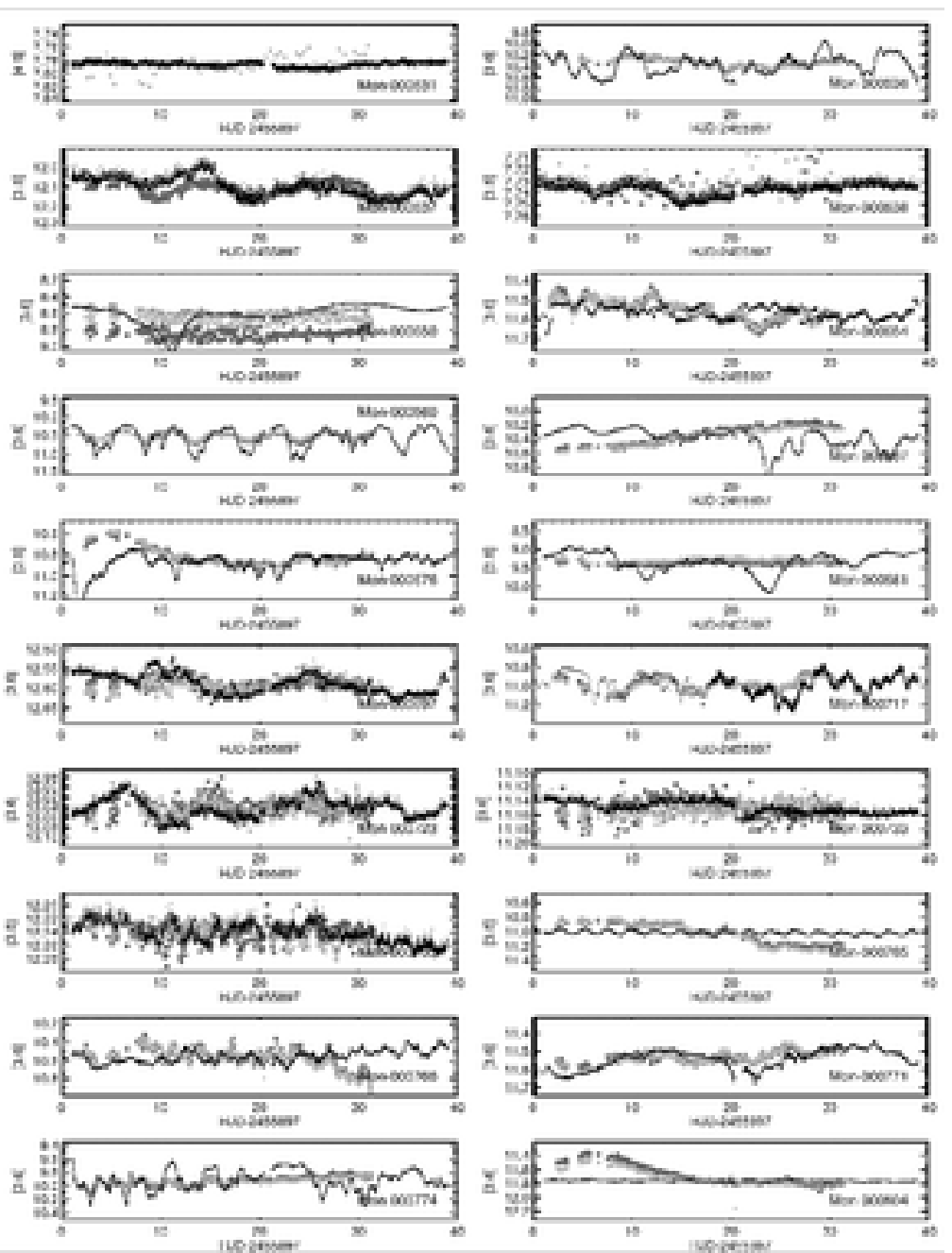}
\end{center}
\caption{--continued.}
\end{figure*}

\addtocounter{figure}{-1}
\begin{figure*}
\begin{center}
\includegraphics[scale=1.5]{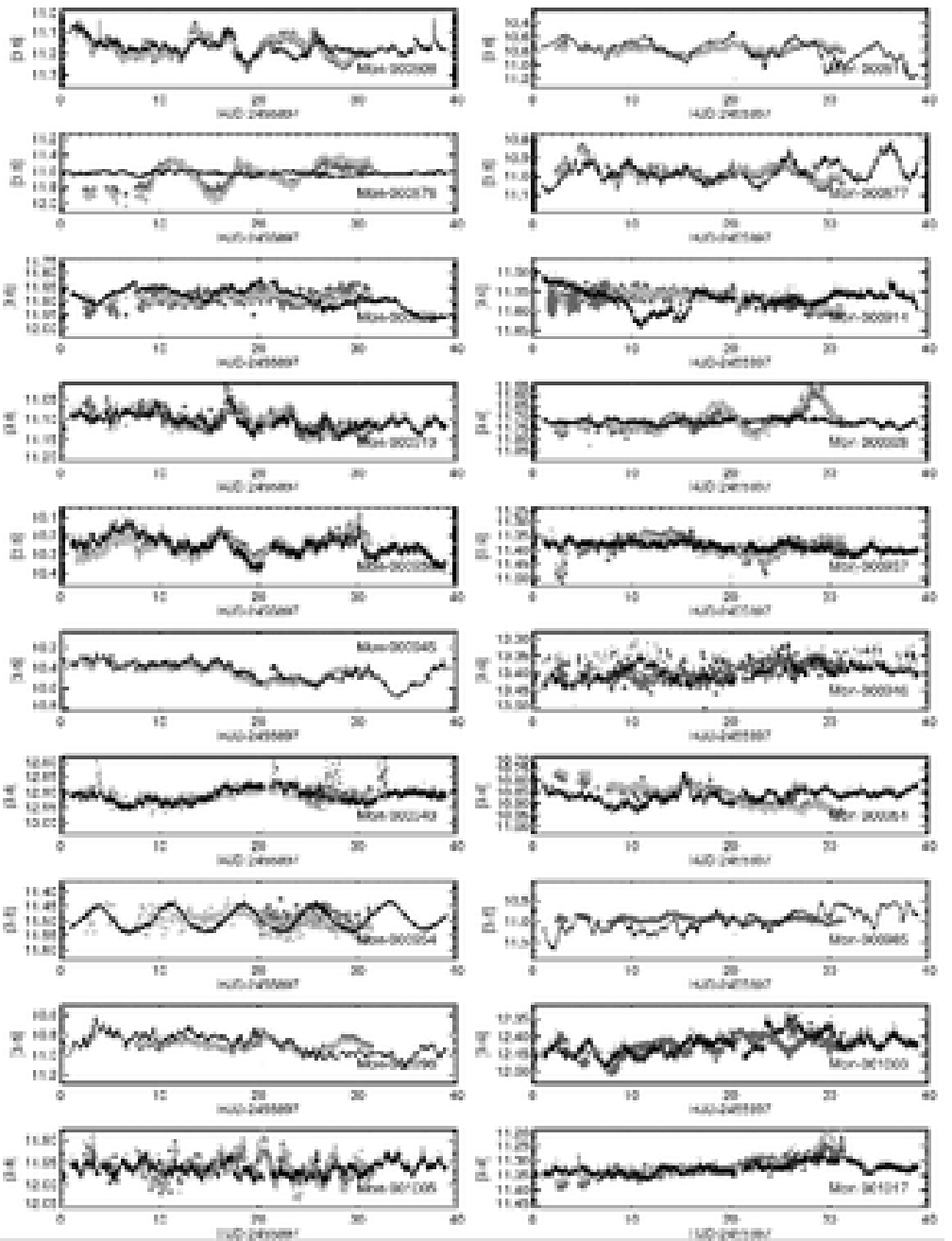}
\end{center}
\caption{--continued.}
\end{figure*}

\addtocounter{figure}{-1}
\begin{figure*}
\begin{center}
\includegraphics[scale=1.5]{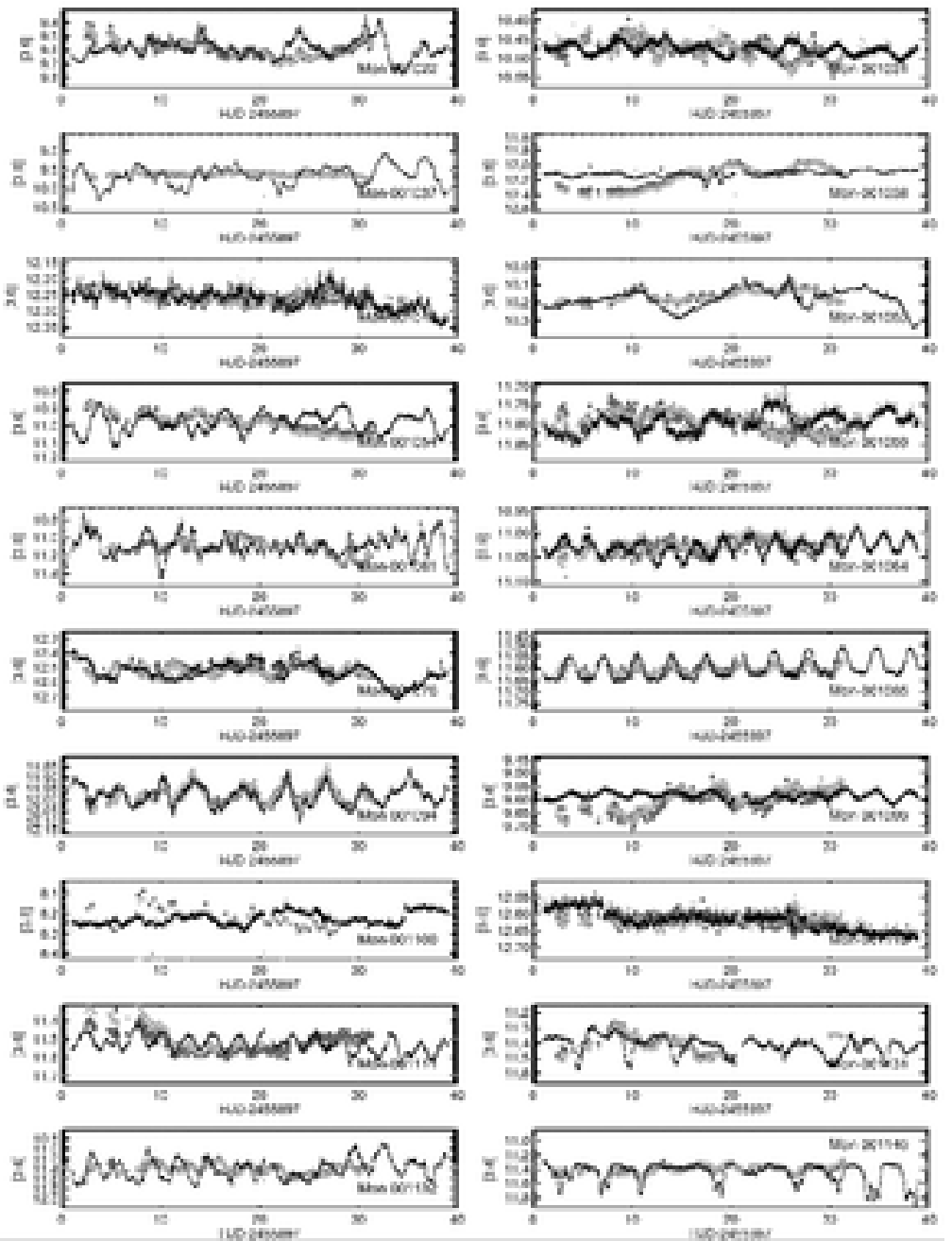}
\end{center}
\caption{--continued.}
\end{figure*}

\addtocounter{figure}{-1}
\begin{figure*}
\begin{center}
\includegraphics[scale=1.5]{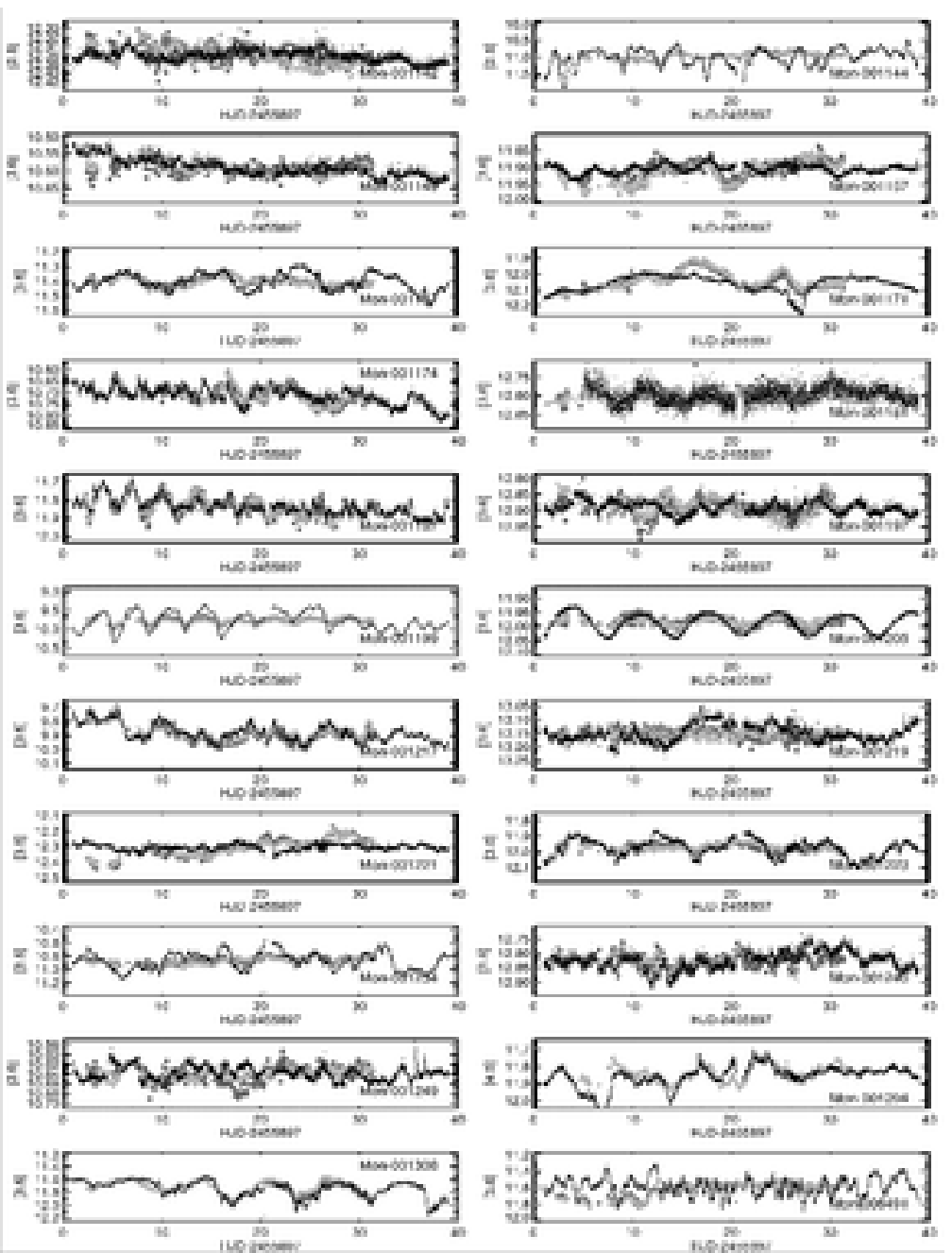}
\end{center}
\caption{--continued.}
\end{figure*}

\acknowledgements{This work is based in part on observations made with the {\em Spitzer} Space Telescope,
which is operated by the Jet Propulsion Laboratory, California Institute of
Technology under a contract with NASA. Support for this work was provided by NASA
through an award issued by JPL/Caltech. SHPA acknowledges support from
CNpq, CAPES and Fapemig. RG gratefully acknowledges funding support from NASA ADAP grants
NNX11AD14G and NNX13AF08G, and Caltech/JPL awards 1373081, 1424329, and
1440160 in support of Spitzer Space Telescope observing programs.
MMG acknowledges support from INCT-A/CNPq. KZ received a Pegasus Marie Curie Fellowship of the Research 
Foundation Flanders (FWO) during part of this work and received funding from the European Research Council 
under the European Community's Seventh Framework Programme (FP7/2007 -- 2013)/ERC grant agreement No. 227224 (PROSPERITY).}

\clearpage
\bibliographystyle{apj}
\bibliography{Morphpaper}

\clearpage
\LongTables
\addtocounter{table}{-3}
\begin{landscape}
\begin{deluxetable}{ccccccccccccccc}
\tabletypesize{\scriptsize}
\tablecolumns{10}
\tablewidth{0pt}
\tablecaption{\bf NGC 2264 members monitored with {\em CoRoT} and {\em
  Spitzer}}
\tablehead{
\colhead{Object} & \colhead{2MASS id} & \colhead{CoRoT id} & \colhead{J} & \colhead{H} &
\colhead{K} & \colhead{[3.6]} & \colhead{[4.5]} & \colhead{[5.8]} & \colhead{[8.0]} &
\colhead{[24]} & \colhead{Class} & \colhead{SpT} &
\colhead{H$\alpha$} & \colhead{Comp?} 
}
\startdata
CSI Mon-000007 & J06415304+0958028 & 223994721 & 12.22$\pm$0.02 & 11.53$\pm$0.03 & 11.18$\pm$0.02 & 10.12$\pm$0.00 & 9.74$\pm$0.00 & 9.32$\pm$0.00 & 8.75$\pm$0.00 & 6.20$\pm$0.03 & II & K7 & - & Y \\
CSI Mon-000011 & J06411725+0954323 & 223985009 & 12.83$\pm$0.02 & 12.05$\pm$0.03 & 11.55$\pm$0.02 & 10.49$\pm$0.00 & 10.07$\pm$0.00 & 9.69$\pm$0.00 & 8.82$\pm$0.00 & 5.60$\pm$0.01 & II & K7 & 58.30 & N \\
CSI Mon-000012 & J06410934+0956081 & 602099712 & 11.91$\pm$0.03 & 10.57$\pm$0.03 & 10.00$\pm$0.03 & 9.54$\pm$0.00 & 9.31$\pm$0.00 & 9.01$\pm$0.00 & 8.52$\pm$0.00 & 6.50$\pm$0.03 & II & K5 & 2.33 & Y \\
CSI Mon-000021 & J06405944+0959454 & 223980412 & 12.78$\pm$0.03 & 12.09$\pm$0.03 & 11.69$\pm$0.02 & 11.25$\pm$0.00 & 10.94$\pm$0.00 & 10.71$\pm$0.01 & 10.05$\pm$0.02 & 7.56$\pm$0.12 & II & K5 & 7.41 & N \\
CSI Mon-000024 & J06415684+0947451 & 603414392 & 14.74$\pm$0.04 & 14.07$\pm$0.05 & 13.81$\pm$0.05 & 13.61$\pm$0.01 & 13.55$\pm$0.01 & 13.39$\pm$0.07 & 12.84$\pm$0.13 & - & II/III & - & - & Y \\
CSI Mon-000056 & J06415315+0950474 & 223994760 & 12.96$\pm$0.02 & 12.30$\pm$0.02 & 12.00$\pm$0.03 & 11.72$\pm$0.00 & 11.39$\pm$0.00 & 11.04$\pm$0.01 & 10.29$\pm$0.01 & 7.87$\pm$0.06 & II & K5 & 1.80 & N \\
CSI Mon-000058 & J06420870+0941212 & 616895632 & - & - & - & 11.68 & 11.27 & 11.24$\pm$0.02 & - & 5.70$\pm$0.03 & flat & - & 94.00 & N \\
CSI Mon-000063 & J06411193+0959412 & 616826518 & 14.20$\pm$0.04 & 13.52$\pm$0.04 & - & 12.81$\pm$0.01 & 12.58$\pm$0.01 & 12.21$\pm$0.03 & 11.46$\pm$0.03 & 9.34$\pm$0.18 & II & M2.5 & 19.35 & N \\
\enddata
\tablecomments{We have assembled a list of NGC~2264 region members,
  candidates, and field stars, called the CSI~Mon catalog. Here we
  show only the Mon identification numbers of objects discussed in
  this paper. The {\em CoRoT} ids are from the SRa05 run. H$\alpha$ refers to the equivalent
 width of the H$\alpha$ emission line, and we adopt the values of \citet{2002AJ....123.1528R} and \citet{2005AJ....129..829D}. 
  Spectral types are from \citet{1956ApJS....2..365W}, \citet{2004AJ....127.2228M}, 
  and \citet{2005AJ....129..829D}, Objects with the disk class II/III
  are have a class III SED based on the slope, but have significant
  evidence of a weak disk based on their [3.6]-[8.0] colors. The ``comp''
  column denotes whether the object appears to have a companion within
  1\arcsec (``Y'' if yes). Detection is based on visual binarity or elongation of the
  PSF reported by \citet{2009AJ....138.1116S}, or spectroscopic indications of
  binarity via variable radial velocity measured by \citet{2006ApJ...648.1090F}.
  Objects with no known companion (``N'') may still be
  undetected binaries. *Objects with IRAC staring data are marked with 
 asterisks.}
\end{deluxetable}

\clearpage
\LongTables
\begin{deluxetable}{cccccccccccc}
\tabletypesize{\scriptsize}
\tablecolumns{10}
\tablewidth{0pt}
\tablecaption{\bf Variability properties of the NGC 2264 members
  monitored with {\em CoRoT} and {\em Spitzer}}
\tablehead{
\colhead{Object} & \colhead{RMS} & \colhead{RMS} & \colhead{RMS} & \colhead{(A)periodic?} &
\colhead{(A)periodic?} & \colhead{Timescale [d]} & \colhead{Timescale [d]}
& \colhead{Morph} & \colhead{Morph} &
\colhead{Stetson} & \colhead{Lamm} \\
\colhead{} & \colhead{(optical)} & \colhead{[3.6]} & \colhead{[4.5]} & \colhead{(optical)} & \colhead{(infrared)} &
\colhead{(optical)} & \colhead{(infrared)} & \colhead{(optical)} & \colhead{(infrared)} &
\colhead{(opt/IR)} & \colhead{class}
}
\startdata
CSI Mon-000007- & 0.080 & 0.100 & AP & AP & AP & 11.70 & 28.89 & B & S & 2.94* & - \\
CSI Mon-000011 & 0.209 & 0.139 & 0.103 & AP & AP & 15.98 & 24.67 & S & L & 2.75 & pv?/iv \\
CSI Mon-000012 & 0.023 & 0.023 & 0.022 & QP & AP & 2.94 & - & QPS & U & 0.65* & - \\
CSI Mon-000021 & 0.033 & 0.104 & - & QP & AP & 3.12 & 35.78 & QPD & L & 0.44 & - \\
CSI Mon-000024 & 0.030 & 0.017 & 0.016 & - & - & - & - & N & N & -0.00 & - \\
CSI Mon-000056 & 0.041 & 0.060 & 0.049 & QP & AP & 5.88 & 35.26 & QPD & L & 0.02 & - \\
CSI Mon-000058 & 0.032 & 0.077 & 0.085 & QP & AP & 2.13 & 15.43 & QPS & U & -0.02 & - \\
CSI Mon-000063 & 0.021 & 0.027 & - & - & AP & - & 16.67 & N & U & -0.01 & - \\
\enddata
\tablecomments{Variability status of stars in this paper. Mon id numbers marked with a - sign
 denote faint ($R>15.5$) variables that were detected only by eye. Columns five and six list whether the stars is aperiodic (``AP'') by the Stetson
  index or RMS measure, periodic (``P''), quasi-periodic (``QP''), or not detected as variable (``-''). Columns 7 and 8 are timescales, the 
  derivation of which depends on whether the object is (quasi-)periodic or  aperiodic. If (quasi-)periodic, then we list the timescale derived 
  from the autocorrelation and periodogram analysis (Section 6.2); if aperiodic, then we list the timescale derived from the PeakFind algorithm 
  (Section 6.5). In columns 9 and 10, we provide the optical and infrared morphologies, with the follow codes:
  S=stochastic, U=unclassifiable variable type, QPS=quasi-periodic
  symmetric, QPD=quasi-periodic dipper, QPB=quasi-periodic burster, L=long-timescale variable, N=non-variable,
  B=burster, D=dipper, EB=eclipsing binary, P=periodic,
  MP=multi-periodic. In column 11, we note the degree of
  optical/infrared correlation via the median running Stetson index;
  here we have averaged the values from 3.6~$\mu$m and 4.5~$\mu$m data
  (both against optical lightcurves). The last column indicates the variability type (``pv'' for periodic, and ``iv'' 
  for irregular) assigned by \citet{2004A&A...417..557L} and compiled in their Table 4. *Stetson indices marked with
  asterisks are larger, because those objects only had HDR mode observations and hence fewer datapoints.}
\end{deluxetable}
\clearpage
\end{landscape}

\end{document}